\pdfoutput=1     
\documentclass[12pt]{JINST}
\usepackage{epsfig}
\usepackage{amsbsy}
\usepackage{rotating}
\usepackage{cite}
\newcommand{\unit}[1]{\ensuremath{\mathrm{\,#1}}}
\renewcommand{\u}[1]{\unit{#1}}

\DeclareFontFamily{U}{euc}{}
\DeclareFontShape{U}{euc}{m}{n}{<-6>eurm5<6-8>eurm7<8->eurm10}{}%
\DeclareSymbolFont{AMSc}{U}{euc}{m}{n} 
\DeclareMathSymbol{\umu}{\mathord}{AMSc}{"16}

\newcommand{\Nhit}{N_\mathrm{hit}}

\def\FD3D{\mathrm{FD_{3D}}}     

\RequirePackage{lineno}
\usepackage{afterpage}
\usepackage{url}

\title{First results of the CALICE SDHCAL technological prototype}

\author{\centering 
\LARGE\bf The CALICE Collaboration
}

\author{\centering
Z.\,Deng,
Y.\,Li,
Y.\,Wang, 
Q.\,Yue, 
Z.\,Yang
\\ \it
Tsinghua University, Department of Engineering Physics.Beijing, 100084, P.R.
China
}

\author{\centering 
D.\,Benchekroun, 
A.\,Hoummada, 
Y.\,Khoulaki
\\ \it
Universit\'{e} Hassan II A\"{\i}n Chock, Facult\'{e} des sciences.\, B.P. 5366 Maarif, Casablanca, Morocco}

\author{\centering
D.\,Boumediene, C.\,Carloganu, V.\,Fran\c cais
\\ \it
Universit\'e Clermont Auvergne, Universit\'e Blaise Pascal, Universit\'e Blaise Pascal, CNRS/IN2P3, LPC, 4 Av. Blaise Pascal, TSA/CS 60026,
F-63178 Aubi\`ere, France
}

\author{\centering 
G.\,Cho, D-W.\,Kim, S.\,C.\,Lee, W.\,Park, S.\,Vallecorsa
\\ \it
Gangneung-Wonju National University, Gangnung 25457, South Korea
}
\author{\centering
S.\,Cauwenbergh, 
M.\,Tytgat,
A.\,Pingault,
N.\,Zaganidis
\\ \it
Ghent University, Department of Physics and Astronomy,
Proeftuinstraat 86, B-9000 Gent, Belgium
}

\author{\centering 
E.\,Brianne, 
A.\,Ebrahimi,
K.\,Gadow, 
P.\,G\"{o}ttlicher, 
C.\,G\"{u}nter, 
O.\,Hartbrich, 
B.\,Hermberg, 
A.\,Irles,
F.\,Krivan, 
K.\,Kr\"{u}ger, 
J.\,Kvasnicka, 
S.\,Lu, 
B.\,Lutz, 
V.\,Morgunov$^8$,
C.\,Neub\"{u}ser
A.\,Provenza,
M.\,Reinecke,  
F.\,Sefkow, 
S.\,Schuwalow,
H.L.\,Tran
\\ \it
DESY, Notkestrasse 85,
D-22603 Hamburg, Germany
}

\author{\centering  
E.\,Garutti, 
S.\,Laurien,
M.\,Matysek,
M.\,Ramilli, 
S.\,Schroeder
\\ \it
Univ. Hamburg,
Physics Department,
Institut f\"ur Experimentalphysik,
Luruper Chaussee 149,
22761 Hamburg, Germany
}

\author{\centering 
B.\,Bilki,
E.\,Norbeck$^8$,
D.\,Northacker,
Y.\,Onel
\\ \it
University of Iowa, Dept. of Physics and Astronomy,
203 Van Allen Hall, Iowa City, IA 52242-1479, USA
}

\author{\centering 
E.\,Cortina Gil,
S.\,Mannai
\\ \it
Center for Cosmology, Particle Physics and Cosmology (CP3)
Universit\'{e} catholique de Louvain, Chemin du cyclotron 2,
1320 Louvain-la-Neuve, Belgium
}

\author{\centering 
V.\,Buridon,
C.\,Combaret, 
L.\,Caponetto,
R.\,Et\'{e}, 
G.\,Garillot,
G.\,Grenier, 
R.\,Han$^1$
J.C.\,Ianigro,
R.\,Kieffer$^2$ ,
I.\,Laktineh,
N.\,Lumb, 
H.\,Mathez, 
L.\,Mirabito, 
A.\,Petrukhin,
A.\,Steen
\\ \it
Univ. Lyon, Universit\'{e} Lyon 1, 
CNRS/IN2P3, IPNL 4 rue E Fermi 69622,
Villeurbanne CEDEX, France
}

\author{\centering 
J.\,Berenguer~Antequera,
E.\,Calvo~Alamillo, 
M.-C.\,Fouz, 
J.\,Marin,
J.\,Puerta-Pelayo, 
A.\,Verdugo
\\ \it
CIEMAT, Centro de Investigaciones Energeticas, Medioambientales y Tecnologicas, Madrid, Spain 
}

\author{\centering 
N.\,Kirikova,  
V.\,Kozlov, 
P.\,Smirnov, 
Y.\,Soloviev
\\ \it
P.\,N.\, Lebedev Physical Institute,
Russian Academy of Sciences,
117924 GSP-1 Moscow, B-333, Russia
}

\author{\centering 
M.\,Chadeeva$^9$,
M.\, Danilov$^9$,
\\ \it
National Research Nuclear University 
MEPhI (Moscow Engineering Physics Institute)
31, Kashirskoye shosse,
115409 Moscow, Russia
}

\author{\centering 
M.\,Gabriel, 
P.\,Goecke,
C.\,Kiesling,
N.\,van\,der\,Kolk, 
F.\,Simon, 
C.\,Soldner, 
M.\,Szalay, 
L.\,Weuste
\\ \it
Max Planck Inst. f\"ur Physik,
F\"ohringer Ring 6,
D-80805 Munich, Germany
}

\author{\centering 
J.\,Bonis, 
B.\,Bouquet,    
P.\,Cornebise, 
Ph.\,Doublet$^5$,
M.\,Faucci-Giannelli$^6$,
T.\,Frisson,
G.\,Guilhem, 
H.\,Li$^7$,
F.\,Richard, 
R.\,P\"oschl, 
J.\,Rou\"en\'e, 
F.\,Wicek, 
Z.\,Zhang
\\ \it
Laboratoire de L'acc\'elerateur Lin\'eaire,
Centre d'Orsay, Universit\'e de Paris-Sud XI,
BP 34, B\^atiment 200,
F-91898 Orsay CEDEX, France
}

\author{\centering 
M.\,Anduze,
V.\,Balagura,
K.\,Belkadhi,
V.\,Boudry, 
J-C.\,Brient, 
R.\,Cornat,
M.\,Frotin,
F.\,Gastaldi, 
Y.\,Haddad$^3$, 
M.\,Ruan$^4$,
K.\,Shpak,
H.\,Videau,
D.\,Yu
\\ \it
 Laboratoire Leprince-Ringuet (LLR)  -- \'{E}cole Polytechnique,
 CNRS/IN2P3,
 Palaiseau, F-91128 France
}
\author{\centering   
S.\,Callier,
S.\, Conforti di Lorenzo, 
F.\,Dulucq, 
G.\,Martin-Chassard, 
Ch.\,de la Taille, 
L.\,Raux, 
N.\,Seguin-Moreau
\\  \it
 Laboratoire OMEGA  -- \'{E}cole Polytechnique,
 CNRS/IN2P3,
 Palaiseau, F-91128 France
}

\author{\centering              
D.\,Jeans,
S.\,Komamiya,
H.\,Nakanishi
\\ \it
Department of Physics, Graduate School of Science, The University of
Tokyo, 7-3-1 Hongo, Bunkyo-ku, Tokyo 113-0033, Japan}
\author{
\it
$^\star$ Corresponding author\newline
E-mail: \email{laktineh@in2p3.fr}
}

\author{\\
\llap{$^1$} Now at CAST, Beijing.\\
\llap{$^2$} Now at CERN, Geneva.\\ 
\llap{$^3$} Now at Imperial College, London.\\
\llap{$^4$} Now at IHEP, Beijing.\\
\llap{$^5$} Now at IUT d'Orsay (Universit\'e Paris-Sud).\\
\llap{$^6$} Now at Royal Holloway, University of London. \\
\llap{$^7$} Now at University of Virginia, Charlottesville, USA.\\
\llap{$^8$} Deceased.\\
\llap{$^9$} Also at P.\,N.\, Lebedev Physical Institute.\\
}

\abstract{ The CALICE Semi-Digital Hadronic Calorimeter (SDHCAL) prototype, built  in 2011,  was exposed to beams of hadrons, electrons and muons  in two short periods in 2012 on two different beam lines of the CERN SPS.  The prototype with its 48  active layers, made of Glass Resistive Plate Chambers  and their embedded readout electronics,  was run in  triggerless and power-pulsing mode.  The performance of the SDHCAL during the test beam was found to be  very satisfactory with an efficiency exceeding 90\% for almost all of the 48 active layers.  
A  linear response (within  $\pm$ 5\%) and a good energy resolution are obtained for a large range of  hadronic energies  (5--80~GeV) by applying appropriate calibration coefficients to the collected data for both the Digital (Binary) and the Semi-Digital (Multi-threshold) modes of the SDHCAL prototype.  The Semi-Digital mode shows better performance at energies exceeding 30~GeV.}

\begin{document}


\section{Introduction}

The CALICE SDHCAL prototype\cite{Prototype} was conceived for two purposes. The first is to confirm that  highly-granular gaseous hadronic calorimeters are capable of measuring  hadronic energy with good resolution while providing an excellent  tracking tool for Particle Flow Algorithms (PFA)~\cite{PFA}~\cite{morgunov}. The second aim is to demonstrate that such calorimeters are compatible with the requirements of the future International Linear Collider (ILC) detectors in terms of efficiency, compactness and power consumption. Prototypes that fulfill such requirements are called technological prototypes. The SDHCAL is the first technological prototype among a family of prototypes of high-granularity calorimeters developed by the CALICE collaboration that includes electromagnetic~\cite{Si-W}~\cite{Sc-W} and hadronic~\cite{AHCAL}~\cite{DHCAL} calorimeters.   

In the following, after a short description of the prototype and of the beams used to test it in the first section, 
the event building procedure and the data quality will be presented in the second section. The simulation tools used to verify the hadronic event selections are briefly discussed in section~3.  In section~4, particle identification is discussed, and  hadronic shower selection is detailed. In section~5,  the stability of the Glass Resistive Plate Chambers (GRPC) used as active medium in the SDHCAL is investigated and a description of a correction method  to equalize the SDHCAL response in time when exposed to  high beam intensity is given. In section~6, the linearity and resolution of different scenarios for  the    energy reconstruction of hadronic showers is presented and results discussed. 

\subsection{Prototype description}

\begin{figure}[htp]
\begin{center}
\includegraphics[width=0.9\textwidth]{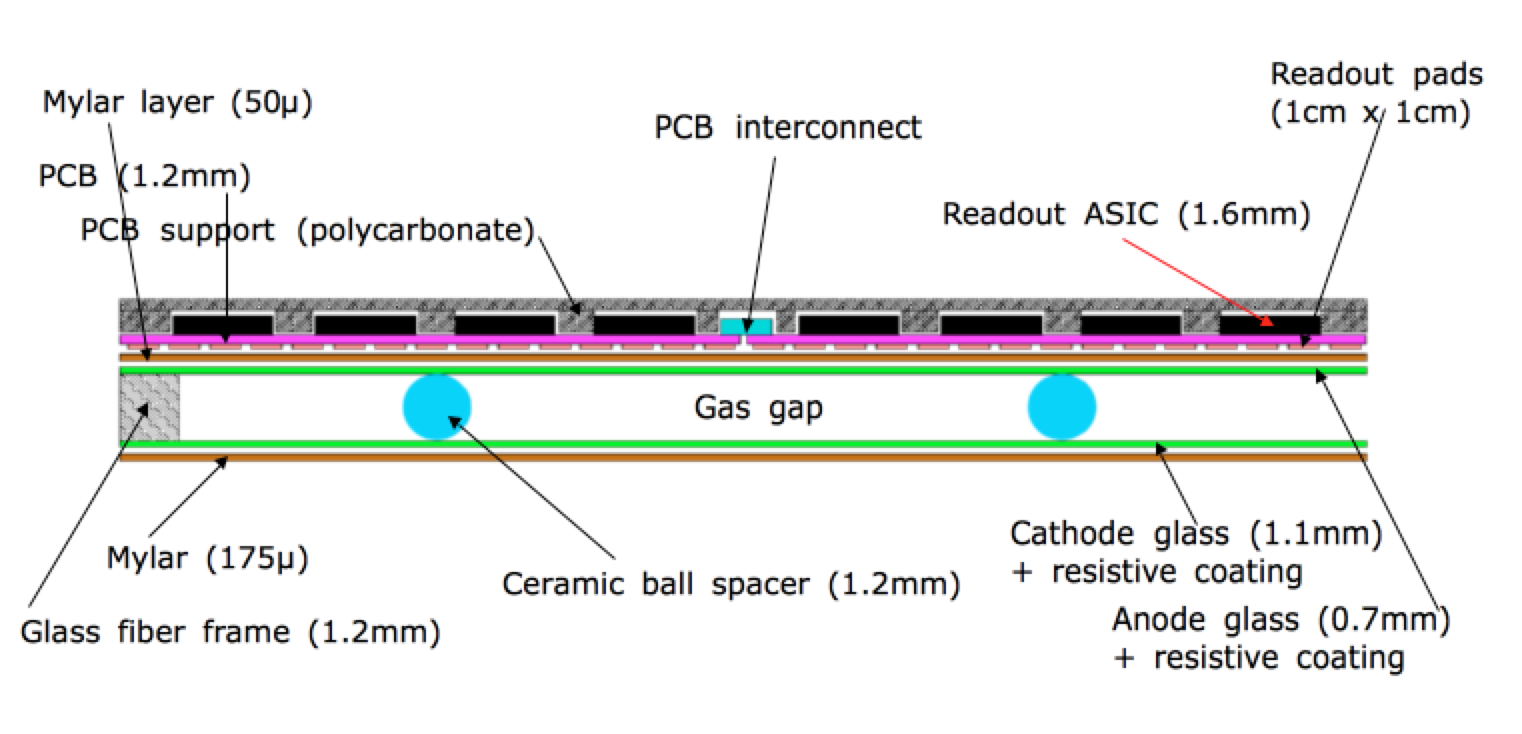}
\caption{A schematic cross-section of  a SDHCAL active layer (not to scale).}
\label{fig.scheme_GRPC}
\end{center}
\end{figure}

The SDHCAL is a sampling hadronic calorimeter. It comprises 48 active layers. Each of these layers is made of a $1\times1$ m$^2$ GRPC.  The GRPC signal  is  read out through 9216 pick-up pads of $1\times1$~cm$^2$ each. 
The pads are located on one face of an electronics board which hosts 144 HARDROC ASICs~\cite{hardroc} on the other face. A HARDROC ASIC has 64 channels each providing a three-threshold readout.
One electronic board is built by soldering three slabs each covering a third of the detector surface.
 The active layer, made of the GRPC and the electronic board, is put inside a cassette made of two 2.5~mm thick stainless steel walls.  This cassette protects the active layer and facilitates its handling. It also keeps the pick-up pads of the electronic board in contact with the GRPC and it constitutes a part of the calorimeter absorber.  
 The total thickness of a cassette is 11~mm of which 6~mm are occupied by the active layer thickness that includes the GRPC (3~mm), and the readout electronics (3~mm). 
A cross-section of a SDHCAL active layer is shown in Fig.~\ref{fig.scheme_GRPC}.

 One edge of the cassette also hosts three Detector InterFace (DIF) cards which transfer the acquisition commands received through HDMI cables  to the ASICs of each slab and collect the data received from these ASICs before forwarding them using the USB protocol to the acquisition stations.
 
 The 48 cassettes are inserted into a self-supporting mechanical structure. The structure is built using 1.5~cm thick stainless steel plates with a gap of 13 mm between two consecutive plates to allow easy insertion of the cassettes. 
A SDHCAL layer made of one cassette and one 1.5~cm thickness stainless steel plate corresponds to about 0.12 interaction lengths ($\lambda_I$) or 1.14 radiation lengths ($X_0$) respectively.
 A triggerless  data acquisition mode was used in the test beam. In this mode, data are collected after an acquisition starts following a command  sent to the ASICs. 
When  the memory of one of the ASICs is full\footnote{The HARDROC ASIC memory allows up to 127 events to be recorded.}, a RAMfull command is sent to all ASICs and the acquisition is stopped to allow the readout of the data recorded in all ASICs. 
The acquisition restarts automatically upon the completion of the data transfer. During the transfer no new data are collected. 
The heating due to the power consumption  of more than 440000 channels of the prototype inevitably leads to an increase of the prototype temperature resulting in a change of the GRPC gain and an increase of the noise. Although the GRPC gain can be controlled to some extent by varying the high voltage applied to the GRPC, the noise  could not be easily reduced. To solve these problems a power-pulsing mode was used. This mode places the electronics in an idle mode during the time period separating two beam spills. In the case of the SPS beam cycle in place in 2012 this amounts to a reduction factor of five of the ASIC power consumption (approximately a  nine second spill every 45 seconds). 
To further reduce the heating effect, a simple cooling system was used. 
Copper tubes, through which water at $10^ \circ$C was circulated, were put in contact with the two lateral sides of the calorimeter.  More details on the SDHCAL prototype can be found in~\cite{Prototype}.

\subsection{CERN SPS beam data samples} 

\begin{figure}[h]
\begin{center}
\includegraphics[width=1.0\textwidth]{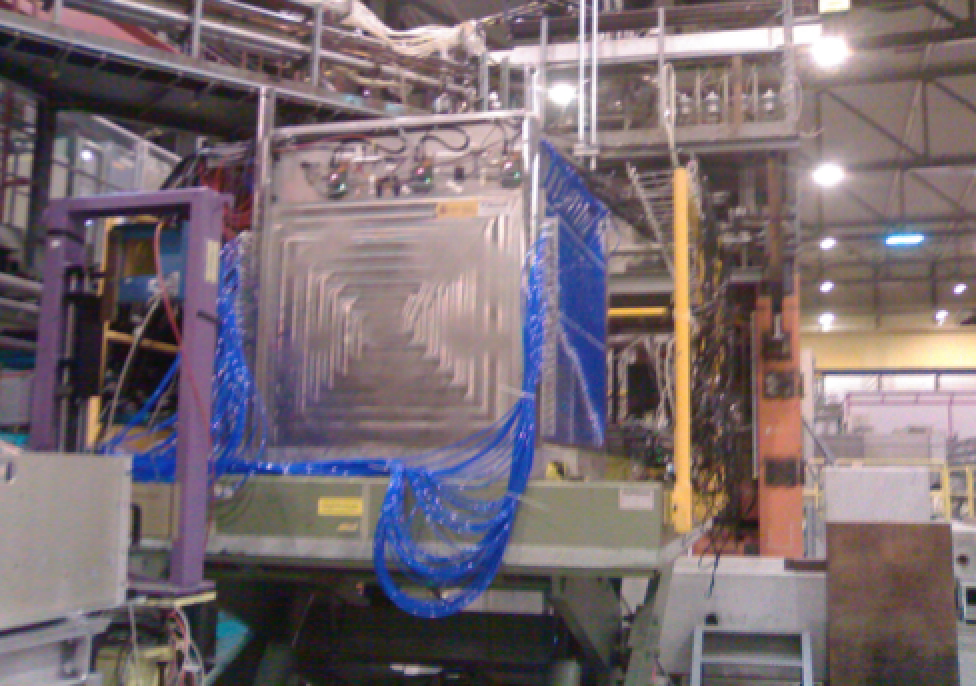} 
\end{center}
\caption{The SDHCAL prototype at the SPS test beam area.}
\label{fig:PrototypeAtSPS}
\end{figure}

The SDHCAL prototype (Fig.\ref{fig:PrototypeAtSPS}) was exposed to pions, muons and electrons  in  the CERN H6 beam line of the SPS for two weeks in August-September 2012 and for  additional two weeks in the H2 beam line in November 2012. 

Since the efficiency of RPCs with electrodes made of float glass,  such as those of the SDHCAL prototype, decreases at high particle rate \cite{jose}  \cite{HRate}  the beam optics was set up to enlarge the beam size and have it as flat as 
possible while reducing its  intensity. This  was intended to  achieve the optimal  GRPC efficiency, and at the same time to collect  
as much statistics as possible.   To check the effect of the beam intensity, the efficiency of reconstructing the beam muons 
was used to ensure that no reduction was observed compared to the efficiency obtained using a very low intensity muon beam as discussed in Sect.~\ref{sec.EvBuild_DQcheck}.  
Consequently,  only runs with particle rates smaller than 1000 particles/spill (i.e. less than $100$~Hz/cm$^2$) were selected.  Within these conditions, several energy points were studied. Positively-charged hadrons of 5, 7.5, 10, 15,  20, 25, 30, 40, 50, 60, 70, 80~GeV  
and positrons of  10, 20, 30, 40, 50, 60 and 70~GeV  were studied in the H6 beam test. In the  H2 beam test  only negative pion and electron beams were used and the energy points studied were 10, 20, 30, 40, 50, 60, 70, 80~GeV. 
Only a few dedicated muon runs were taken to check  the detector stability during the beam test. The pion runs contained also an important contribution of muons resulting from pion decays in flight which could be used to monitor the detector performance. 
To reduce the electron contamination of the hadron runs, a 4 mm upstream lead filter was used.  
The use of this filter was found to be rather effective at high energy (E > 20~GeV). At lower energies, where the contamination of the  hadron beam by electrons was high, it was observed from examining the event shapes\footnote{Showers associated to electron events are more compact both longitudinally and transversely than those of the hadron ones.  They start showering in the first layers and are almost free of any track segments while hadron events have frequently such segments and start, on average, showering after the electron ones due to the fact that in the SDHCAL prototype the $\lambda_I$ value is almost ten times higher than the $X_0$ one.} that some contamination from electrons was still present.  In the electron samples almost no contamination by pions was observed. 

In the H6  beam line only positive pions were available. The contamination of this pion beam by protons was present at energies above  20~GeV~\cite{tilecal1}.  Although the electromagnetic component of the showers produced by protons differs slightly from that of pions, one does not expect to observe sizable differences in the behavior of the hadronic showers produced by the two species at low and moderate energies. At high energies differences could also arise because of the shorter interaction length of protons compared to that of pions~\cite{interaction-length1}~\cite{interaction-length2}.  Showers produced by protons are on average more contained than those produced by pions.  More hits are thus expected in showers produced by protons compared to pions of the same energy in the SDHCAL prototype.

 
\subsection{Prototype running operation} 
\label{sec.prot}

An important feature of the SDHCAL readout is the presence of three thresholds. The aim of using the threshold information is not to measure the energy deposit in each pad but an attempt to distinguish between pads crossed by few, many or very many charged particles.  The information of the three thresholds is coded in two bits.  The threshold values were fixed to 110~fC, 5~pC and 15~pC respectively, the average MIP-induced charge being around 1.2~pC with a rather large spread~\cite{ranhan}. The choice of these values was motivated by an optimisation of the energy reconstruction of hadronic showers based on preliminary and rather simple simulation models of the SDHCAL prototype~\cite{calor2010}.   Of the prototype's 6912 ASICs, seven were switched off.  Three of them failed during the preparation tests in the laboratory, and were not replaced. The other four ASICs were found to be noisy and were masked.  No additional ASIC failed during the data taking or between the test beam periods.  In addition, the same electronics  gain of unity was used for all the channels.  This was done in order to see the performance that this technology can reach without any correction. The gas mixture used to run the GRPC was made of  TetraFluoroEthane (TFE, 93\u{\%}),  $\mathrm{CO_2}$ (5\u{\%}) and $\mathrm{SF_6}$ (2\u{\%}). The high voltage applied on the GRPC was 6.9 kV independently of the ambient pressure and temperature\footnote{Variations of ambient pressure and temperature lead to gas gain variations similar to that of the HV.}.

\section{Event building and data quality monitoring}
\label{sec.EvBuild_DQcheck}

\begin{figure}[htp]
\begin{center}
\includegraphics[width=0.6\textwidth]{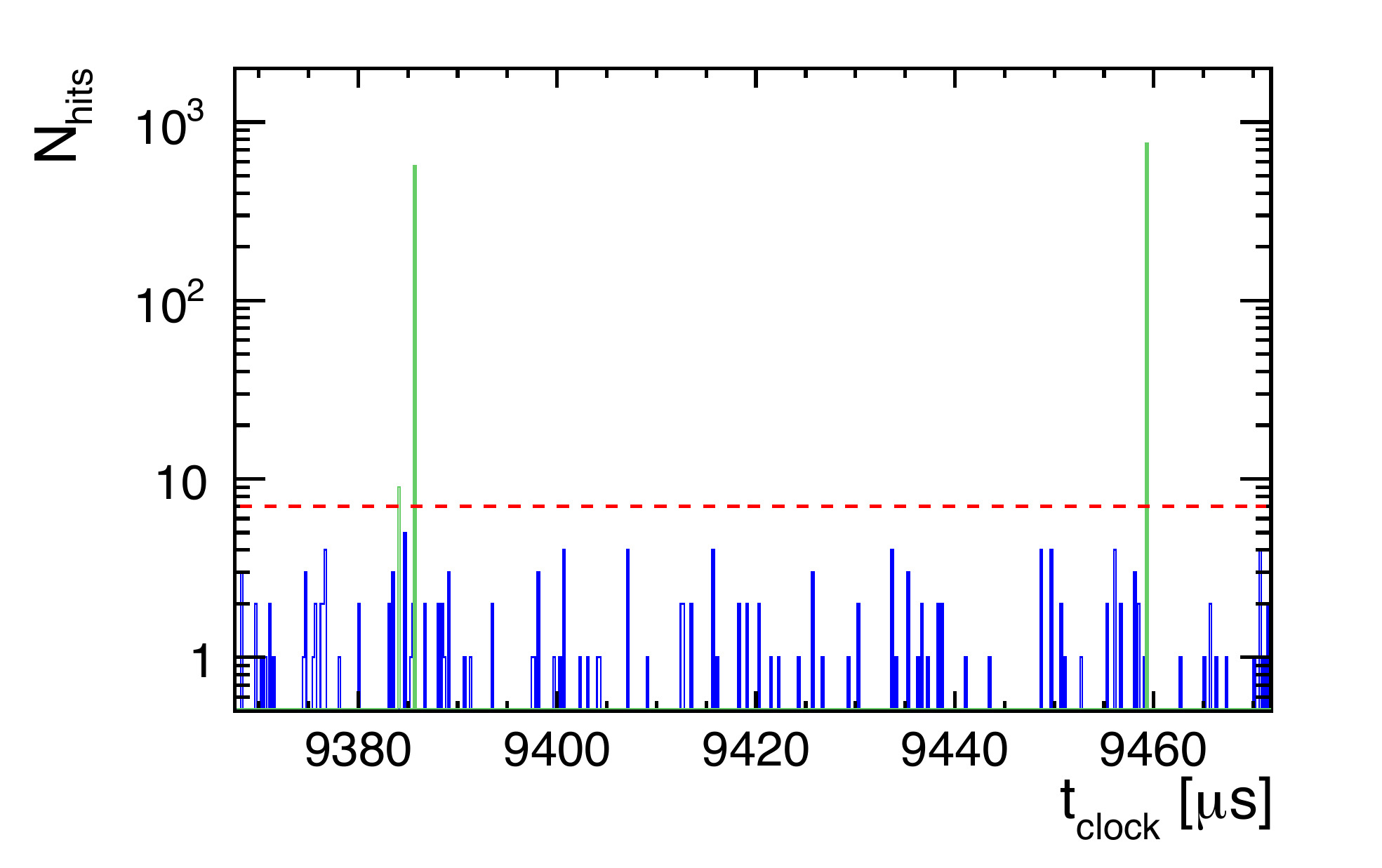}
\caption{Hit time spectrum for a 40~GeV pion beam run. Each bin corresponds to one clock tick of the detector's electronics (200 ns).
The green lines show the physics events selected by time clustering, i.e. events above the red dashed line. 
}
\label{fig.time_histo}
\end{center}
\end{figure}

With the use of the triggerless acquisition mode,  the collected data include not only the information of the fired pads (hits) 
resulting from the interaction of the beam particles (pions, electrons, muons, ...) but also those due to cosmic rays and noise. 
To select hits related to the incoming beam particles, a time clustering procedure is used.  The time of the hits is recorded  using a time-stamp whose counter is incremented every 200 ns (the ASIC internal clock period).  A histogram of hit time is built for each acquisition readout with a bin-width set to the time-stamp value of 200 ns. In Fig.\ \ref{fig.time_histo}  such a histogram is shown.  It includes 40~GeV pions, muons from the beam and cosmic rays as well as noise.   Only histogram bins with more than seven hits are then used to initiate the time clustering process.  
This choice allows noise events to be rejected while eliminating only a negligible fraction of hadronic showers produced by pions 
of energy larger than 5~GeV. The time clustering  of muon, electron and pion events shows that hits belonging to a local maximum as well as those of the two adjacent  time bins are sufficient to build the whole event.  No hit is allowed to belong to two different events and events with common hits are rejected.
 Besides the time occurrence the only information to be used in the following analysis is the space coordinates of the hits 
determined by the location of the fired pad and the threshold coding (either 1, 2 or 3).

\begin{figure}[htp]
\begin{center}
\includegraphics[width=0.6\textwidth]{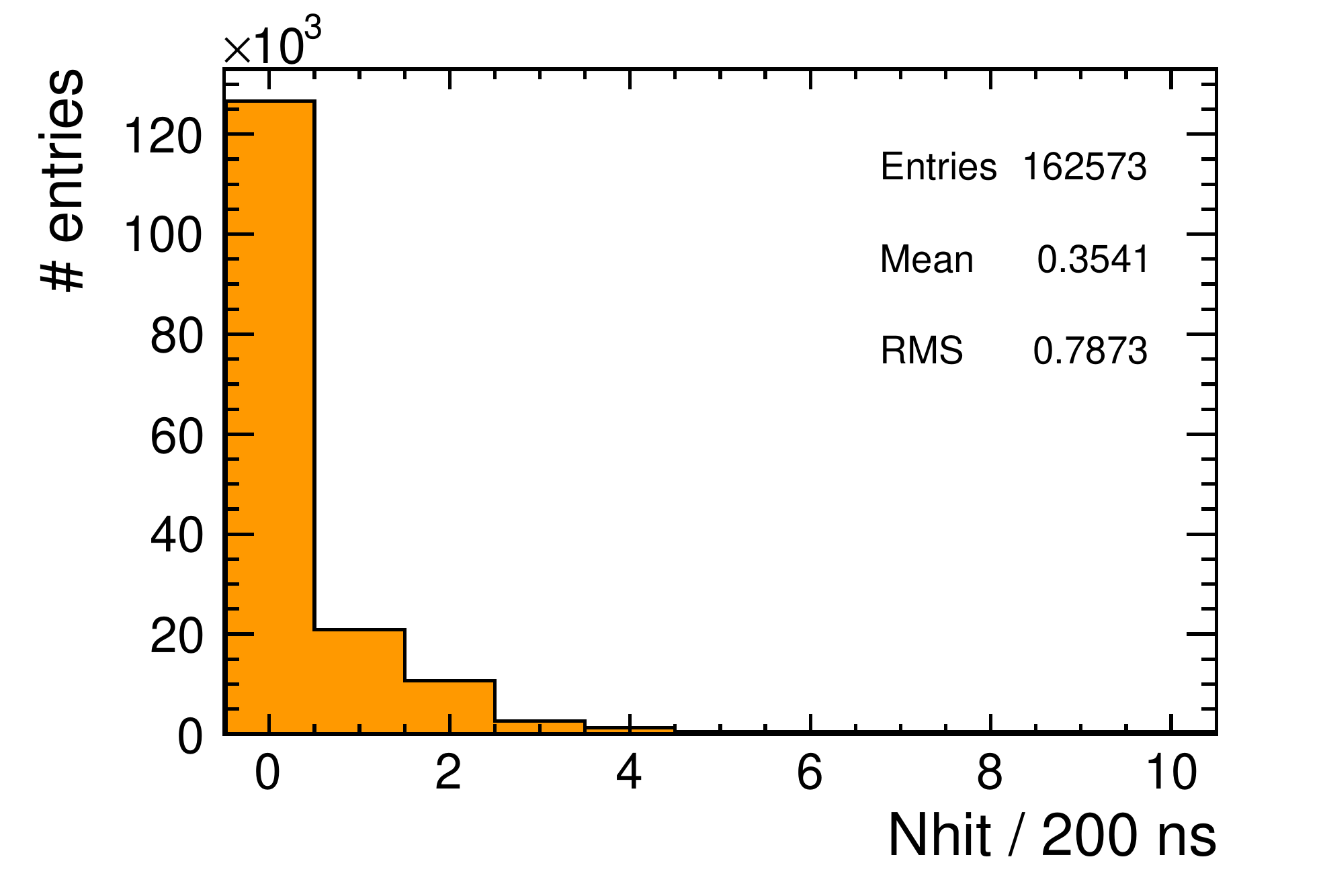}
\caption{Distribution of the number of noise hits in a time slot of 200 ns (one clock tick) for the whole detector. 
  An average of 0.35 hits / 200\u{ns} is found for the complete detector.}
\label{fig.histo_noise}
\end{center}
\end{figure}

Among the selected events there are a few  which are clearly due to electronic noise. 
These events are characterized by the occurrence of many hits  belonging to the same electronic slab, and sometimes to the whole electronic board made 
of three slabs.  These coherent-noise events are easily identified since the hits are concentrated in one layer, in contrast to a particle interaction in the calorimeter. They are probably related to grounding problems in some layers, and are removed from the selection.
Once the hits of the physical events and those associated to the coherent noise  are identified the remaining hits are used to estimate the noise rate. 
Figure~\ref{fig.histo_noise} shows a distribution of the number of noise hits recorded by the entire prototype per time slot (200~ns). From this distribution, one can estimate the average number of incoherent noise hits in one physical event (three time slots) to be around $3 \times 0.35 $  hits. 

To monitor the calorimeter performance, the efficiency ($\varepsilon$) and  hit multiplicity ($\mu$) of each of the 48 layers are estimated using the muons present in hadron samples. 
To study the efficiency of one layer, tracks are built using the hits of the other layers.  To build a track, hits of each layer are grouped in 
clusters if their cells share an edge. 
Isolated clusters which are at least 12~cm away from other clusters of any layer are dropped. 
Tracks are then built from all the selected clusters excluding the layer under study.
Tracks are required to have clusters in at least seven layers\footnote{\samepage The tracks associated to the muons used in this study cross almost all the layers of the prototype.}.
The layers should be on both sides of the studied layer except for the first and last layer.
The $\chi^2$ of the constructed track is then estimated \footnote{\samepage The cluster $x$ (resp. $y$) position uncertainty used to compute the $\chi^2$ is taken as  $N_{c}/\sqrt{12}$~cm where $N_c$ is the number of the cluster's hits projected on the $x$ (resp. $y$) axis.  x and y axes are respectively the horizontal and the vertical axis parallel to the prototype layers.}.

\begin{figure}[htp]
\begin{center}
\includegraphics[width=0.43\textwidth]{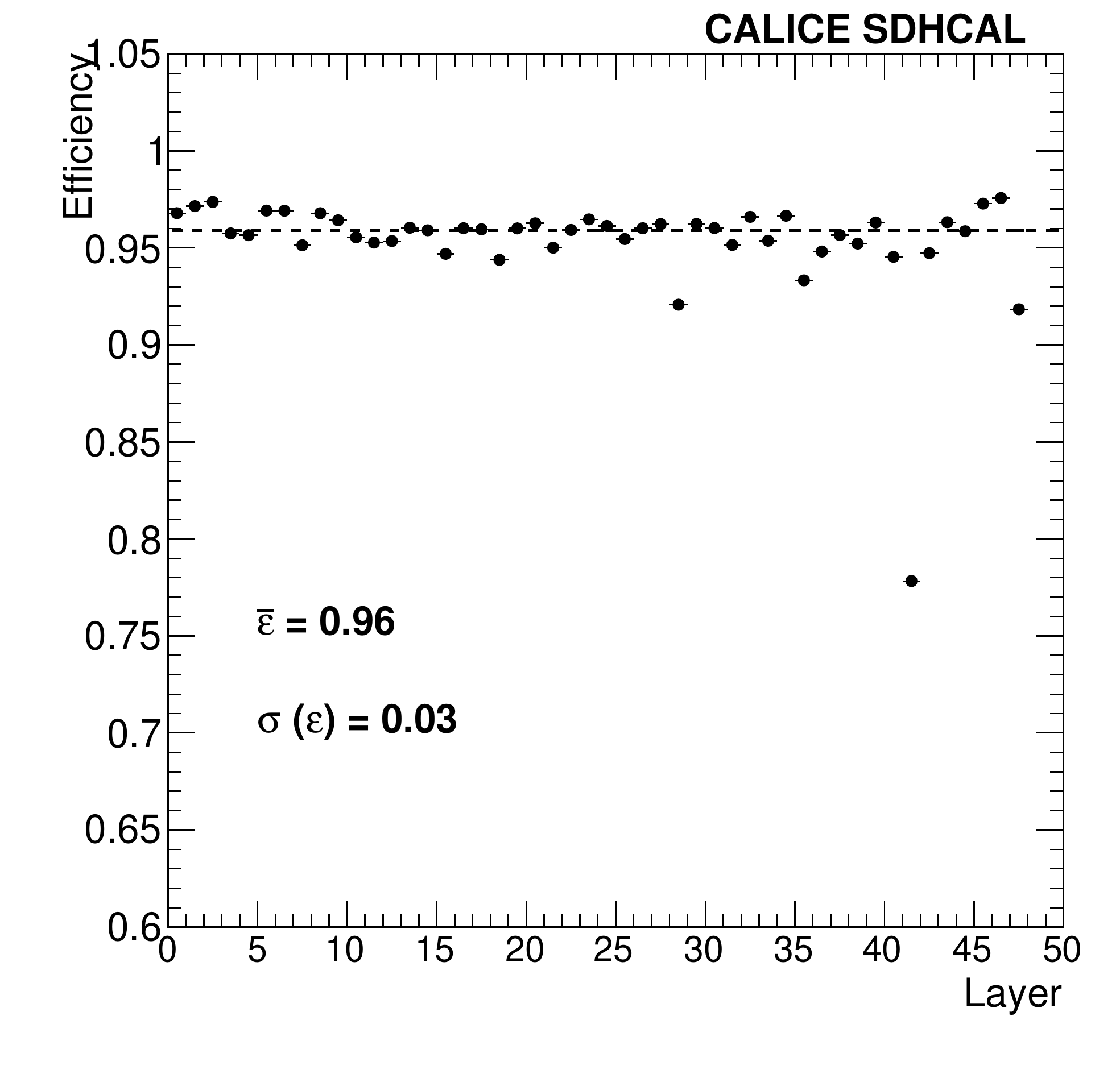}
\includegraphics[width=0.43\textwidth]{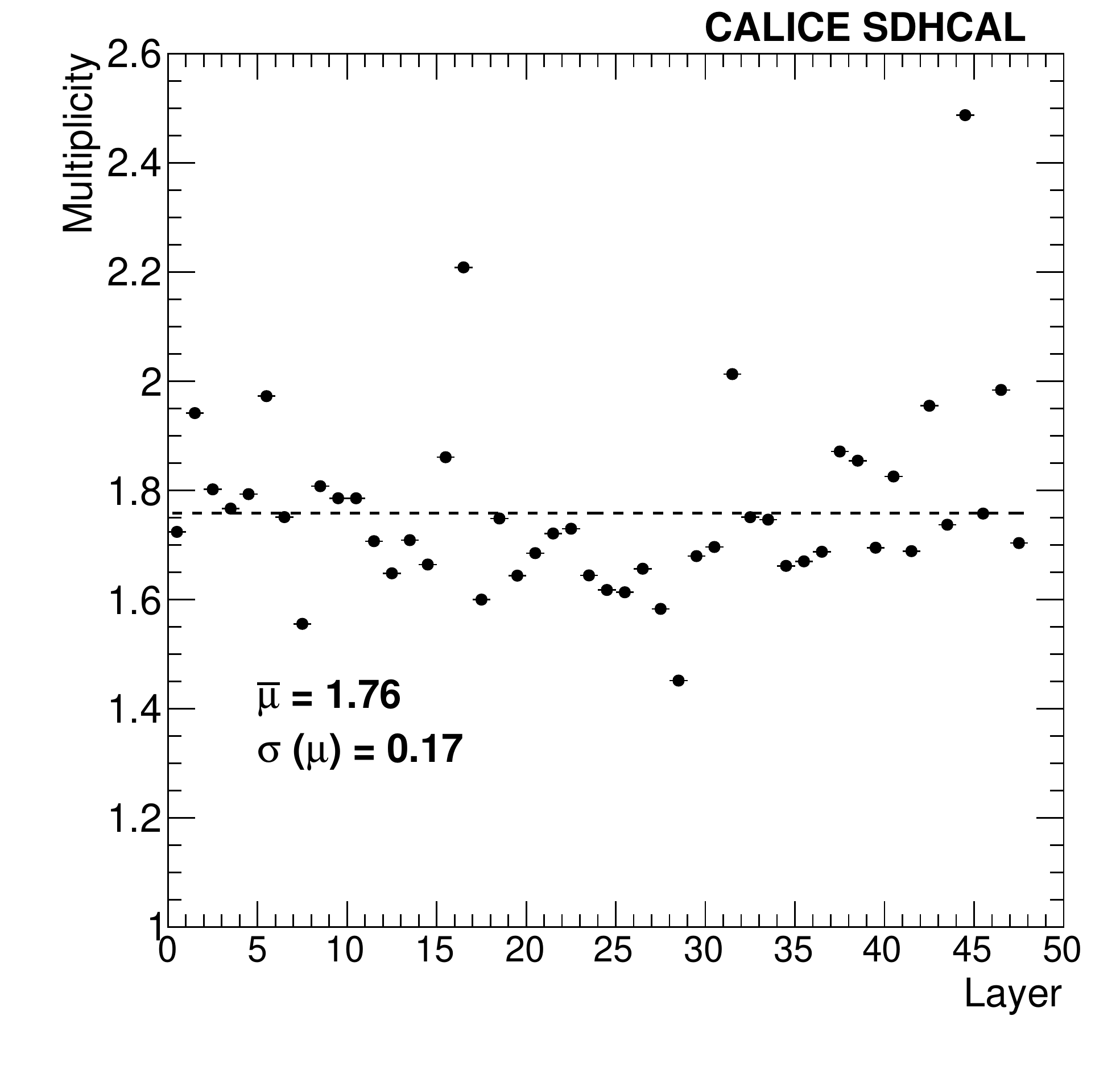}
\caption{Efficiency (left) and hit multiplicity (right) of the 48 layers estimated using the data of the 2012 H6 beam test. 
The dashed black line is the average efficiency (left) and average multiplicity (right). Values of the average efficiency ($\bar \epsilon$) and of the average hit multiplicity ($\bar \mu$) as well as the  RMS of the efficiency distribution ($\sigma(\epsilon)$) and  that of the hit multiplicity distribution $(\sigma(\mu)$) of the 48 layers are also given.}
\label{fig.eff_august}
\end{center}
\end{figure}

\begin{figure}[htp]
\begin{center}
\includegraphics[width=0.43\textwidth]{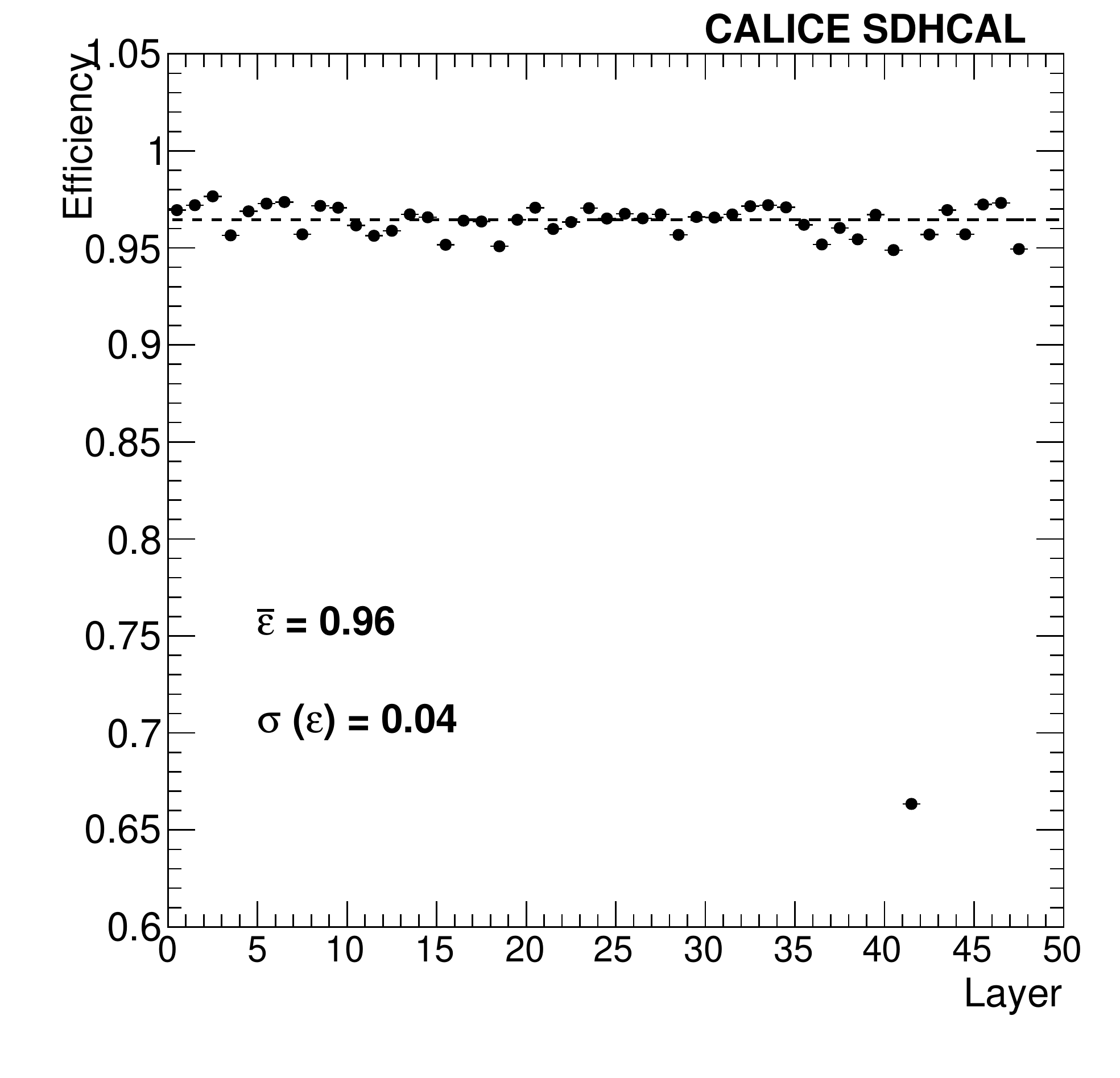}
\includegraphics[width=0.43\textwidth]{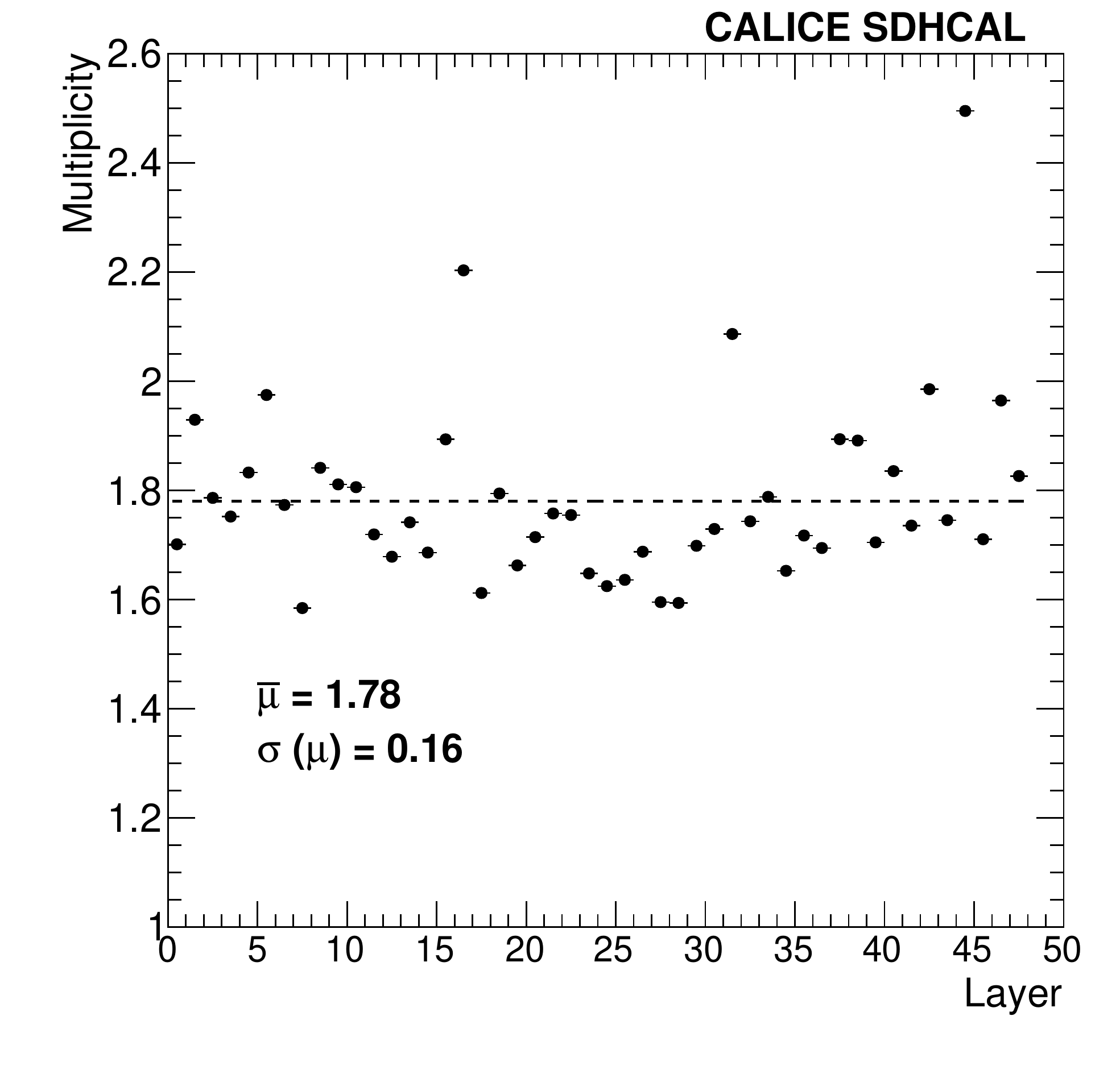}
\caption{Efficiency (left) and hit multiplicity $\bar \mu$  (right) of the 48 layers estimated using the data of the 2012 H2 beam test. The dashed black line is the average efficiency  (left) and average multiplicity (right) of the 48 layers. Values of the average efficiency ($\bar \epsilon$) and of the average hit multiplicity ($\bar \mu$) as well as the  RMS of the efficiency distribution ($\sigma(\epsilon)$) and  that of the hit multiplicity distribution $(\sigma(\mu)$) of the 48 layers are also given.} 

\label{fig.eff_november}
\end{center}
\end{figure}

Only tracks with a $\chi^2<20$ are used to estimate the efficiency. 
The expected impact point of the track in the layer under study is determined. %
The efficiency is then estimated as being the fraction of tracks for which at least one hit is found at a distance of less than 3~cm around the expected position. 
The  hit multiplicity for a track in one layer is also estimated by counting the number of hits, if any, in the cluster built around the closest hit to the track's impact. %
The hit multiplicity is then computed by averaging that of all the tracks going through the sensitive region under study.
Fig.~\ref{fig.eff_august} shows the average efficiency and hit multiplicity of the layers during the H6  beam test.
Other methods to estimate the efficiency and particle multiplicity were tested, and confirm the results presented here.  As can be seen in Fig.~\ref{fig.eff_august} (left) layer number 41 shows lower efficiency.  A problem with one  third of the electronics in this layer was observed, most probably related to a power line on one of the three slabs. This problem could not be cured during the beam test.  Similar results were obtained in the H2 beam test\footnote{The faulty slab of layer 41 was switched off in this beam test.} as can be seen in Fig.~\ref{fig.eff_november}.   
A few layers show a significant deviation of their  average hit multiplicity compared to that obtained by averaging on that of all the layers. This could be the result of a lower surface resistivity of the electrodes painting used to apply the electric field in the GRPC~\cite{small}. Another possibility is a slightly smaller average gas gap in these layers compared  to the nominal one (1.2~mm) resulting in a higher gain and thus more charge produced by the avalanche leading to a higher hit multiplicity.    


\section{Simulation}
\begin{figure}[htp]
\begin{center}
\includegraphics[width=0.5\textwidth]{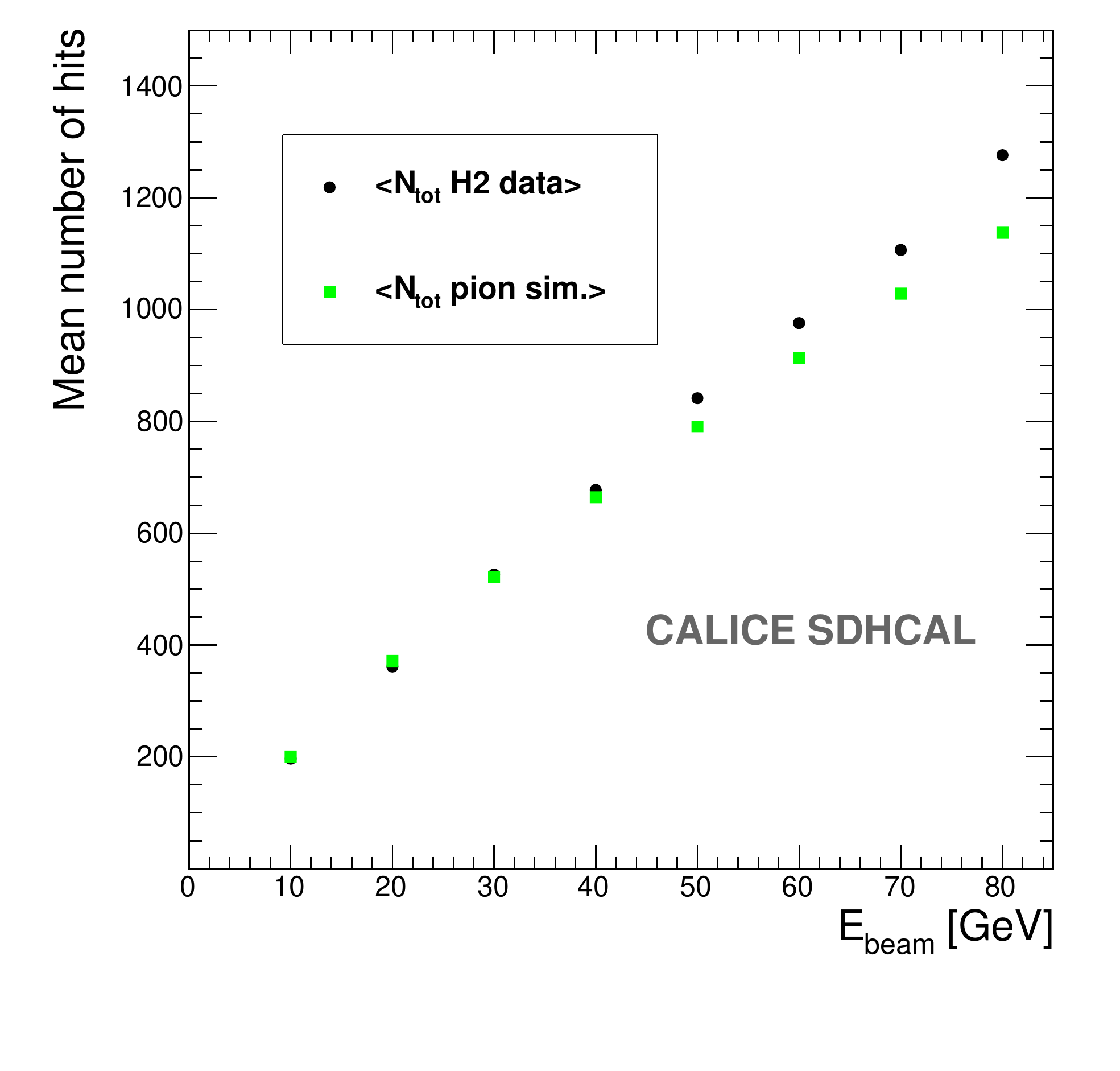}
\caption{Mean number of hits for  pion showers  as a function of the beam energy in the  2012 H2 runs and that of simulated pion events in the SDHCAL based on the FTFP-BERT-HP physics list.
}
\label{SIMU}
\end{center}
\end{figure}

A  simulation model based on the GEANT4 toolkit~\cite{GEANT4P} was developed and the interactions of different kinds of particles including pions, protons, muons and electrons of different energies  within the SDHCAL were simulated. To simulate the GRPC response to the passage of these particles, a digitizer that exploits  the steps provided by  GEANT4 was developed~\cite{arnaud}.  The digitizer  parameters were tuned to reproduce the response  of the SDHCAL prototype to muons and electrons. Using the FTFP-BERT-HP physics list of GEANT4 a rather good agreement  was found when comparing the simulation and the SDHCAL data. More details about the digitizer and the comparison between the SDHCAL data and a few hadronic models is given in~\cite{arnaud}.  In this work the simulation is used only to verify the event selection quality and to estimate the possible biases that may affect the energy reconstruction of hadronic showers. Figure~\ref{SIMU} shows a comparison of the mean number of hits between data and the simulation, for  pion showers of different energies, after applying the same selection.\\
\section{Pion shower selection}\label{selections}
\subsection{Proton contamination}
\begin{figure}[htp]
\begin{center}
\includegraphics[width=0.5\textwidth]{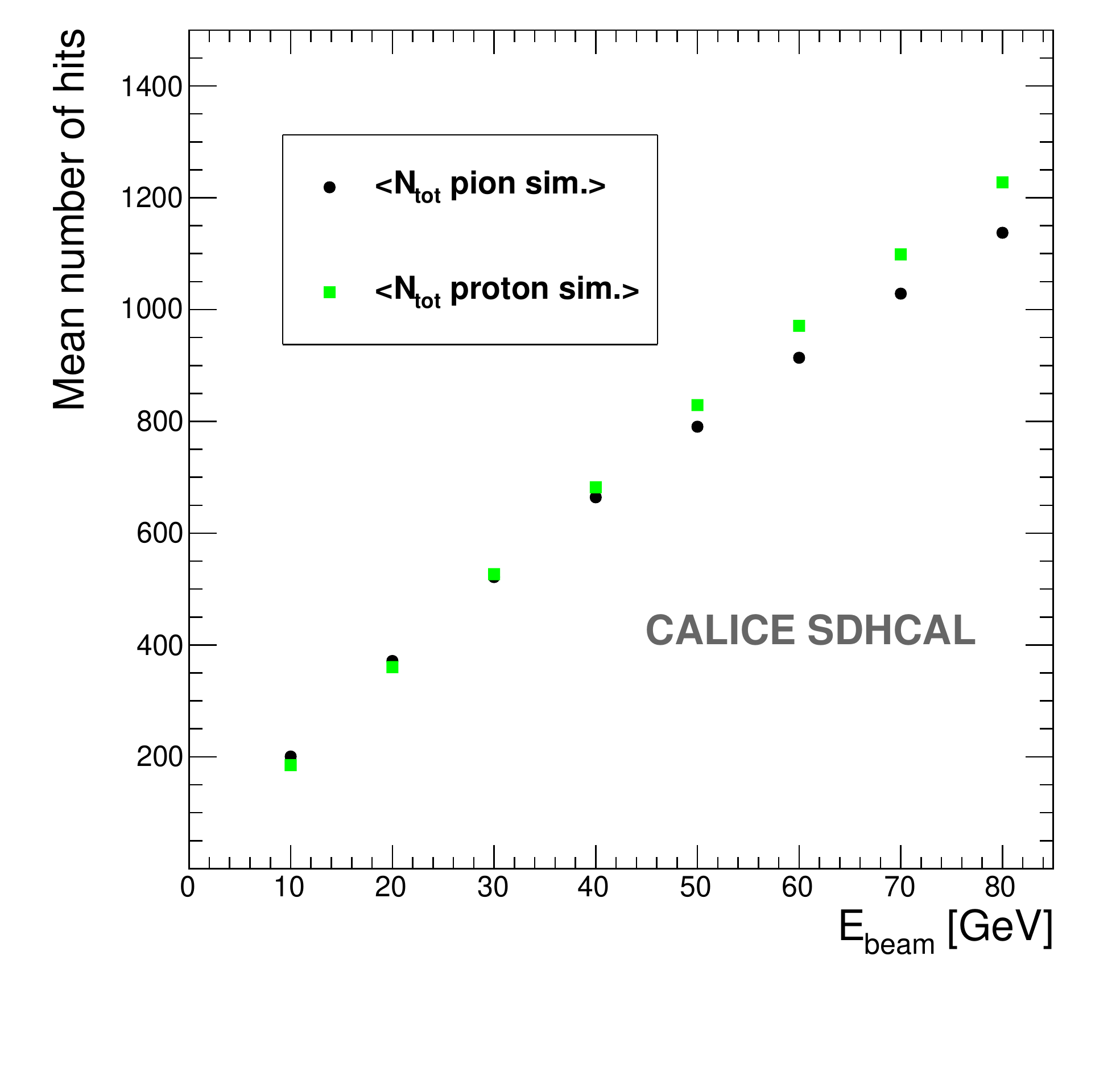}
\caption{Mean number of hits produced by  pions (black circles) and protons (green squares) in the SDHCAL prototype for several energy points as given by the simulation based on the FTFP-BERT-HP list.}
\label{proton-pion}
\end{center}
\end{figure}

The H6 positive pion beam is contaminated by  different kinds of hadronic particles with protons as the  dominant contaminant. The presence of protons in the H6 beam was measured by  the ATLAS  Collaboration   \cite{tilecal1} which found that  the contamination of protons is present  above 20 GeV, and  becomes important  (45-60\%) in the energy range from 50 to 100 GeV.  In the absence of a Cherenkov detector in front of our prototype to identify the different species, both pion and proton events were collected.  At high momentum where the effect of the mass difference of the two species is negligible, the hadronic showers of both pion and proton are expected to be similar in infinite depth calorimeters\footnote{A slight difference of the electromagnetic component of the hadronic showers between the two species may introduce a small difference in non-compensating calorimeters.}. Unfortunately, in our limited depth prototype, a difference of the two species is expected at high momentum (energy) where leakage is present, since the proton interaction length (16.8~cm) is smaller than that of the pion (20.4~cm) in the steel absorber. This means that protons which start showering on average slightly before pions, have comparatively more hits since they have less leakage.   This is confirmed by the simulation of pion  and proton events in the SDHCAL as shown in Fig.~\ref{proton-pion}. The H2 beam was made of negative pions and hence proton contamination is absent.  To study the energy resolution in the SDHCAL we need to keep the two  samples separated. The study of data collected of the H2 beam test will then provide the energy resolution obtained in our prototype for pions and will be used as a reference, while the data collected in H6 will include both species. 

\subsection{Electron contamination}
\begin{figure}[!ht]
\begin{center}
\includegraphics[width=0.5\textwidth]{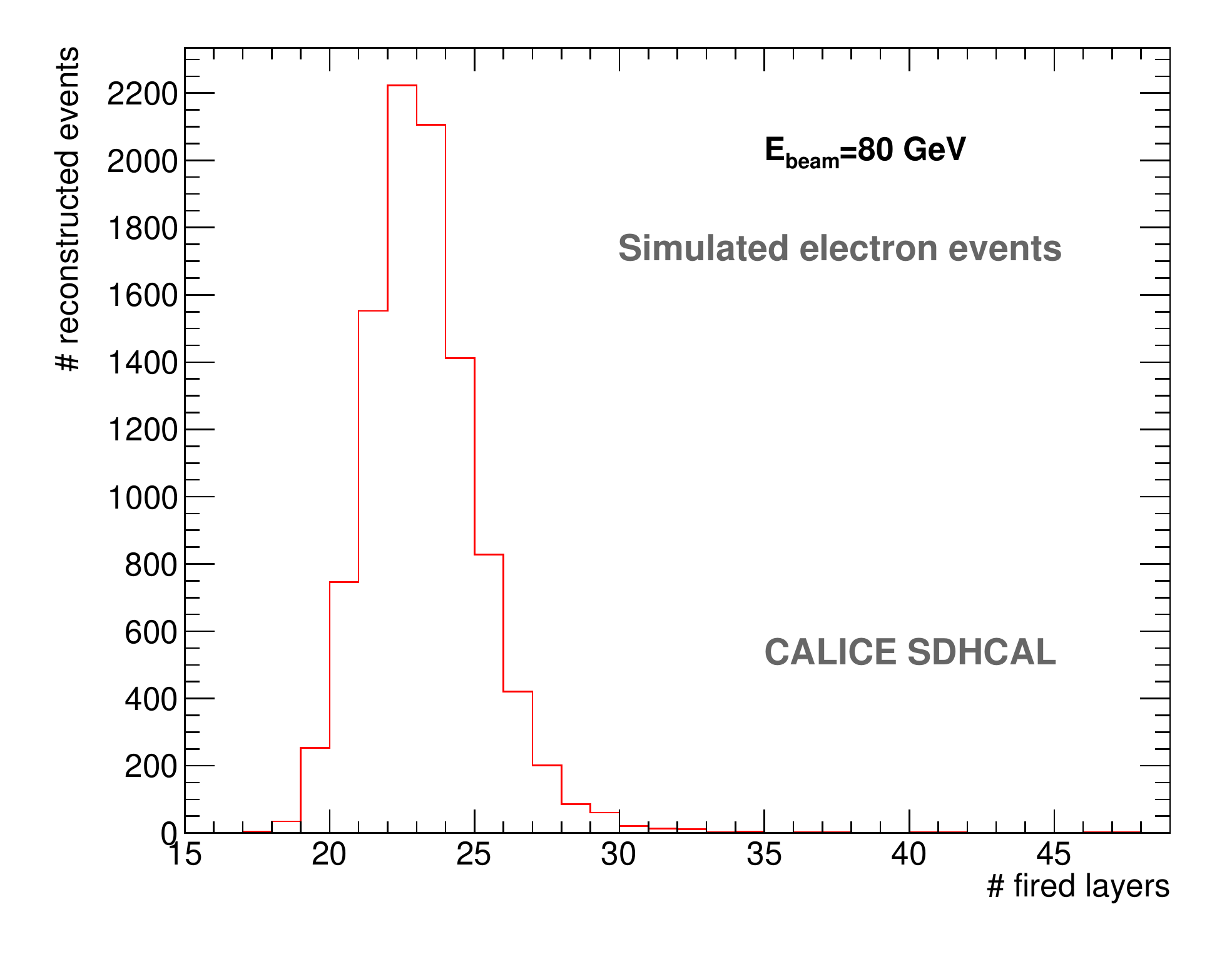}
\caption{Distribution of number of fired layers by 80~GeV electrons as given by the simulation.
}
\label{fig:ElecLayer}
\end{center}
\end{figure}

Electrons are also present in the pion beam despite the use of a lead filter to reduce their number.  As mentioned before, the absence of a Cherenkov counter or any other detector able to discriminate  electrons from pions makes it necessary to find other means to eliminate the electrons in our hadronic samples. We use the fact that electrons start their electromagnetic shower in the first plates of the prototype. This is due to the fact that the radiation length in steel is 1.76~cm.
For data which include electromagnetic and hadronic showers, requiring that the shower starts in the fifth layer or later should in principle remove almost all  electrons since this represents about $6\ X_0$. To define the start of the shower we look for the first layer with more than 4 fired pads. To eliminate fake shower starts due to accidental noise or a locally high multiplicity, in addition to the first layer the three following layers were also required to have more than 4 fired pads.  Electromagnetic showers in the energy range between 5 and 80~GeV are longitudinally contained in less than 30 layers of our SDHCAL prototype, as shown by simulated electron events in Fig.~\ref{fig:ElecLayer}.  This electron rejection criterion is therefore applied only for events in which no more than 30 layers containing each more than 4 fired pads are found.  This limitation helps to minimize the loss of true pion hadronic showers at high energy where the number of fired layers generally exceeds 30. In this way high energy  pions starting  their shower in the first layers are not rejected. Low energy hadrons (pions and protons) fire a relatively smaller number of layers and are mostly fully contained in the SDHCAL.

\begin{figure}[!h]
\begin{center}
\includegraphics[width=0.325\textwidth]{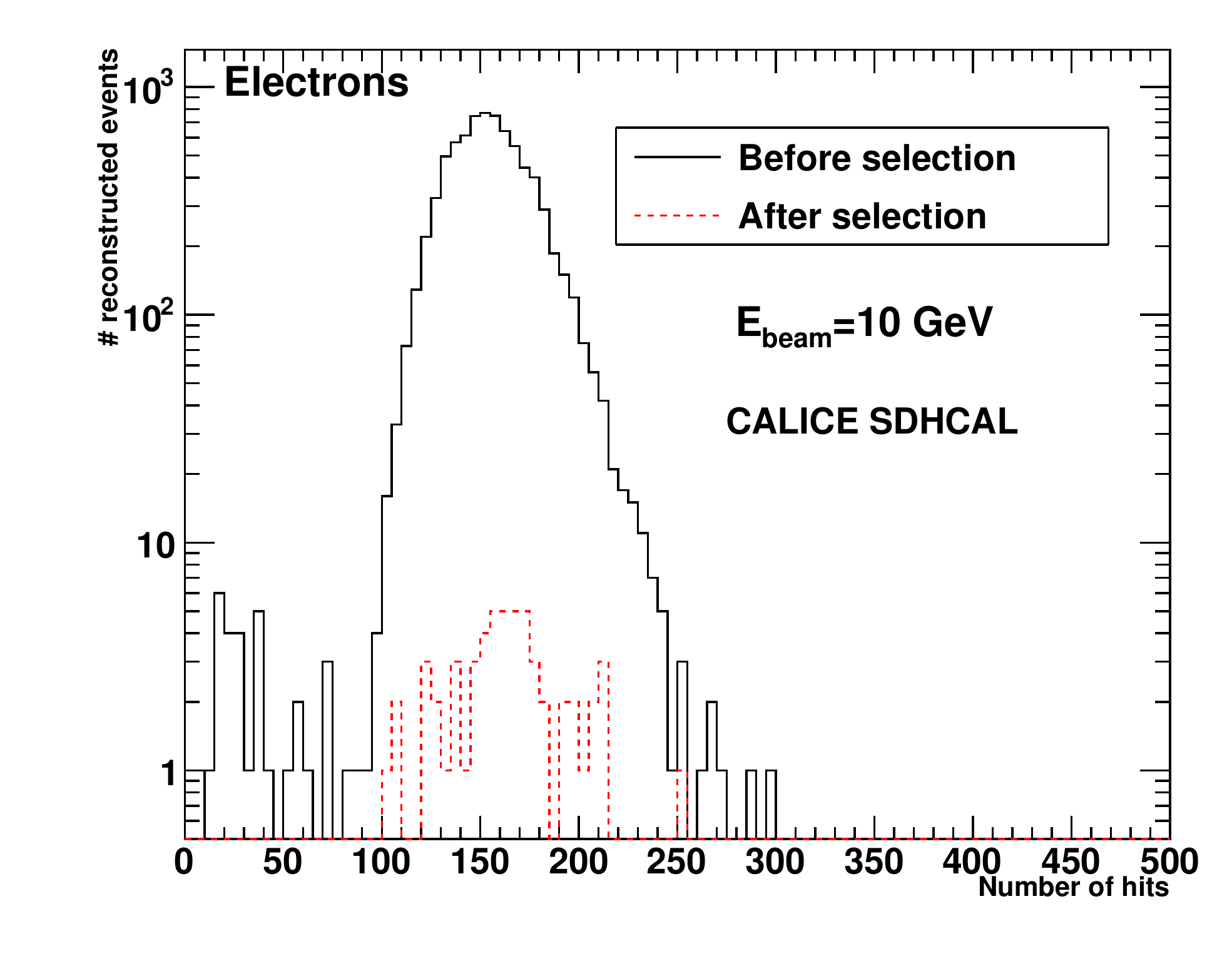}
\includegraphics[width=0.325\textwidth]{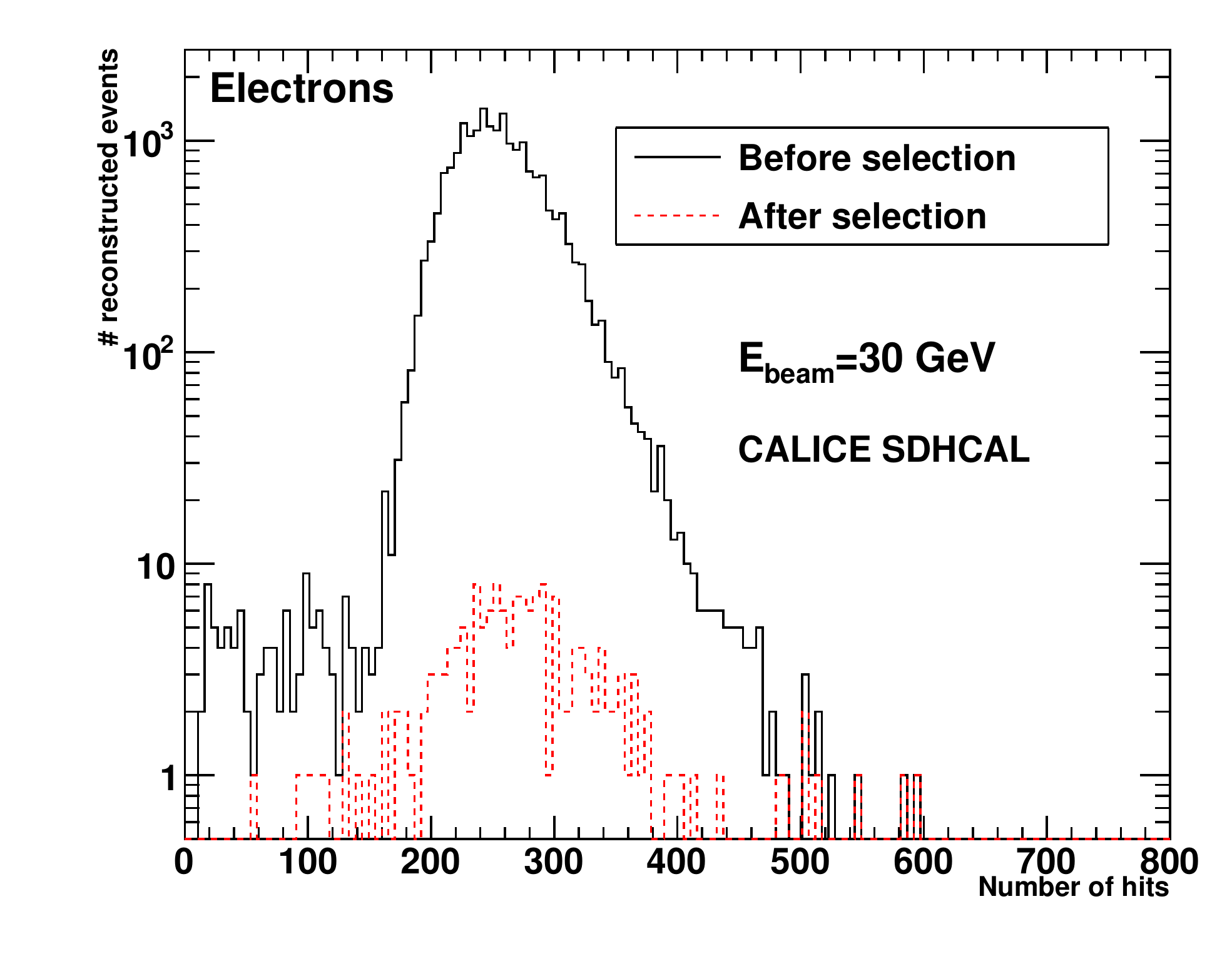}
\includegraphics[width=0.325\textwidth]{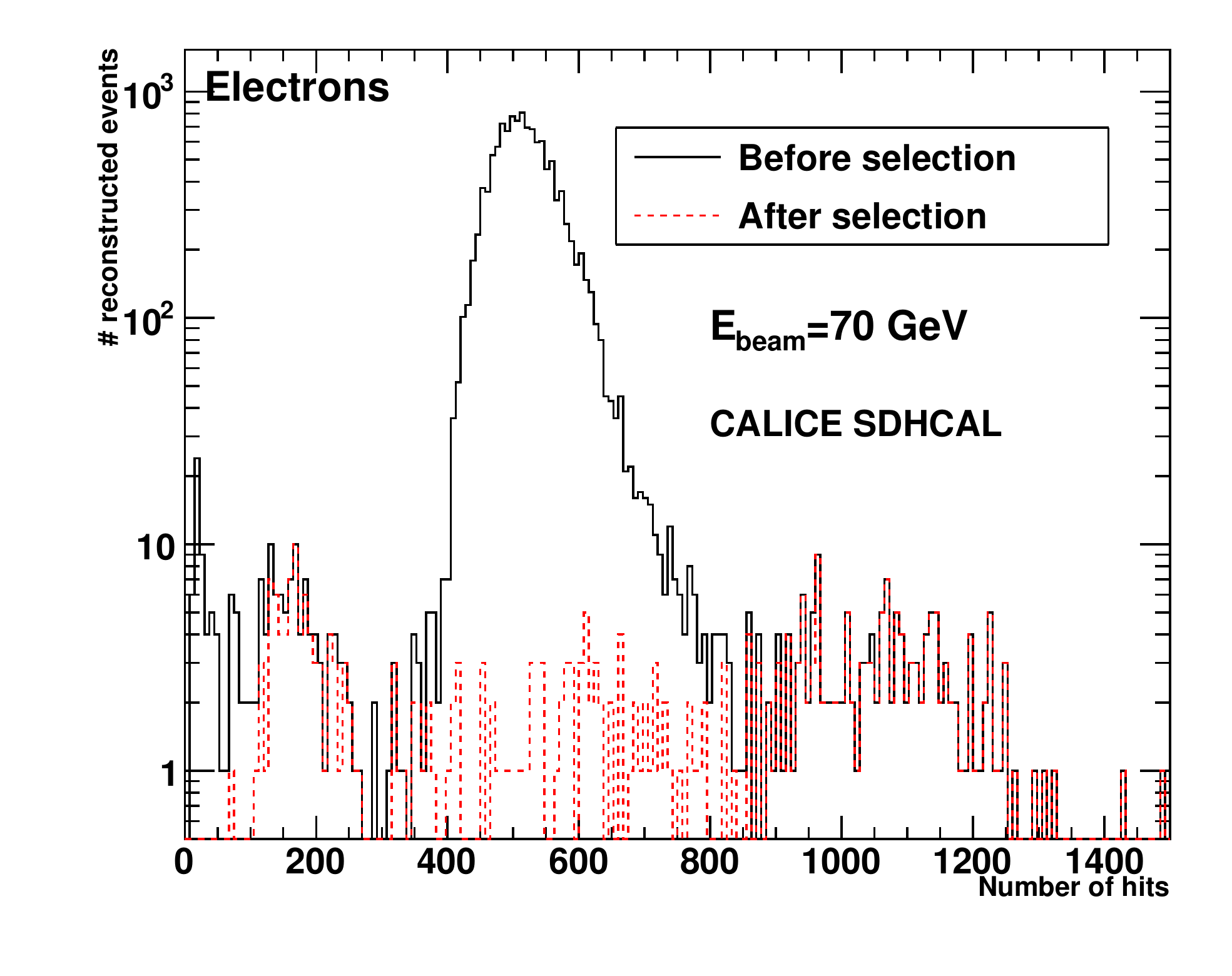}
\caption{Distribution of number of hits for 10, 30 and 70~GeV electron runs from the H6 data sample before (solid black line) and after (dashed red line) electron rejection.}
\label{fig:elec-rejection}
\end{center}
\end{figure}

To verify such a selection, simulated pion and proton events of different energies were studied.  It was found that, except for a reduction of their number,  only hadronic showers of 5 GeV are impacted in terms of energy reconstruction. To check the rejection power of this selection, it was applied to electron runs at different energies collected during the same campaign. Figure~\ref{fig:elec-rejection} shows the distribution of number of hits before and after the selection for 10, 30 and 70~GeV electron runs. It shows clearly the rejection power of this selection which is higher than 99.5\% for energies higher than 10 GeV.

\subsection{Beam- and cosmic muons contamination}
\begin{figure}[!h]
\begin{center}
\includegraphics[width=0.45\textwidth]{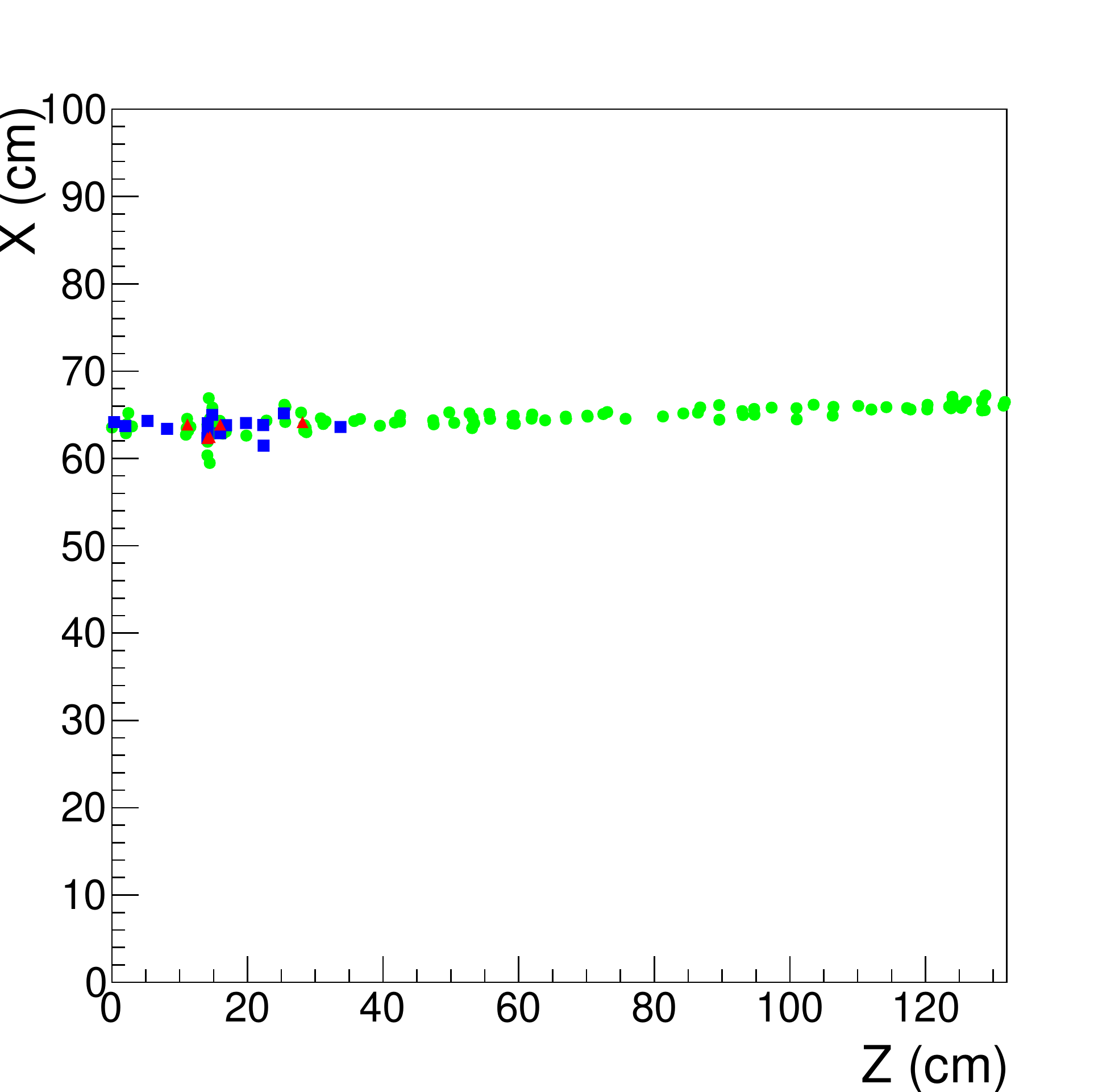}
\includegraphics[width=0.45\textwidth]{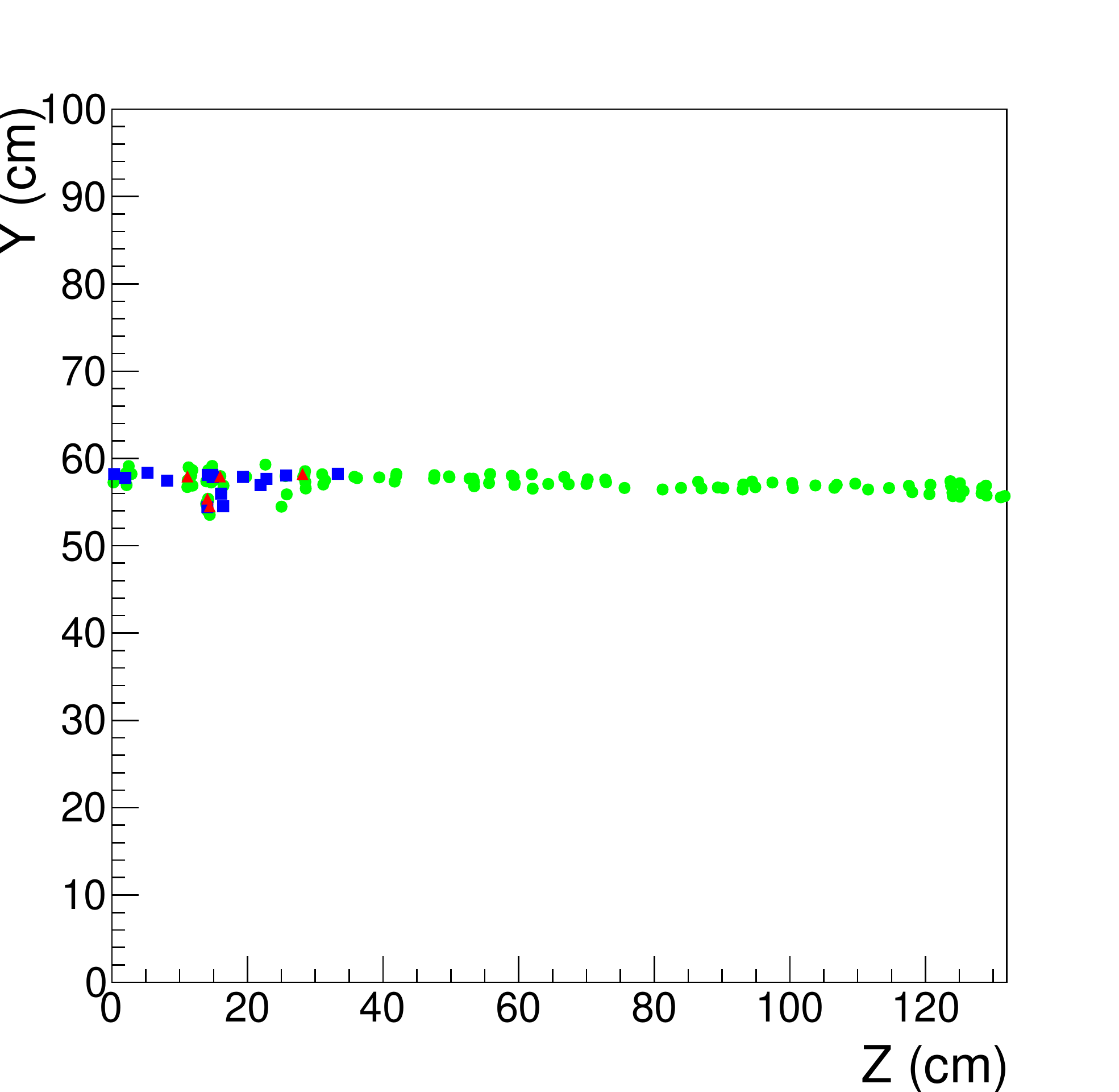}
\caption{Event display of a 50~GeV radiative muon with red color indicating the highest threshold fired pads, blue color indicates the middle threshold, and green color is for the lowest one.}
\label{fig:radmuon}
\end{center}

\end{figure}

\begin{figure}[!h]
\begin{center}
\includegraphics[width=0.325\textwidth]{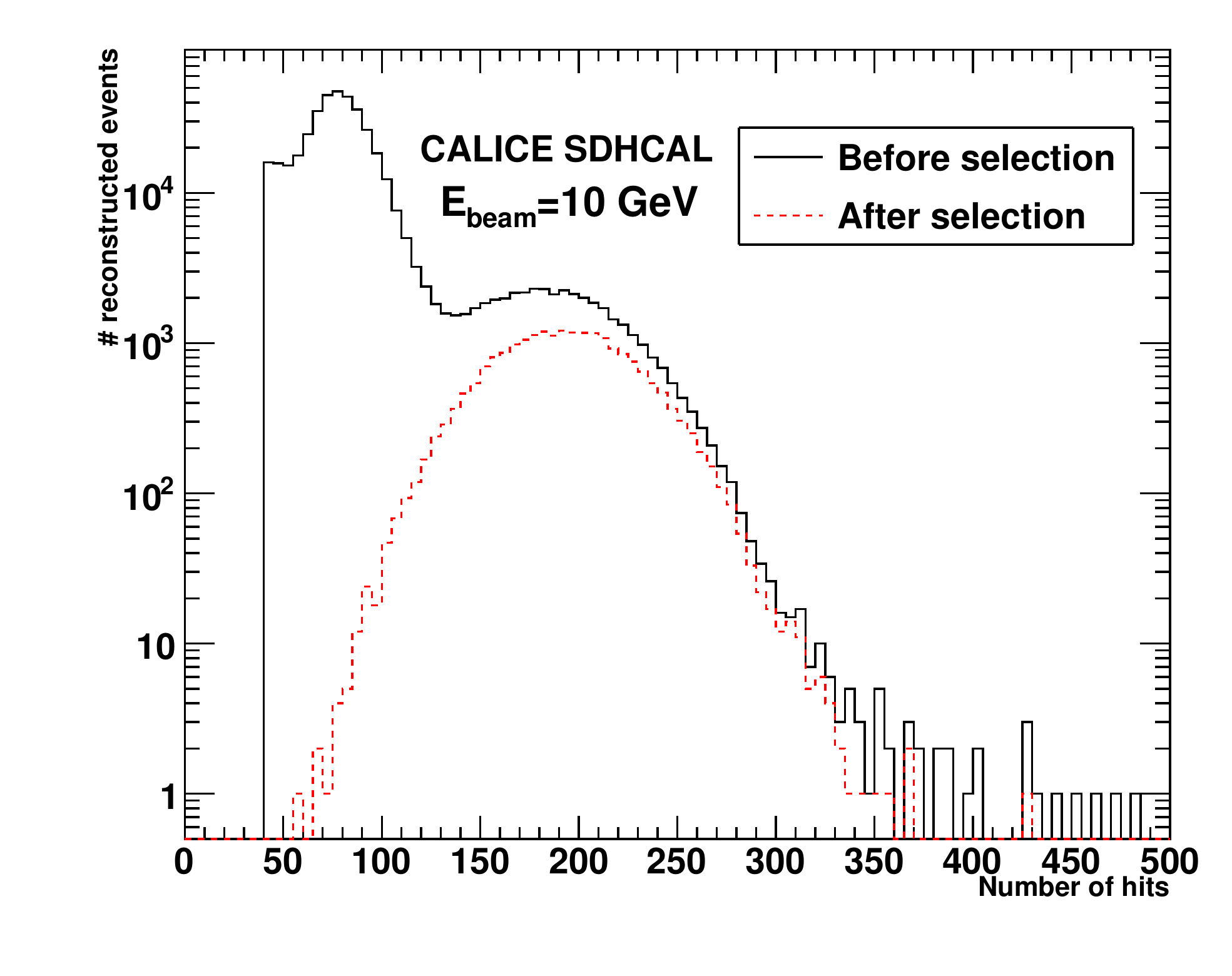}
\includegraphics[width=0.325\textwidth]{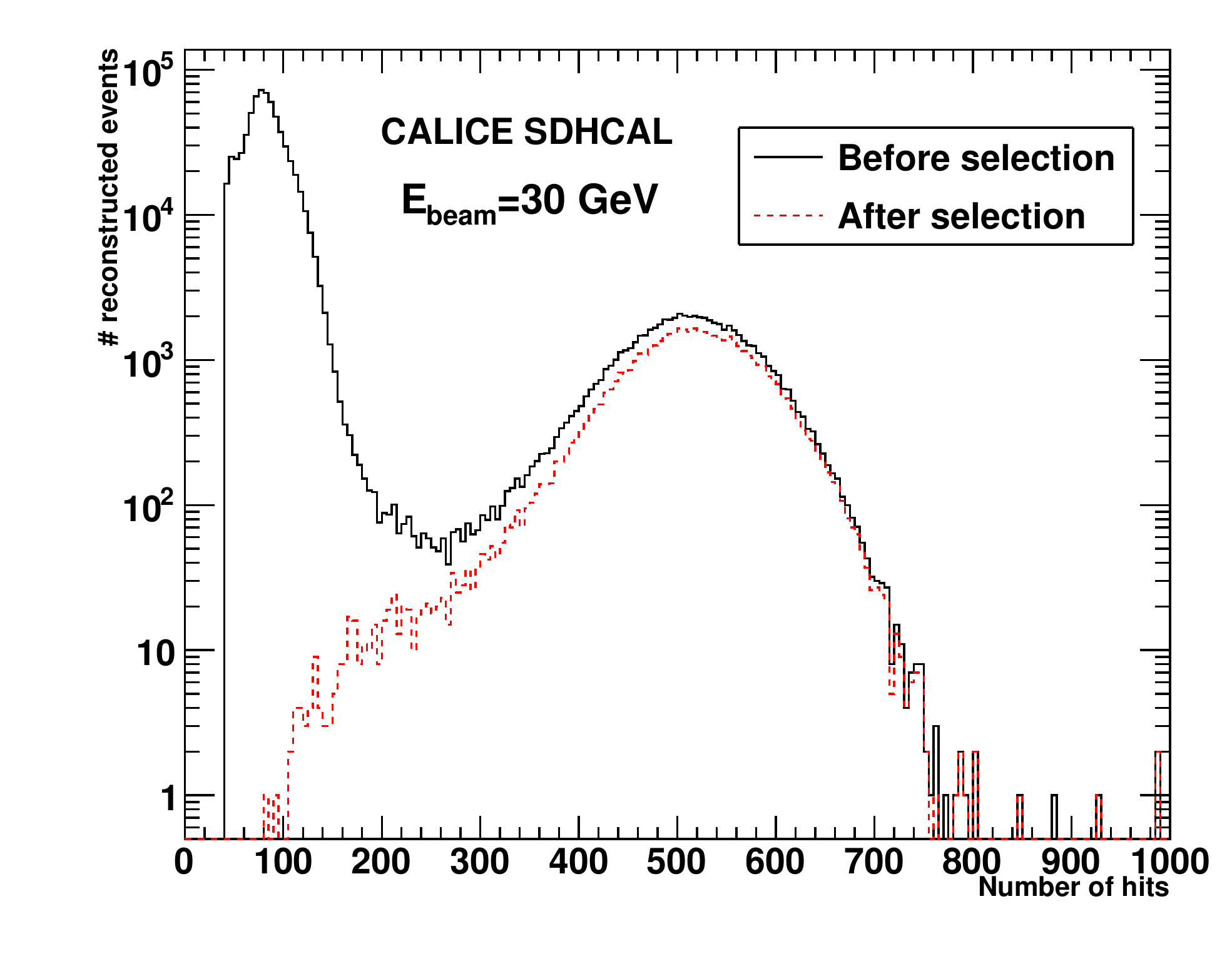}
\includegraphics[width=0.325\textwidth]{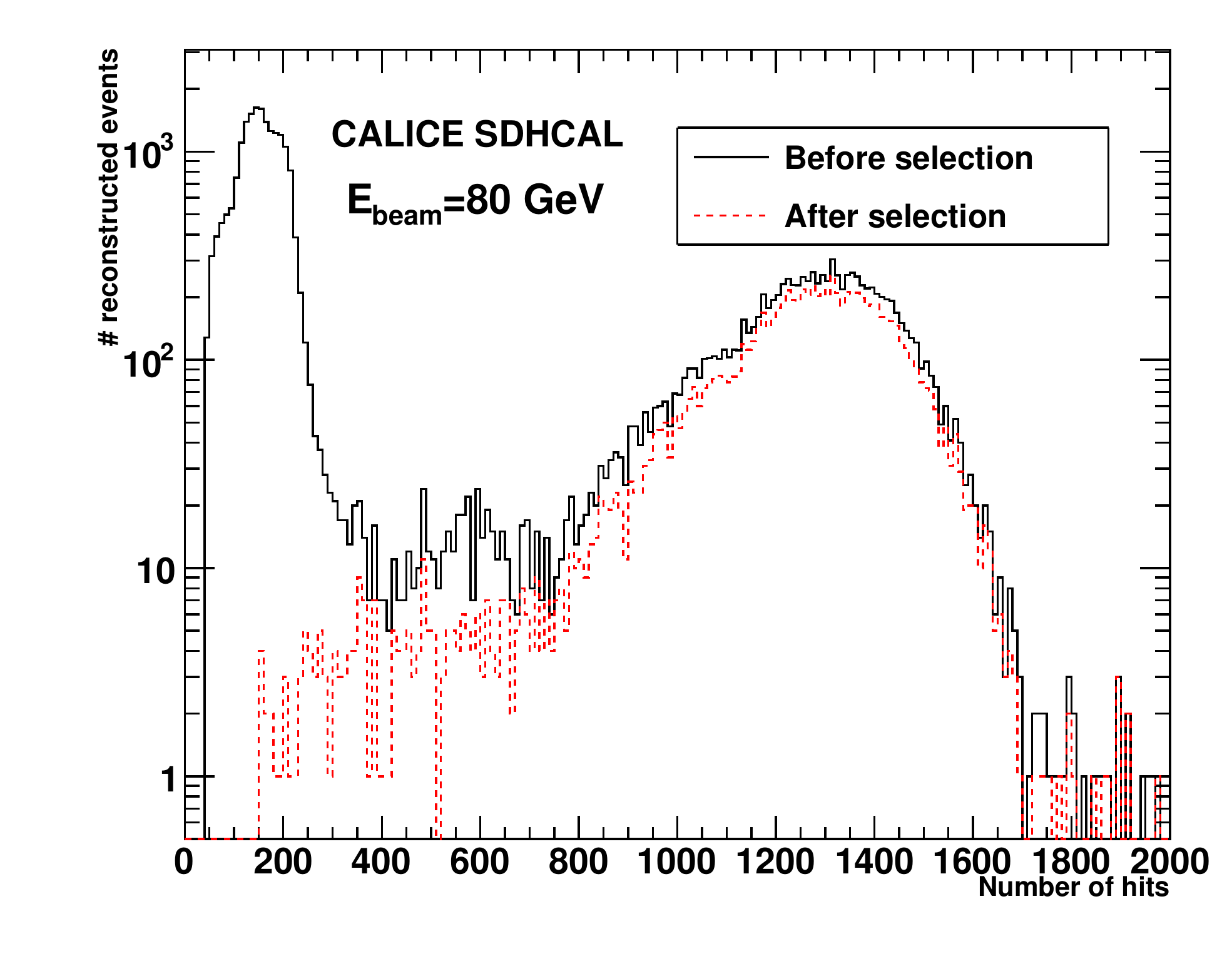}
\caption{Number of hits for 10, 30 and 80~GeV pion runs from the H6 data sample before (black line) and after (dashed red line) full selection.}
\label{fig:pion-selection}
\end{center}
\end{figure}

Muons are also present in the pion beam. They are produced by pions decaying before reaching the prototype. To eliminate these muons as well as the  contamination by cosmic muons, the average number of hits per fired layer is requested to be greater than $2.2$. This is higher than the average pad multiplicity which was found to be around $1.76$ (Fig.\ \ref{fig.eff_august}).  This selection eliminates the non-radiating muons. To eliminate most of the radiating ones, such as the one featured in Fig.~\ref{fig:radmuon},  we require that the ratio between the number of layers in which the root mean square of the hits' position in the $x-y$ plane  exceeds  5 cm in both x and y directions and the total number of layers with at least one fired hit  is larger than 20\%.  The effect of this selection on  showering pions is  limited. Indeed the ratio of 20\%  corresponds in this case essentially to pions which start showering in the last ten layers of our SDHCAL prototype and whose number is negligible.   
In addition, to eliminate contamination by neutral hadrons in our data sample, events are required to have at least 4 hits in the first 5 layers.\\ 
The result of the selection is shown in Fig.~\ref{fig:pion-selection} for data runs of three energies: 10, 30 and 80~GeV where the total number of hits of the collected events is shown before and after the selection.  

Remaining electrons are not important as can be seen in Fig.~\ref{fig:elec-rejection}. One can see in Fig.~\ref{fig:pion-selection}  that for the events of  30 and 80~GeV the selection has very little effect on the hadron component (right tail).   For the 10~GeV run, there is a loss in the total number of events  but the right tail still has the same shape.    This is confirmed by applying the same selection on simulated events of hadrons where no electron contamination is present.

The selection criteria are summarized in Table \ref{tab:allCuts}.\\
\begin{table}[!ht]
\begin{center}
\begin{tabular}{c||c}
  Electron rejection       & Shower start $\geq 5$ or $N_\mathrm{layer}> 30$           \\\hline
  Muon rejection           & $\frac{N_\mathrm{hit}}{N_\mathrm{layer}}>2.2$                      \\\hline
  Radiative muon rejection & $\frac{N_\mathrm{layer} \setminus \mathrm{RMS}>5cm}{N_\mathrm{layer}}>20\%$ \\\hline
  Neutral rejection        & $N_\mathrm{hit\in First\ 5\ layers}\geq4$
\end{tabular}
\caption{Summary of the different cuts applied to select the pions. $N_\mathrm{hit}$ is the total number of hits and $N_\mathrm{layer}$ is the number of fired layers.} 
\label{tab:allCuts}
\end{center}
\end{table}

\section{Beam intensity correction}\label{CORRECTION}

Even though the beam parameters during the two data-taking periods were optimized to get spills containing less than 1000 particles, it was observed that, for some runs of both periods with relatively high number of particles per spill, the number of hits associated to hadronic showers was decreasing during the spill time.  The decrease was more apparent for the number of hits associated to the second and third thresholds of the semi-digital readout as can be seen in Fig.~\ref{fig:den0}.  The effect was also more frequent  in runs of high energy pions. This effect is clearly due to the limitation of the GRPC rate capability since the particle rate increases during the spill time due to the structure of the spill.  The consequence of such behavior is a degradation of the energy resolution for the hadronic showers.   In order to correct for this effect, two special calibration techniques were developed. The first one is a linear fit calibration. To achieve this the average number of hits associated to each threshold of each hadronic shower is plotted as a function of  their time with respect to the  spill start. Then the slope of a  linear fit to the number of hits distribution is determined. The corrected number of hits $N_{\mathrm{corr}_j}$ for each run and for each threshold $j$ is defined according to the following  formula:

\begin{figure}[!t]
\begin{center}
\includegraphics[width=0.325\textwidth]{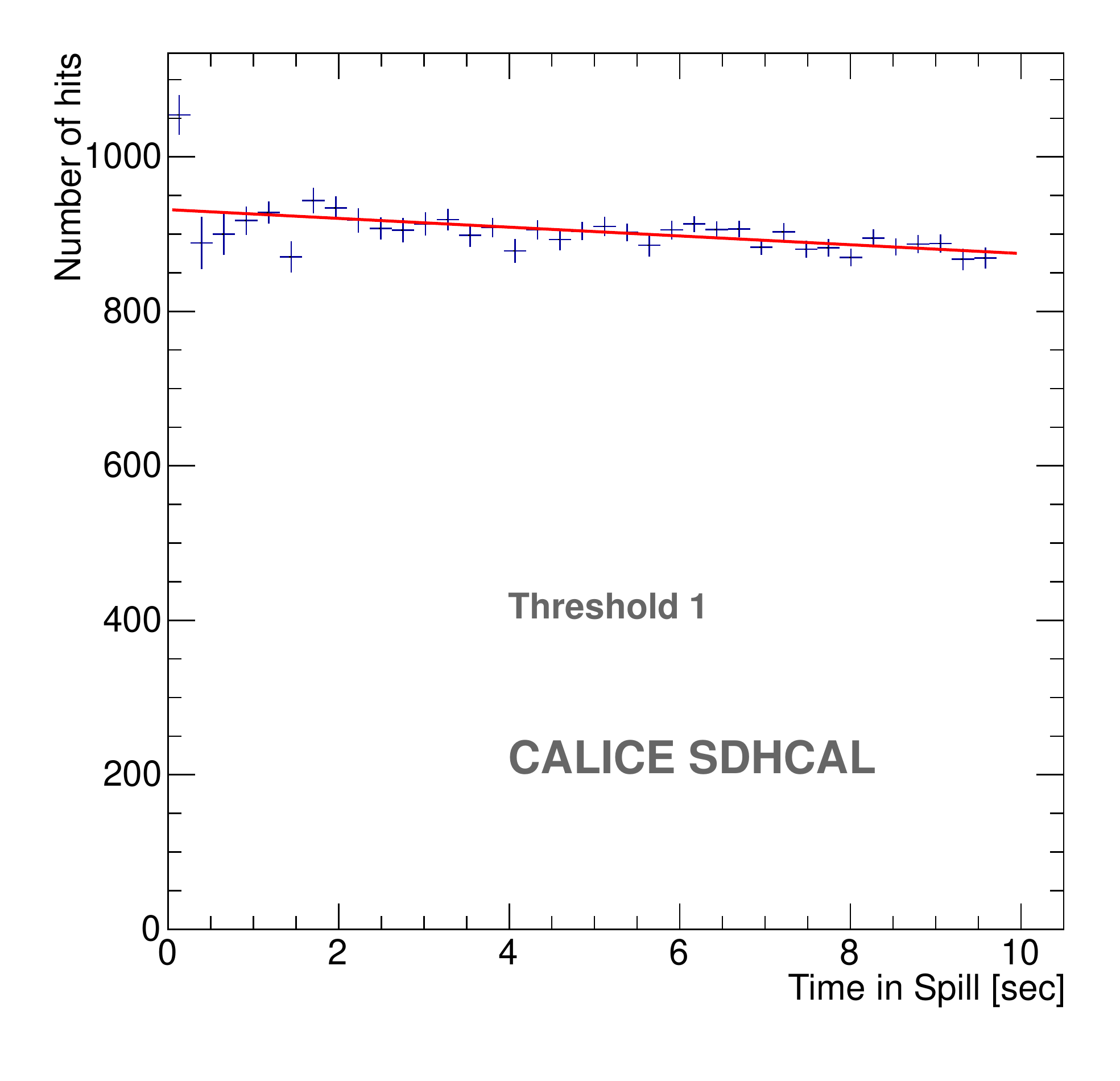}
\includegraphics[width=0.325\textwidth]{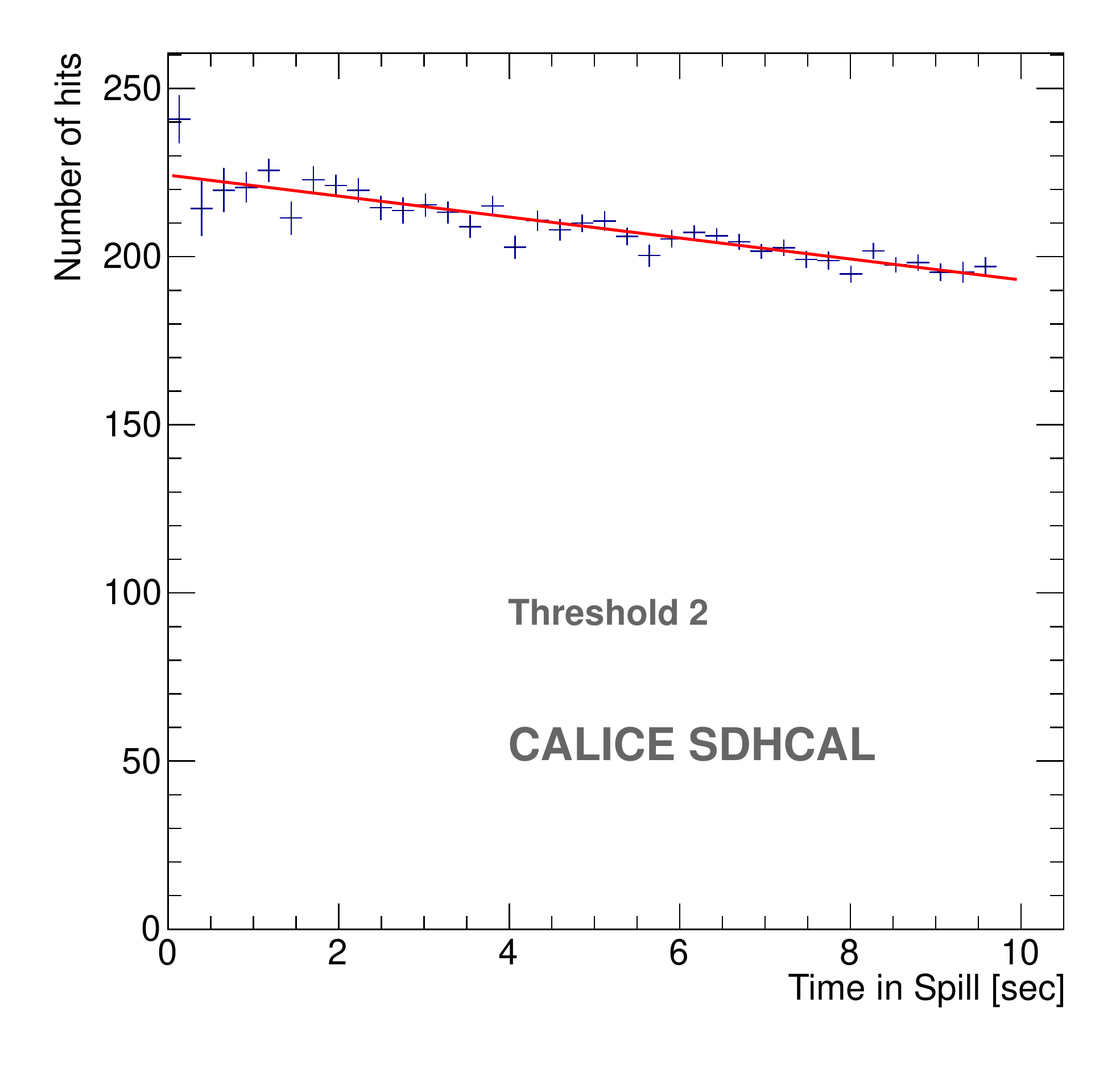}
\includegraphics[width=0.325\textwidth]{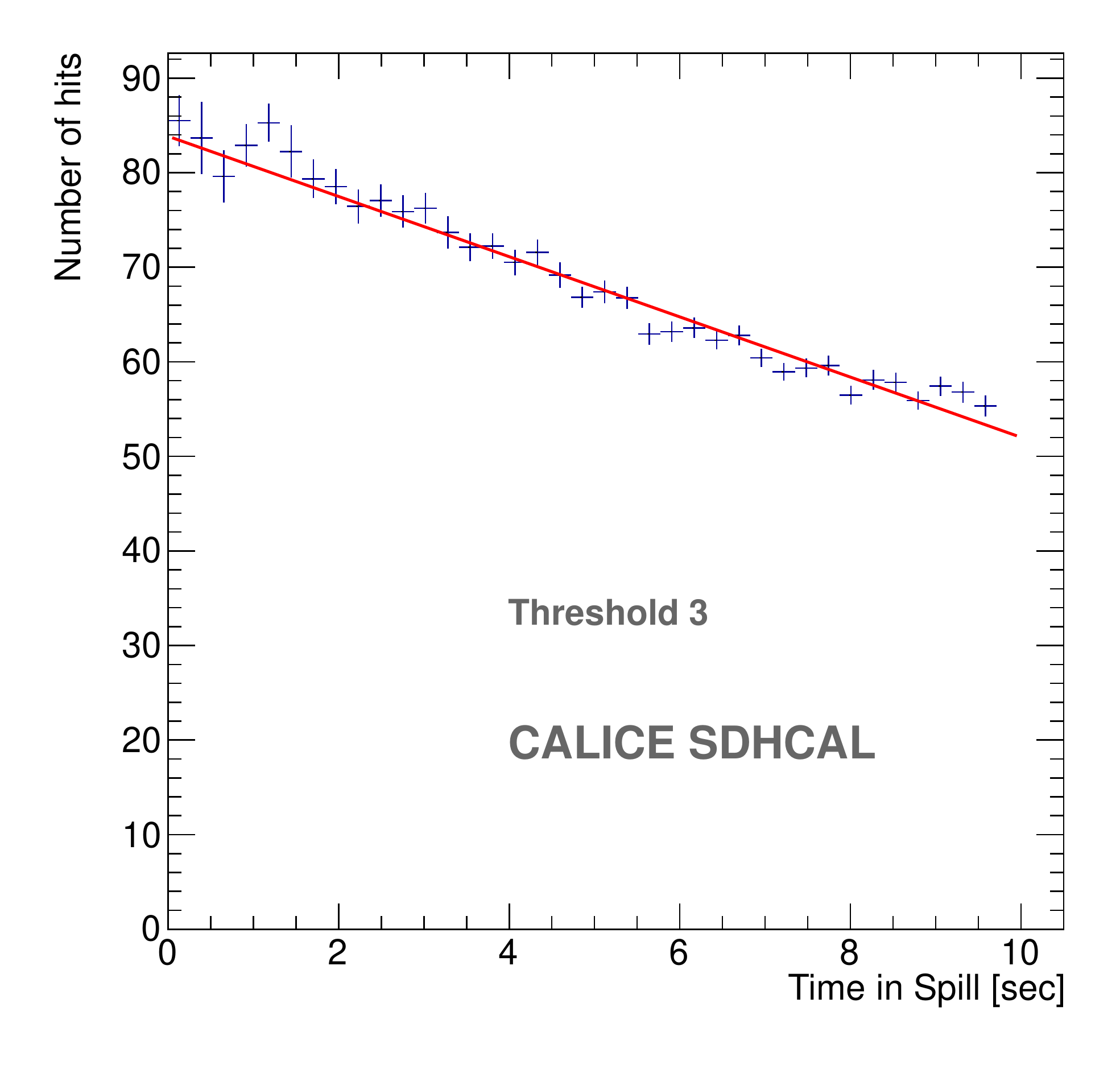}
\caption{From left to right: mean number of hits as a function of spill time for first, second, and third threshold in a 80~GeV  pion run from the H6 data sample. A linear fit is also shown.}
\label{fig:den0}
\end{center}
\end{figure}

\begin{figure}[!t]
\begin{center}
\includegraphics[width=0.325\textwidth]{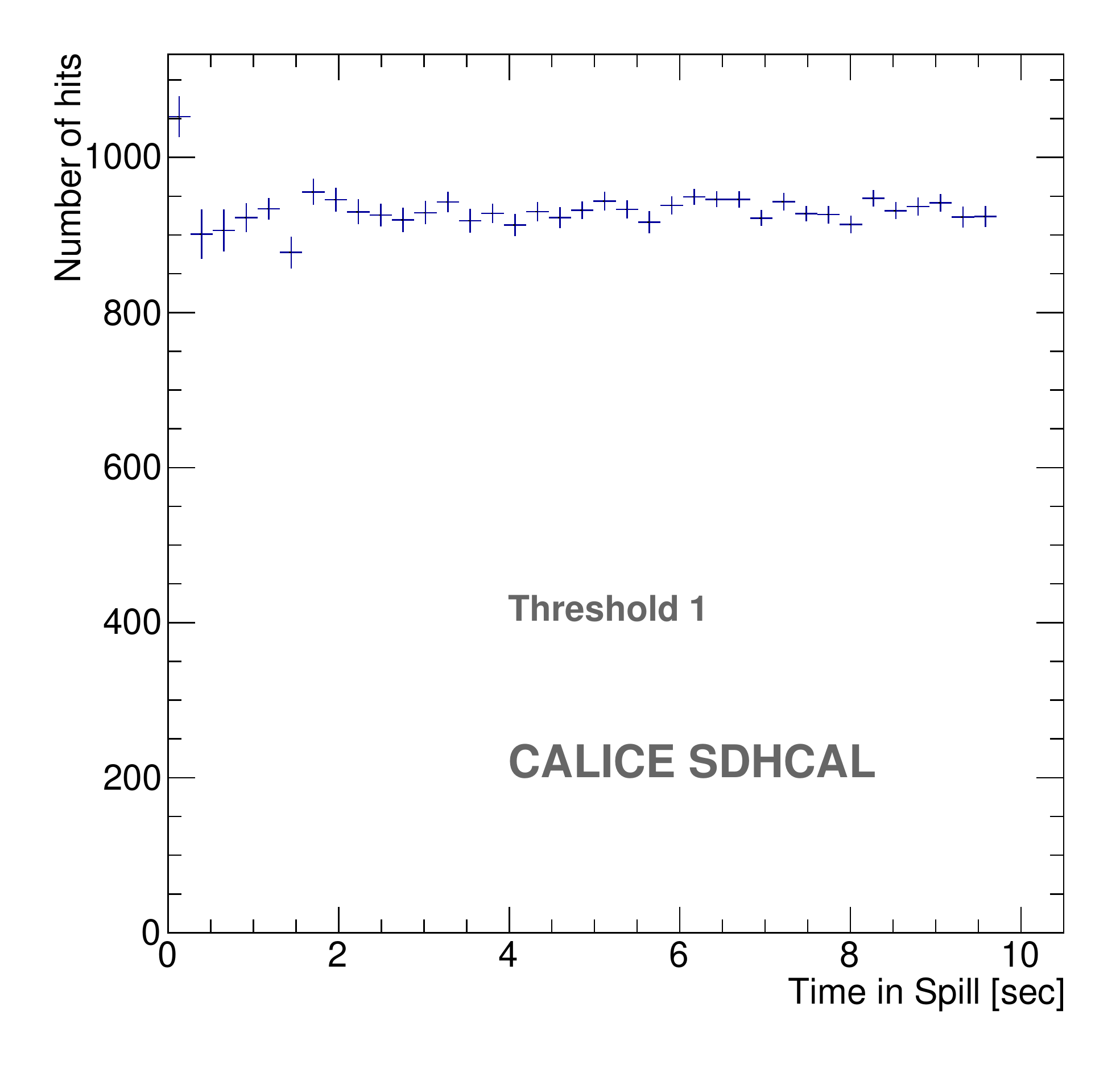}
\includegraphics[width=0.325\textwidth]{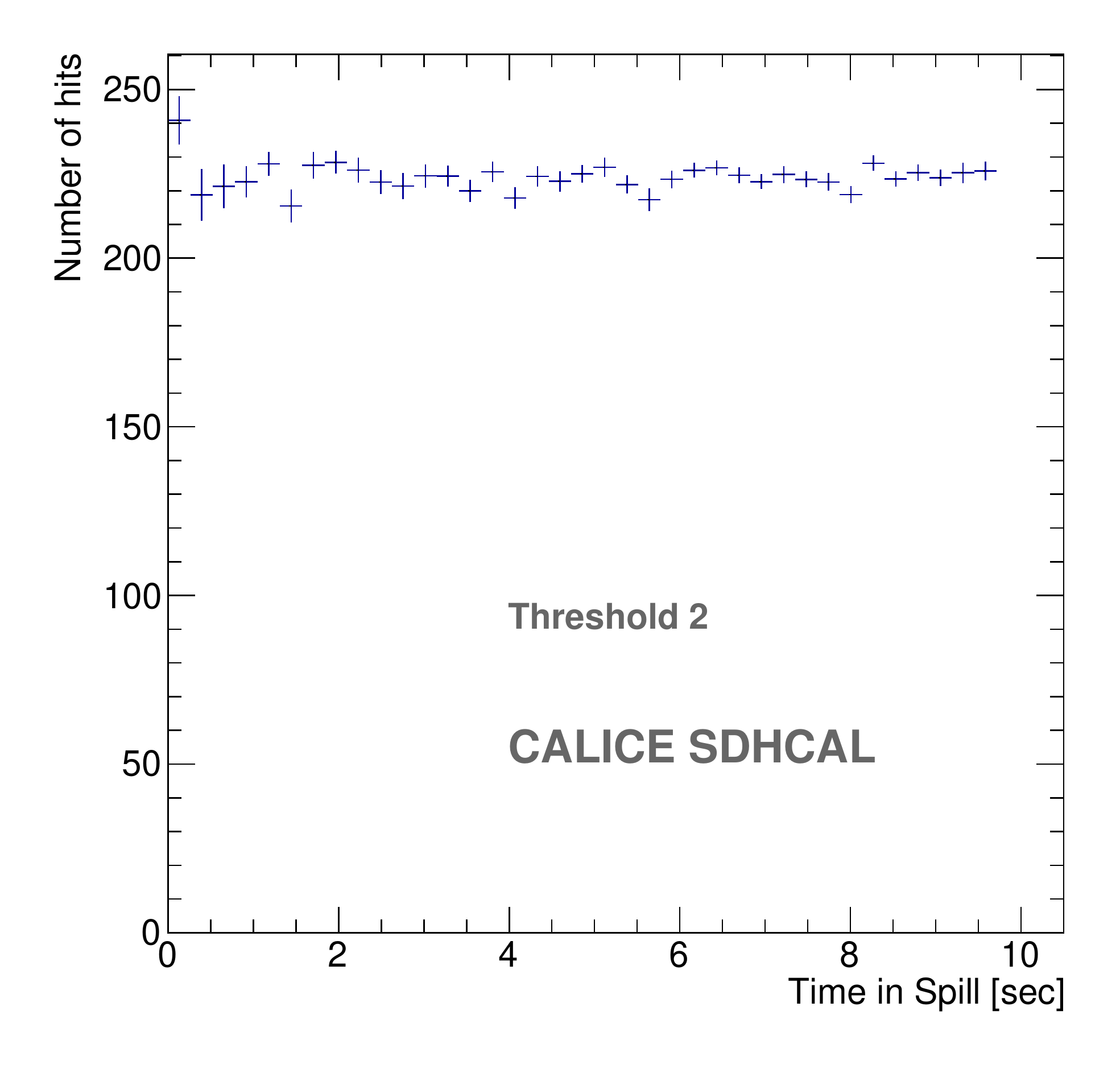}
\includegraphics[width=0.325\textwidth]{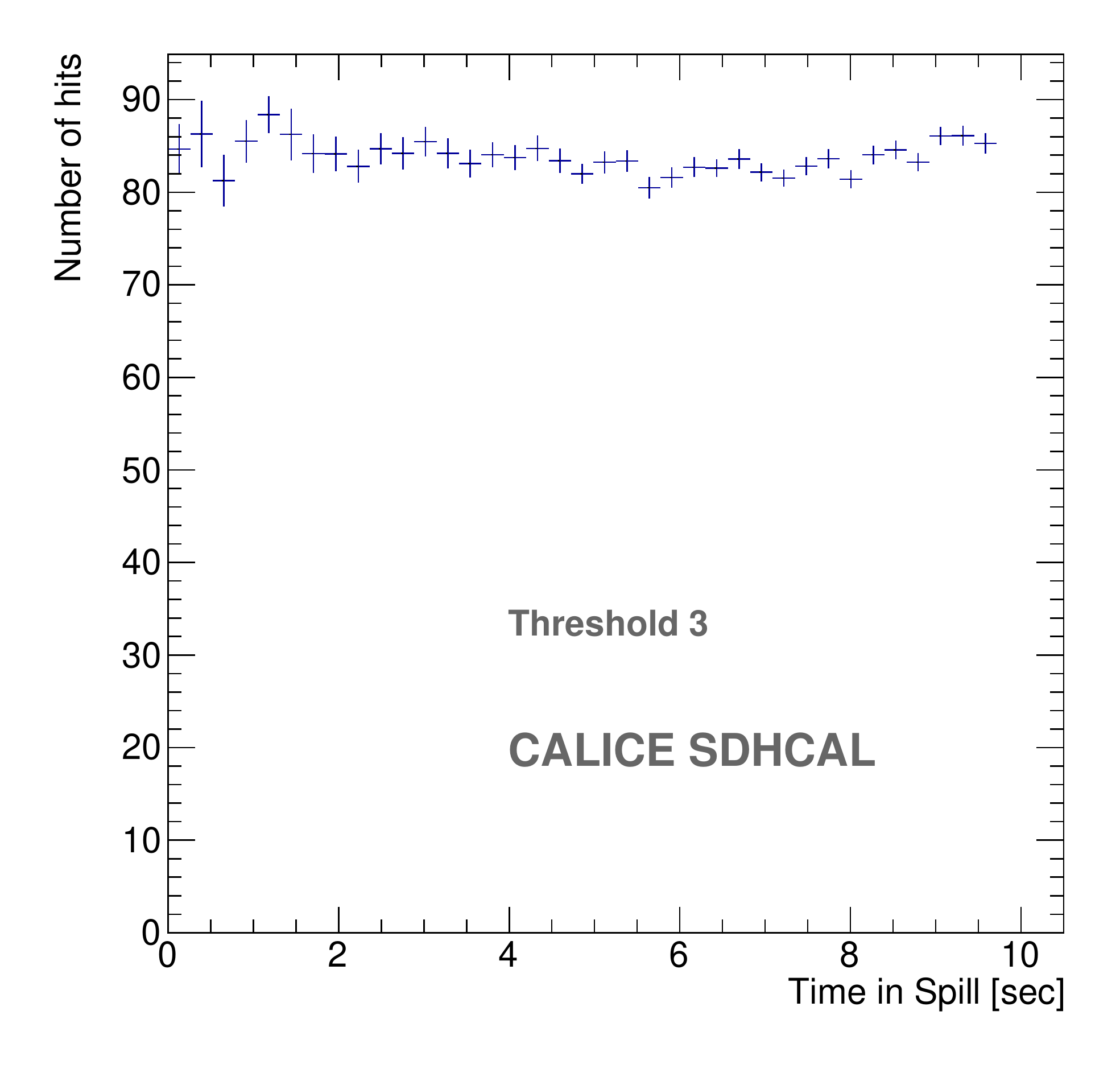}
\caption{From left to right: corrected mean number of hits as a function of spill time for first, second and third threshold in a  80~GeV pion run from the H6 data sample.}
\label{fig:den1}
\end{center}
\end{figure}
\begin{figure}[!h]
\begin{center}
\includegraphics[width=0.4\textwidth]{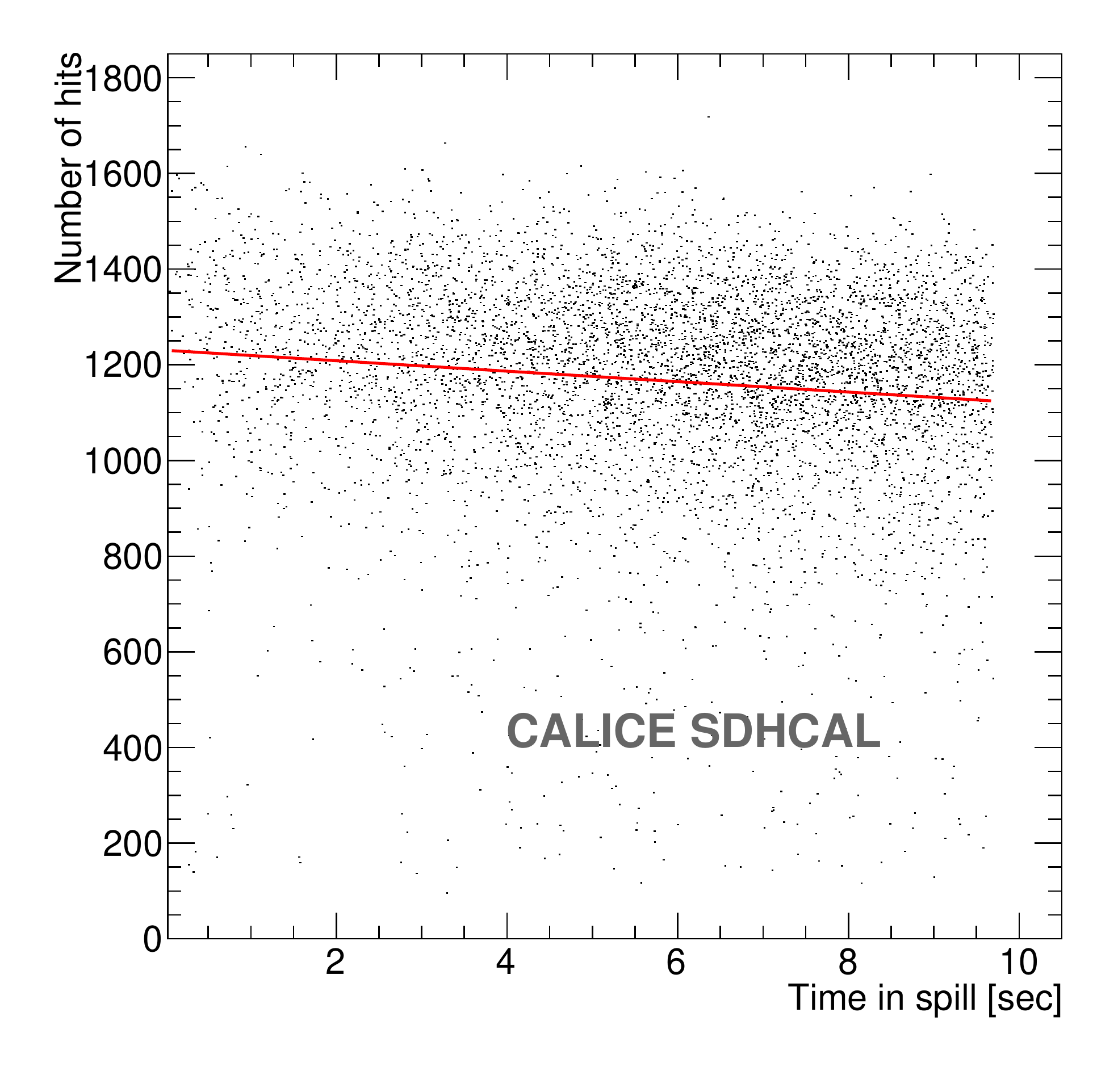}
\includegraphics[width=0.4\textwidth]{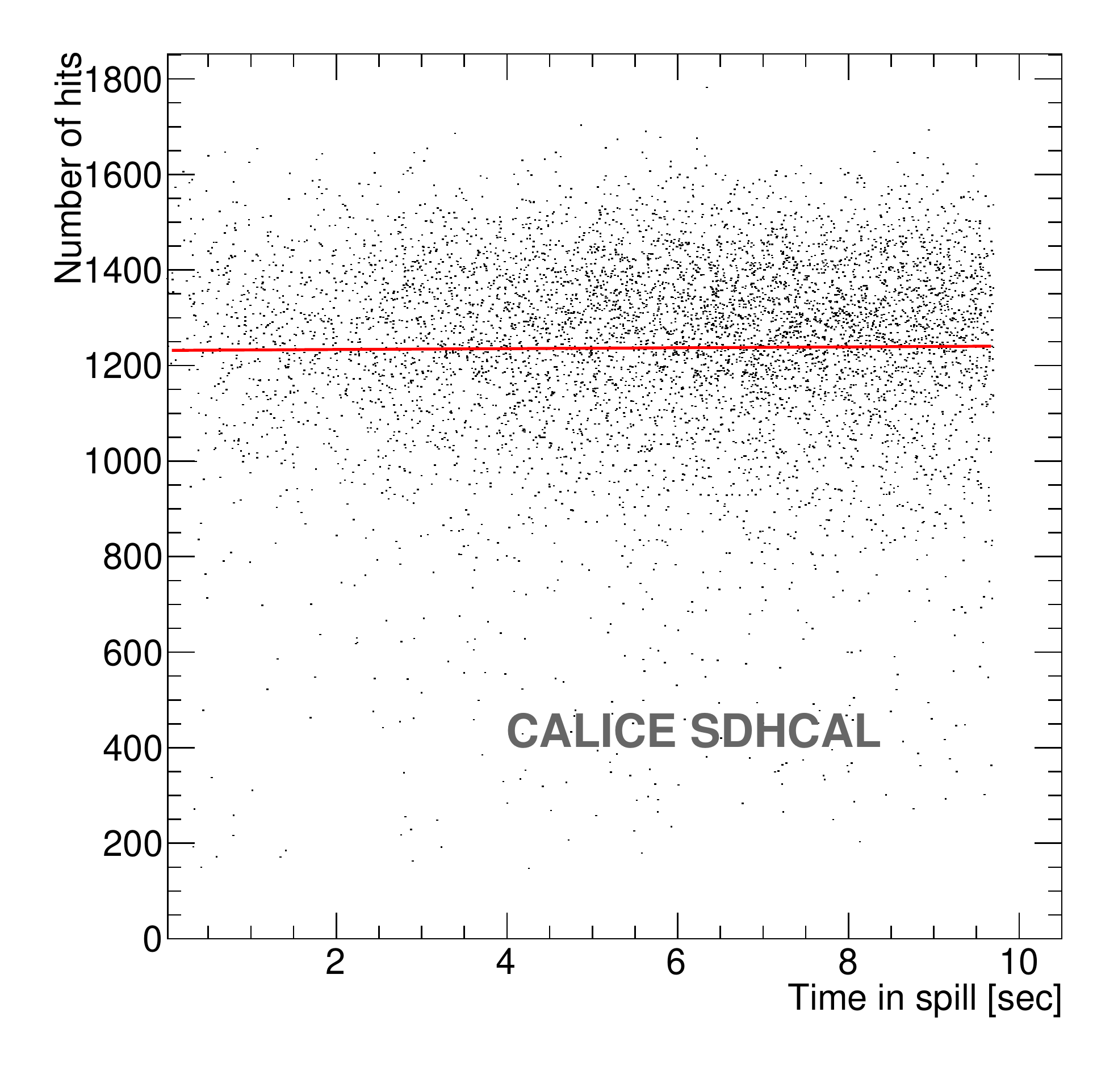}
\caption{Total number of hits for  80~GeV pions from the H6 runs as a function of spill time before (left) and after (right) linear fit calibration. A linear fit that includes the contribution of the events with low number of hits is also shown.}
\label{fig:linfit}
\end{center}
\end{figure}

\begin{center}
\begin{equation}
 N_{\mathrm{corr}_j}= N_j - \lambda_j * T
\end{equation}
\end{center}    

\noindent
where  $N_1$ is the number of hits which are above the first threshold and below the second. $N_2$ is  the number of hits  which are above both the first and the second but below the third threshold, and $N_3$ is the number of hits that are above the third threshold.   $\lambda_j$ is the correction slope for  threshold~$j$ and $T$ is the time since  start of the spill.   The correction allows the conditions of low particle rate at the beginning of the spill to be retrieved as can be seen in Fig.~\ref{fig:den1}.   The results after the linear fit calibration for an 80~GeV run from the H6 data for the total number of hits can be seen in Fig.~\ref{fig:linfit}. 

  The alternative way of doing the correction is a calibration using several time slots. For each run and for each threshold, the spill time is divided into 5 slots.  A Gaussian fit is performed separately to the number of hits distribution of each threshold for each time slot. The mean value  from the fit to the first distribution (at the beginning of the spill) is taken as a reference. The correction factors for the other 4 time slots $(i=2\cdots 5)$ and for each threshold $(j=1\cdots3)$ are defined as $C_{i,j}= \bar{N}_{1,j}/\bar{N}_{i,j}$ where $\bar{N}_{i,j}$ is the mean number of hits for a given time slot $(i)$ and a given threshold $(j)$. The corrected number of hits associated to the $(j)$ threshold  $N_{\mathrm{corr}_j}$ of a given event occurring in the time slot $(i)$ is then defined as following:
\begin{center}
\begin{equation}
  N_{\mathrm{corr}_j}= {N_j* C_{i,j}}.
\end{equation}
\end{center}    
\noindent
For both methods the corrected total number of hits $N_{\mathrm{hit}}$ is given by:
\begin{center}
\begin{equation}
 N_{\mathrm{hit}} =\sum_{j=1}^{3} N_{\mathrm{corr}_j}.
\end{equation}
\end{center}    
\noindent
The two calibration methods presented above are able to correct for the beam intensity effect. We observe that the energy resolution is slightly better for the linear fit calibration method while the linearity is found to be a little worse in this case.  Nevertheless, for both the linearity and the resolution, the relative difference of the results obtained with the two methods is found  to be within a few percent.  Finally the lack of statistics for some runs lead us to choose the  linear fit calibration method as the default one for both H2 and H6 data samples and the difference of the results obtained with the two methods is included in the systematics. Another cross check is done by using the data sample from the first time slots only. The result in resolution is similar to that after spill time correction and the difference of the two methods output is found to be within statistical uncertainties. 

\section{Energy resolution}

The selection of hadronic showers based on the criteria presented in section~\ref{selections} is important for the study of the  linearity and the energy 
resolution of the hadronic showers measured in the SDHCAL prototype in the running conditions presented in the introduction.
The selected hadronic showers belonging to runs of the same  energy are combined for H2 and for H6 data separately 
and the distribution of the total number of hits ($\Nhit$) is plotted for each energy.

\begin{figure}[htp]
\begin{center}
\begin{tabular}{cc}
\includegraphics[width=0.325\textwidth]{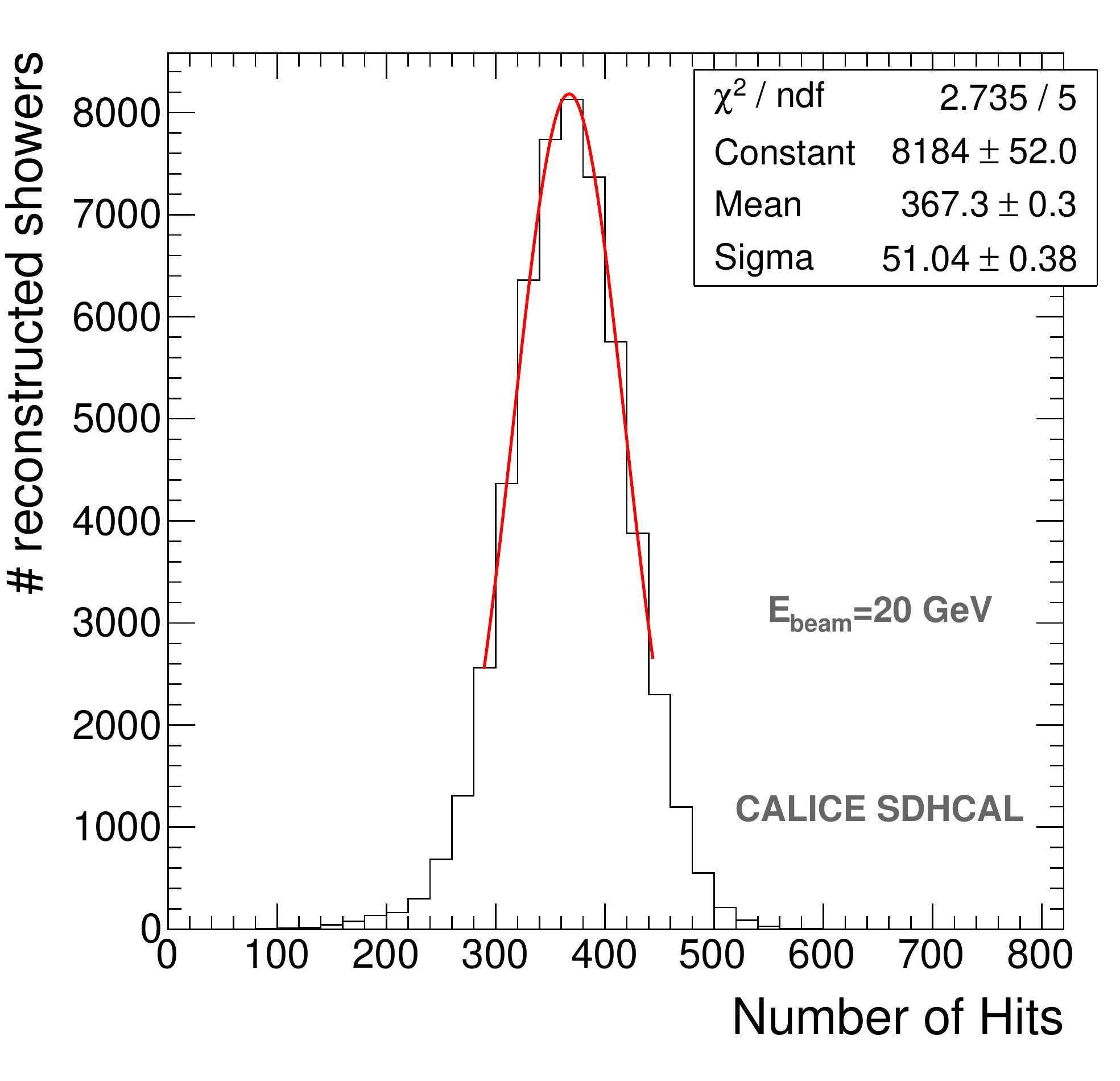} 
\includegraphics[width=0.325\textwidth]{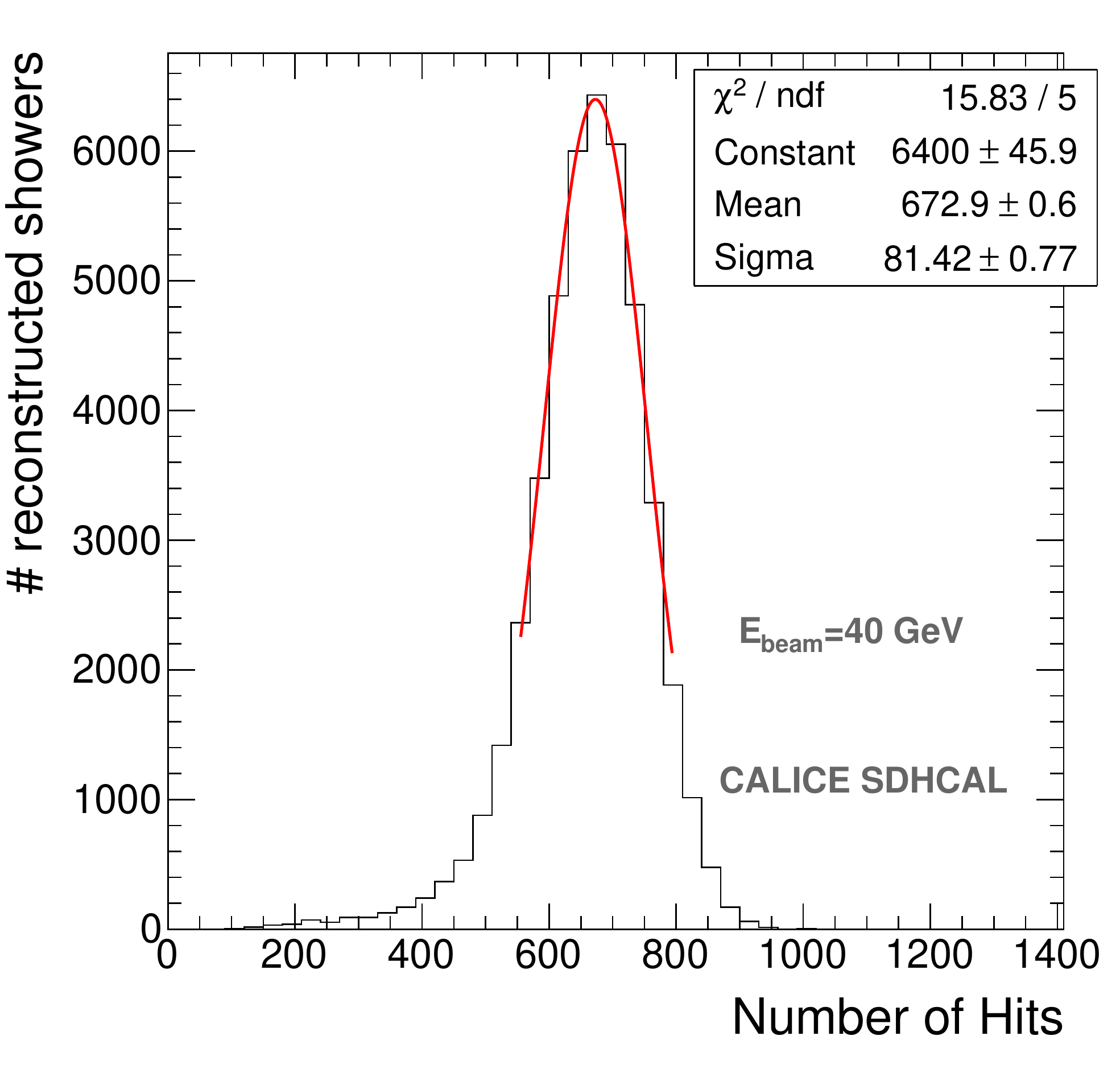} 
\includegraphics[width=0.325\textwidth]{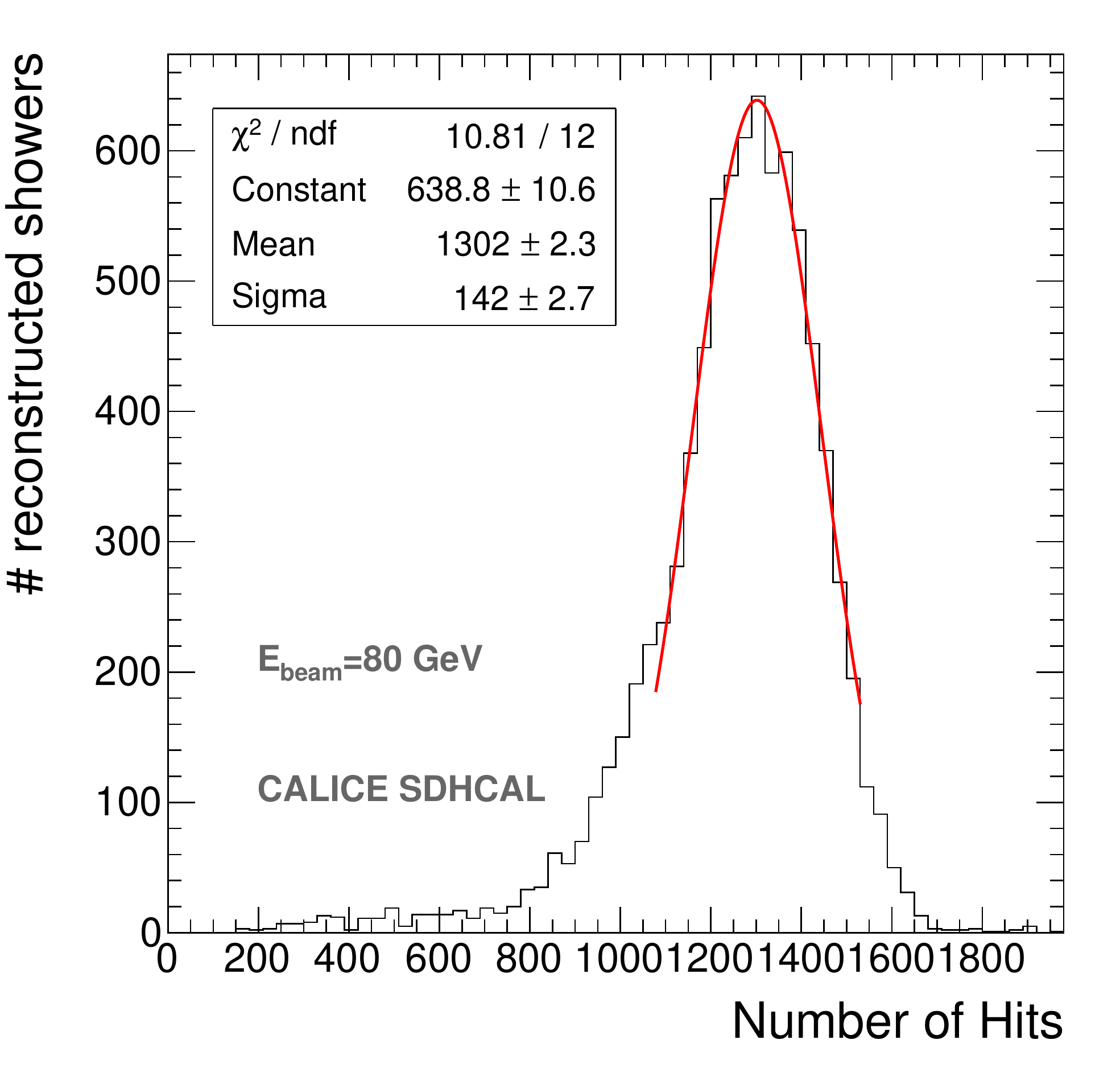} 
\end{tabular}
\caption{Histogram of the total number of hits for pion showers of 20~GeV (left), 40~GeV (middle) and  80~GeV (right)  of the 2012 H6 runs. 
The distributions are fitted with a Gaussian function in a $\pm 1.5 \sigma$ range 
around the mean.
}
\label{fig.Nhits_Energies_Gauss}
\end{center}
\end{figure}

Two kinds of fits are used to estimate the average number of hits for a given energy.  The first uses a two-step Gaussian fit procedure. First, a Gaussian is used to fit over the full range of the distribution. Secondly, a Gaussian is fitted only in the range of $\pm1.5 \sigma$ of the mean value of the first fit.  This limitation is imposed because of the tail at small number of $\Nhit$ that is due essentially to hadrons showering late in the prototype as well as remaining radiating muons.  The $\sigma$ of the second fit was used for the estimation of the resolution. The second fit uses  the Crystal Ball~(CB) function~\cite{CrystalBall} to take into consideration the presence of this tail.
The Crystal Ball function has been used for the estimation of systematic uncertainties as described in section~\ref{SYS}.     
Fig.~\ref{fig.Nhits_Energies_Gauss} shows the distributions of $\Nhit$ with the Gaussian fit for three energies of the H6 runs.\\

\begin{figure}[htp]
\begin{center}
\includegraphics[width=0.45\textwidth]{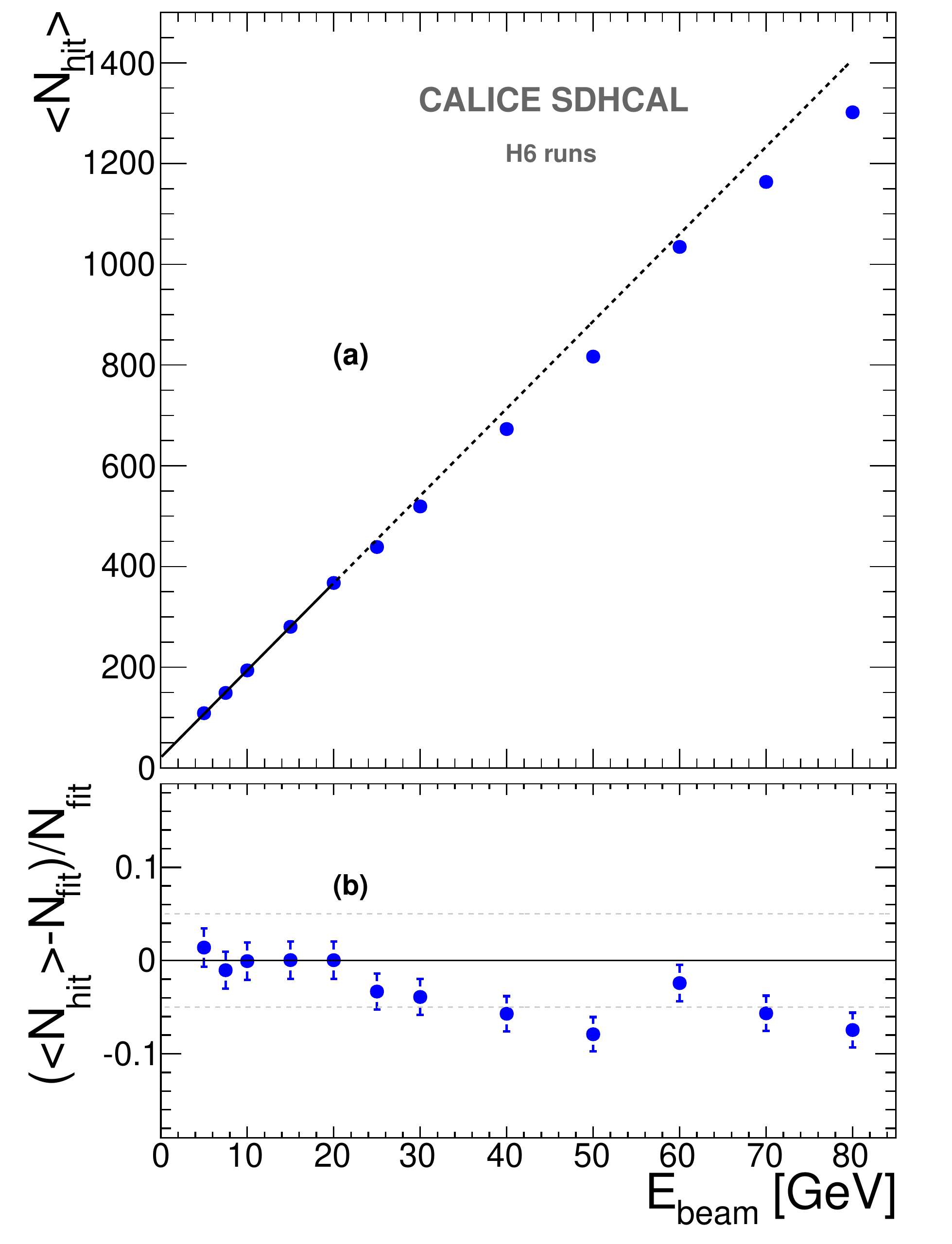}
\includegraphics[width=0.45\textwidth]{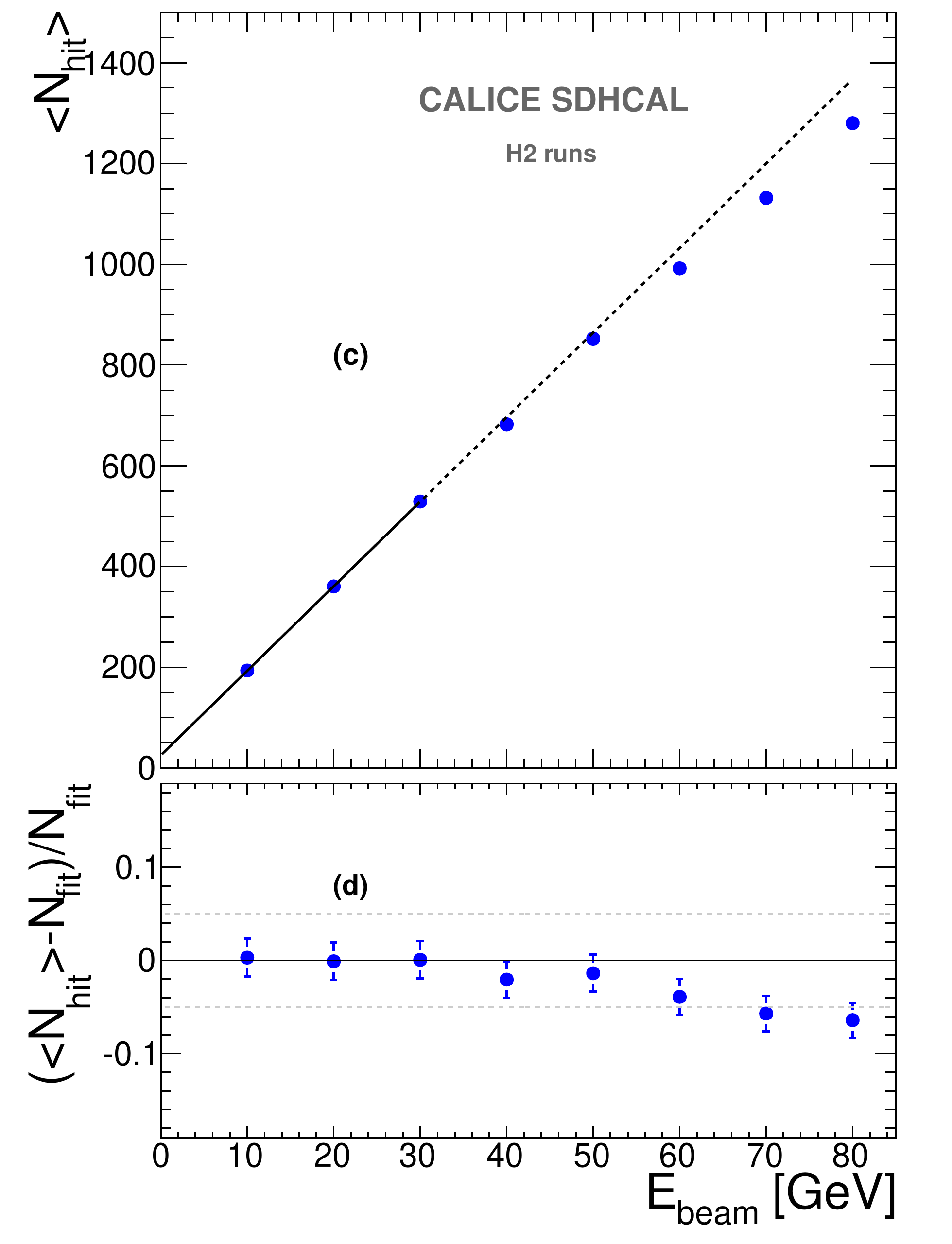}
\caption{ Mean number of hits as a function of the beam energy (a) for reconstructed hadron showers of the 2012 H6 data  and for pion showers (c) of the 2012 H2 data. The line indicates the result of a linear fit for energies up to 20 (30) GeV(solid section of the line) to the H6 (H2) data.  Relative deviation of the observed mean number of hits to the fitted line (b) of the 2012 H6 and (d) of the 2012 H2 data as a function of the beam energy for reconstructed hadronic showers.}
\label{fig.Nhit_linearity_ecart}
\end{center}
\end{figure}

The results  for the $\Nhit$ values obtained using the Gaussian  fit are summarized in Fig.~\ref{fig.Nhit_linearity_ecart}.    In this figure and in the following, all results include systematic uncertainties estimated as explained in Sect.~\ref{SYS}. Fig.~\ref{fig.Nhit_linearity_ecart}  also shows the deviation of the detector response with respect to the straight line  
obtained by  fitting the data points  between 5-20~GeV  in case of the 2012 H6 runs and  those between 10-30~GeV for the 2012 H2 runs.   

\noindent

\subsection{SDHCAL: Binary mode}\label{sec.BinaryMode}
The observed behavior of the mean number of hits ($\Nhit$) as a function of the energy (Fig.~\ref{fig.Nhit_linearity_ecart})
suggests that one can estimate the energy of the hadronic shower up to 20-30~GeV  to a good approximation by using the formula~: $ E_\mathrm{reco} = C \Nhit +D $ with parameters $C$ and $D$ determined from data.   For higher energy, saturation effects show up. The number of hits lies increasingly  lower than the linear extrapolation from low energy data.   The increase of the number of particles in the core of the shower as the energy increases does not result in a proportional  increase of the fired pads.  To account for the deviation with respect to the linear behavior, several parameterizations were tested. A good linearity is obtained using the formula~: 

 \begin{center}
\begin{equation}
 E_\mathrm{reco} = A_1 N_{\mathrm{hit}}  + A_2  N_{\mathrm{hit}}^2 + A_3 N_{\mathrm{hit}}^3.  
\end{equation}
\label{eq0}
\end{center}

\begin{figure}[htp]
\begin{center}
\begin{tabular}{cc}
\includegraphics[width=0.45\textwidth]{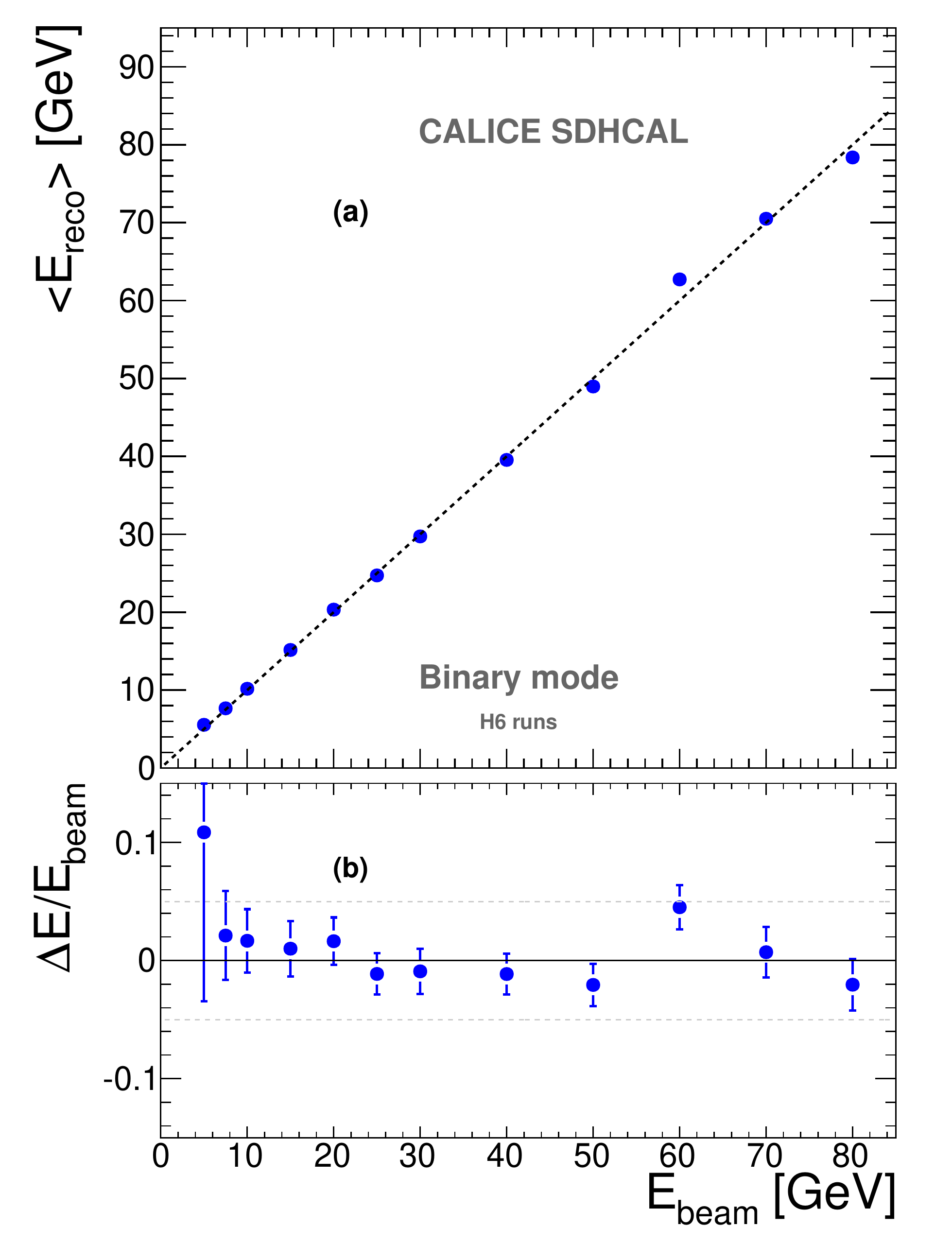} &
\includegraphics[width=0.45\textwidth]{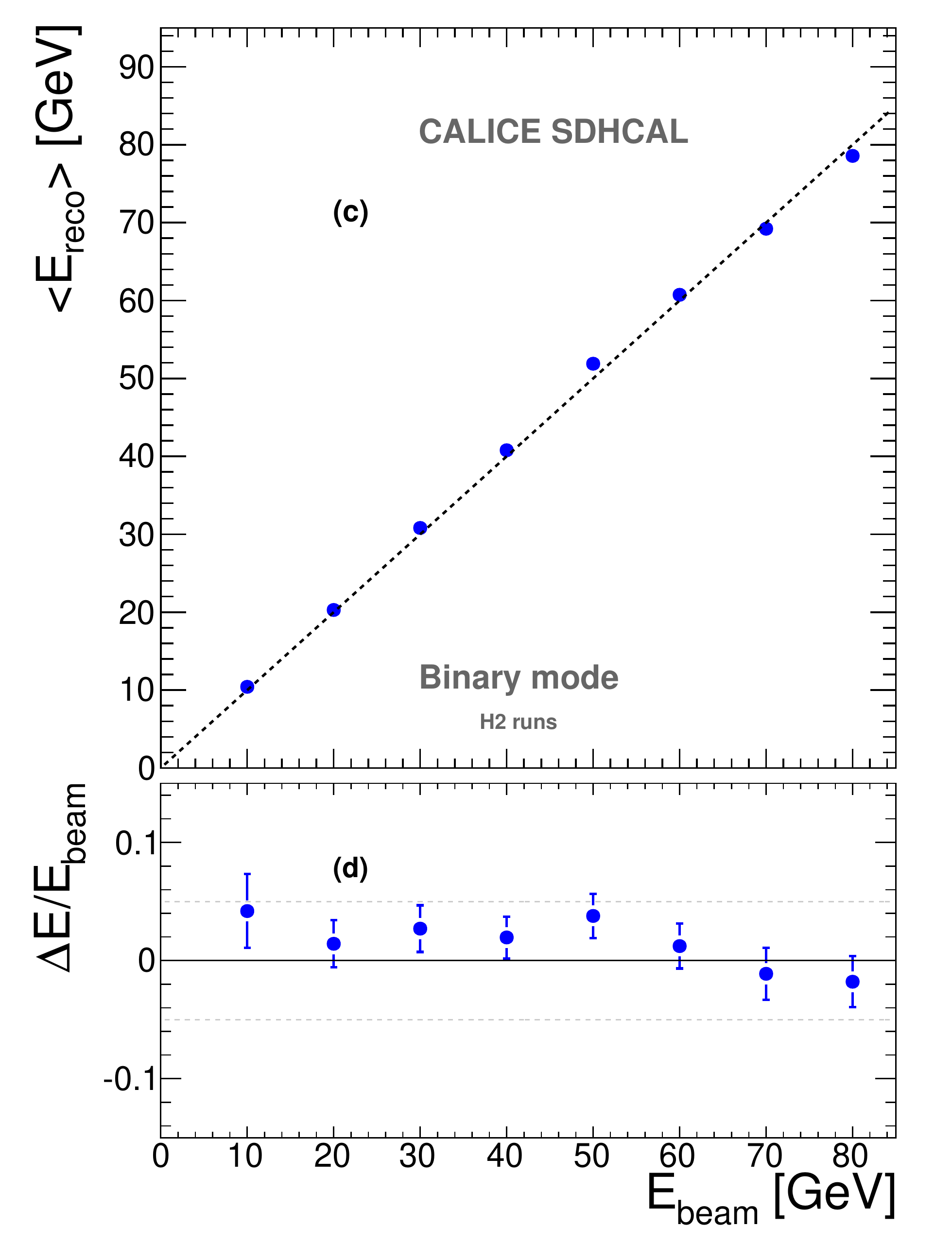}
\end{tabular}

\caption{ Mean reconstructed energy for hadron showers (a) in H6  and (c) in H2 data  as a function of the beam energy.  Errors bars are shown but are smaller than the marker points. The dashed line passes through the origin with unit gradient.
Relative deviation of the  mean reconstructed energy with respect to the beam energy as a function of the beam energy for hadron showers in H6 (b) and H2 (d) data. The reconstructed energy is computed using  only the total number of hits ($\Nhit$)  and the linearity-restoring algorithm described in section~\protect\ref{sec.BinaryMode}. }
\label{fig.DHCAL_linearity_ecart}
\end{center}
\end{figure}

\begin{figure}[htp]
\begin{center}
\begin{tabular}{cc}
\includegraphics[width=0.45\textwidth]{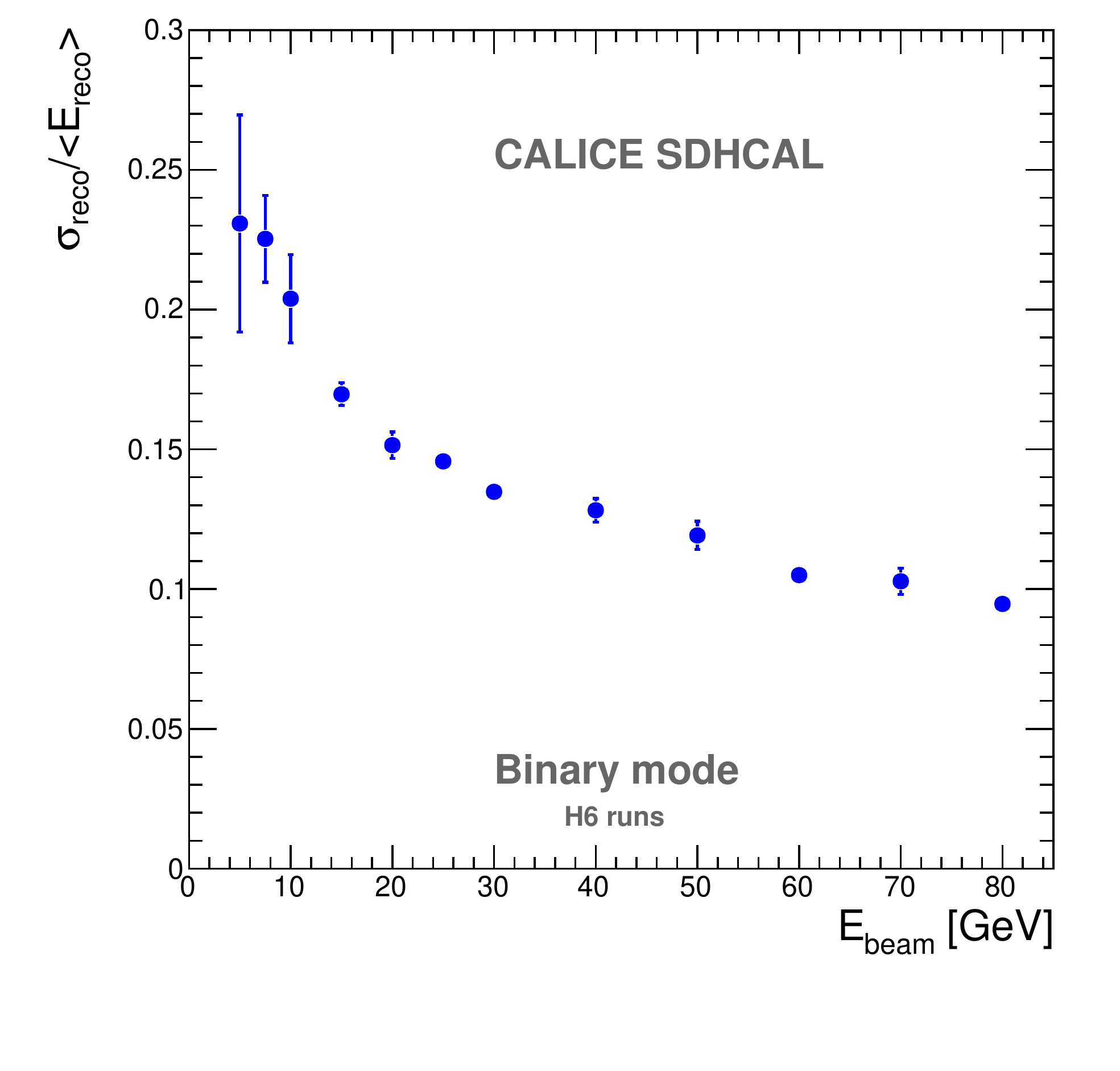} 
\includegraphics[width=0.45\textwidth]{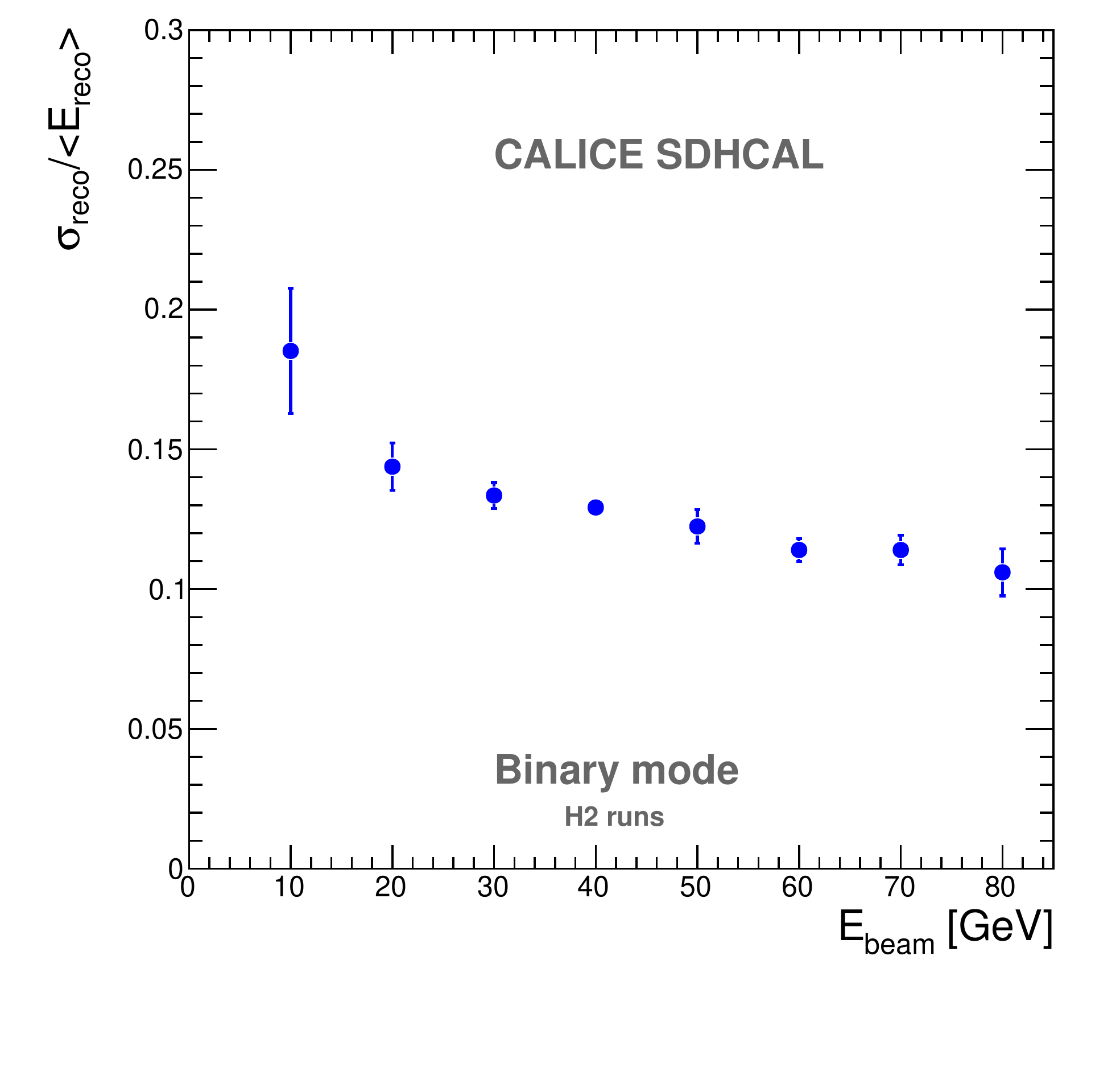}
\end{tabular}

\caption{$\frac{\sigma_{{\mathrm{reco}}}}{<E_{\mathrm{reco}}>}$ is the relative resolution of the reconstructed hadron energy as a function of the beam energy at H6 (left) and H2 (right) runs. The reconstructed energy is computed using  only the total number of hits ($\Nhit$)  and the linearity-restoring algorithm described in section~\protect\ref{sec.BinaryMode}. }

\label{fig.DHCAL_res}
\end{center}
\end{figure}

The method used to extract  the different coefficients $A_1$, $A_2$ and $A_3$ from data, based on a $\chi^2$ minimization technique will be explained in the next section (Formula 6.3) along with the multi-threshold mode. Once the parameters $A_1$, $A_2$ and $A_3$ are fixed, the energy of each hadronic shower is determined using the Formula 6.1. The energy distributions obtained in this way are fitted using the same two-step Gaussian fit procedure employed for the total number of hits. This provides an estimation of the energy of the different runs as well as the associated resolution.
As expected, this method of shower energy reconstruction restores linearity as can be seen in Fig.\ref{fig.DHCAL_linearity_ecart} for H6 and H2 runs.   
 Fig.~\ref{fig.DHCAL_res} shows the relative resolution as a function of the beam energy for the two runs.\\

\subsection{SDHCAL: Multi-threshold mode}
\label{sec.MultiThreshMode}

\begin{figure}[htp]
\begin{center}
\begin{tabular}{cc}
\includegraphics[width=0.42\textwidth]{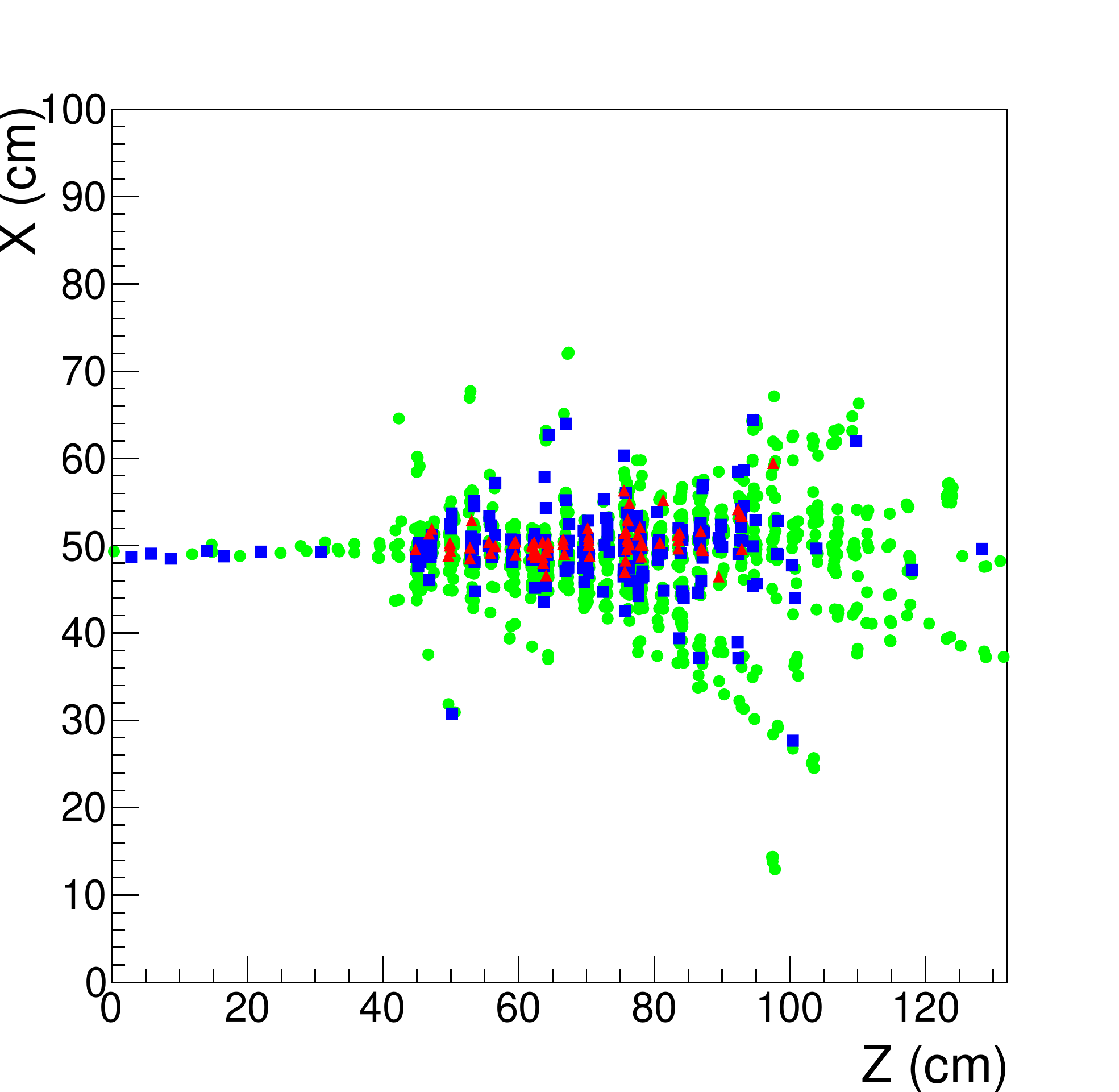} &
\includegraphics[width=0.42\textwidth]{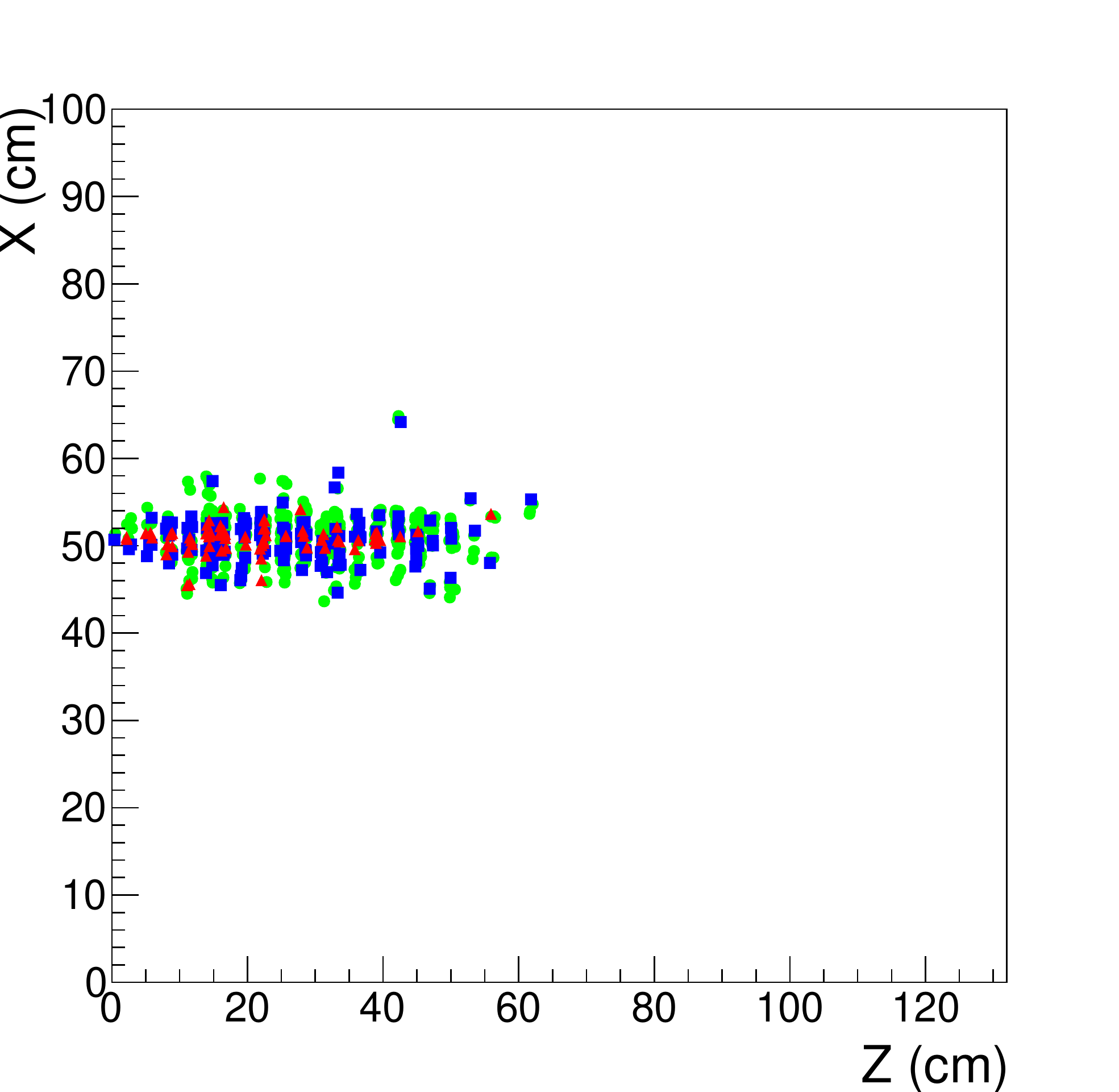}
\end{tabular}
\caption{Left: 70~GeV pion event display with red color triangles indicating highest threshold fired pads, blue color squares  indicating  the middle threshold, and green color circles indicating that of the lowest one. Right: 70~GeV electron event display with the same color coding. }
\label{fig.event_display_thresholds_pion}
\end{center}
\end{figure}

To fully exploit the data provided by the SDHCAL, the  information related  to the three thresholds can be used. 
As mentioned in section~\protect\ref{sec.prot} this information may help to better estimate the total number of charged particles 
produced in a hadronic shower.  Indeed, pads crossed by two particles at the same time and  separated  
by a  distance larger than that of the avalanche  size (1--2~mm)~\cite{chinese} will have their induced charge added. 
The MIP charge spectrum of the GRPC being broad, the precise measurement of the charge can not indicate the exact number of charged particles 
crossing the pad. However it can help to indicate whether this number is low, large or very large. Although the simulation of electromagnetic and hadronic showers  in the SDHCAL using a realistic GRPC response corroborates this idea,  the first validation of this  is the observation in the core of hadronic and electromagnetic showers, where more particles are expected, of a higher density of hits above the second and  third thresholds as can be seen from the event displays of Fig.~\ref{fig.event_display_thresholds_pion}. 

\begin{figure}[htp]
\begin{center}
\includegraphics[width=0.45\textwidth]{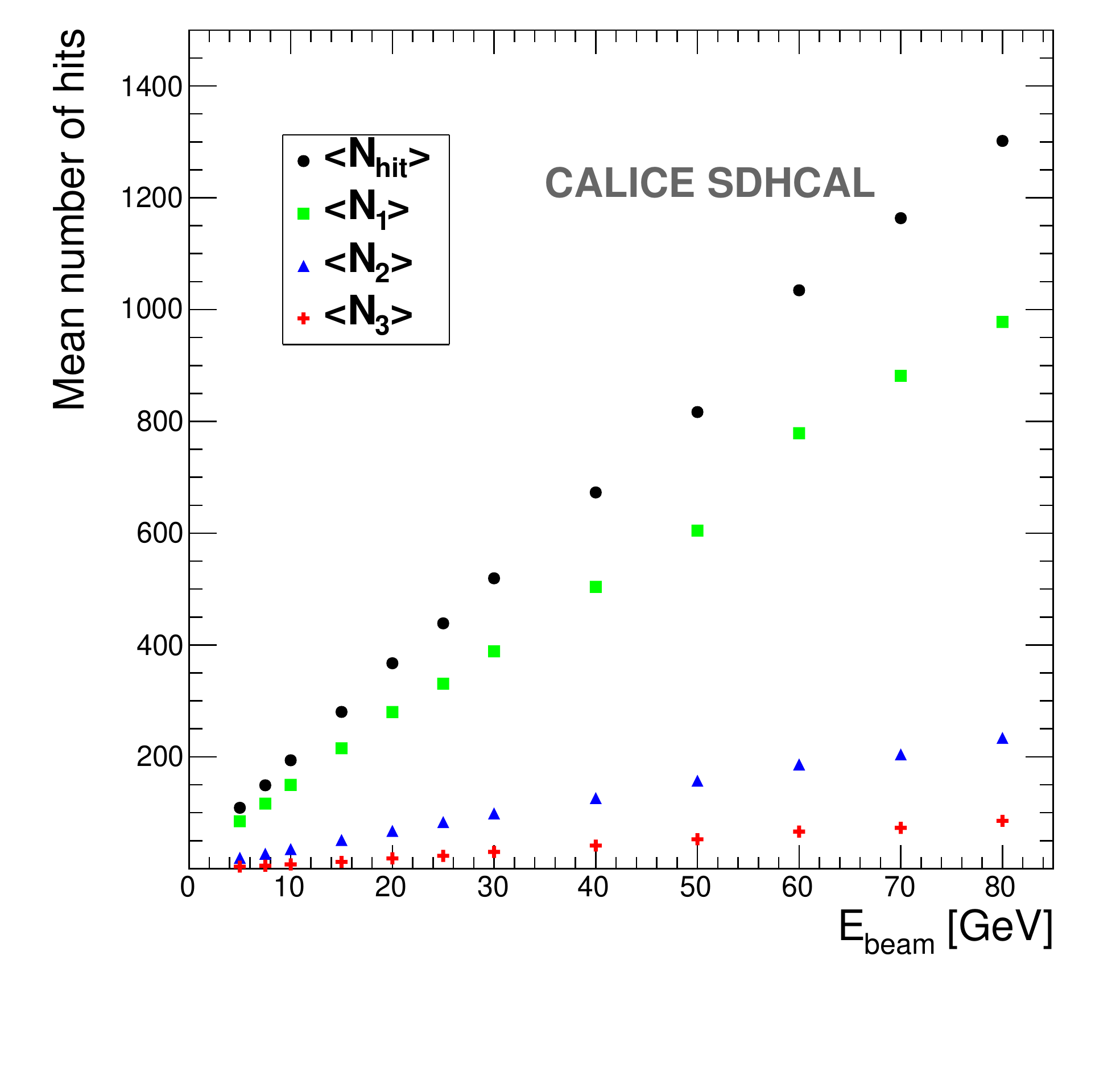} 
\includegraphics[width=0.45\textwidth]{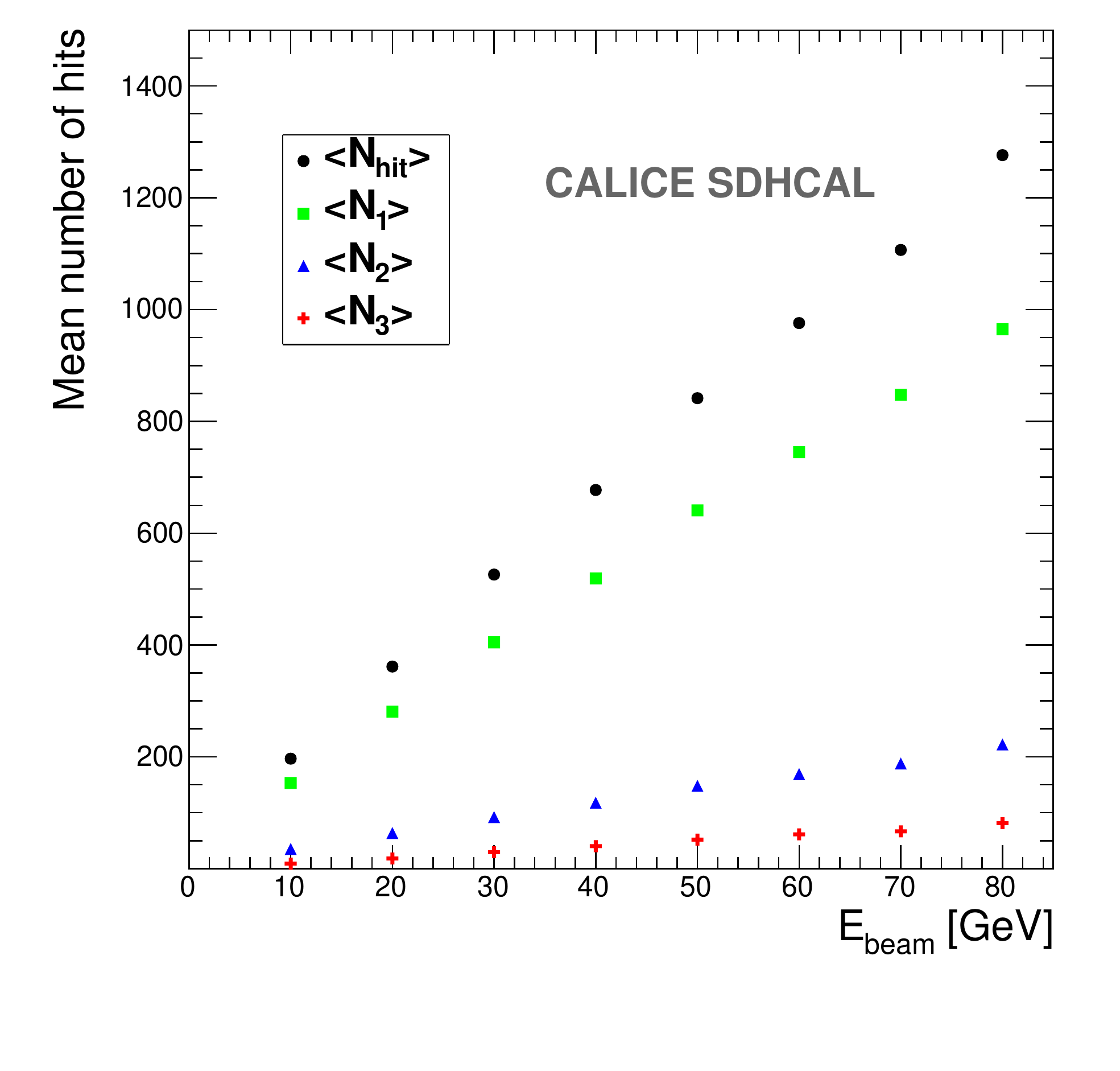}
\caption{Average number of hits in the hadronic shower sample corresponding to the first threshold only (green squares), to the second threshold but not the third  one (blue triangles), to the third threshold (red crosses), and  to the total number (black circles) as a function of the beam energy in the 2012 H6 runs (left) and the 2012 H2 runs (right).}
\label{fig.N123}
\end{center}
\end{figure}

\begin{figure}[htp]
\begin{center}
\includegraphics[width=0.45\textwidth]{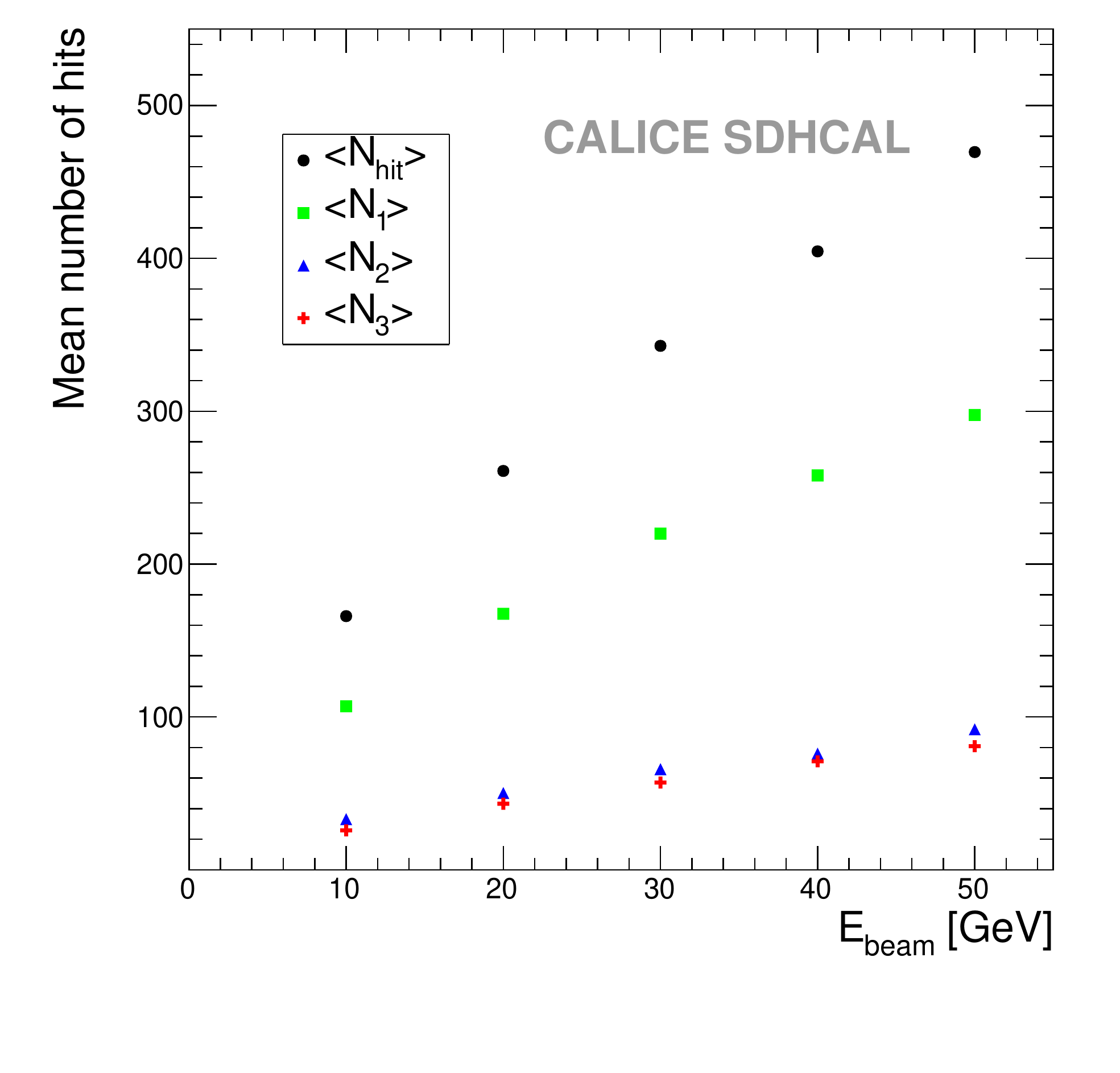} 
\caption{Average number of hits in the electromagnetic shower sample corresponding to the first threshold only (green squares), to the second threshold but not the third  one (blue triangles), to the third threshold (red crosses), and  to the total number (black circles) as a function of the beam energy of electron samples collected in the  2012 H6 beam test.}
\label{fig.N123e}
\end{center}
\end{figure}

\begin{figure}[htp]
\begin{center}
\includegraphics[width=0.7\textwidth]{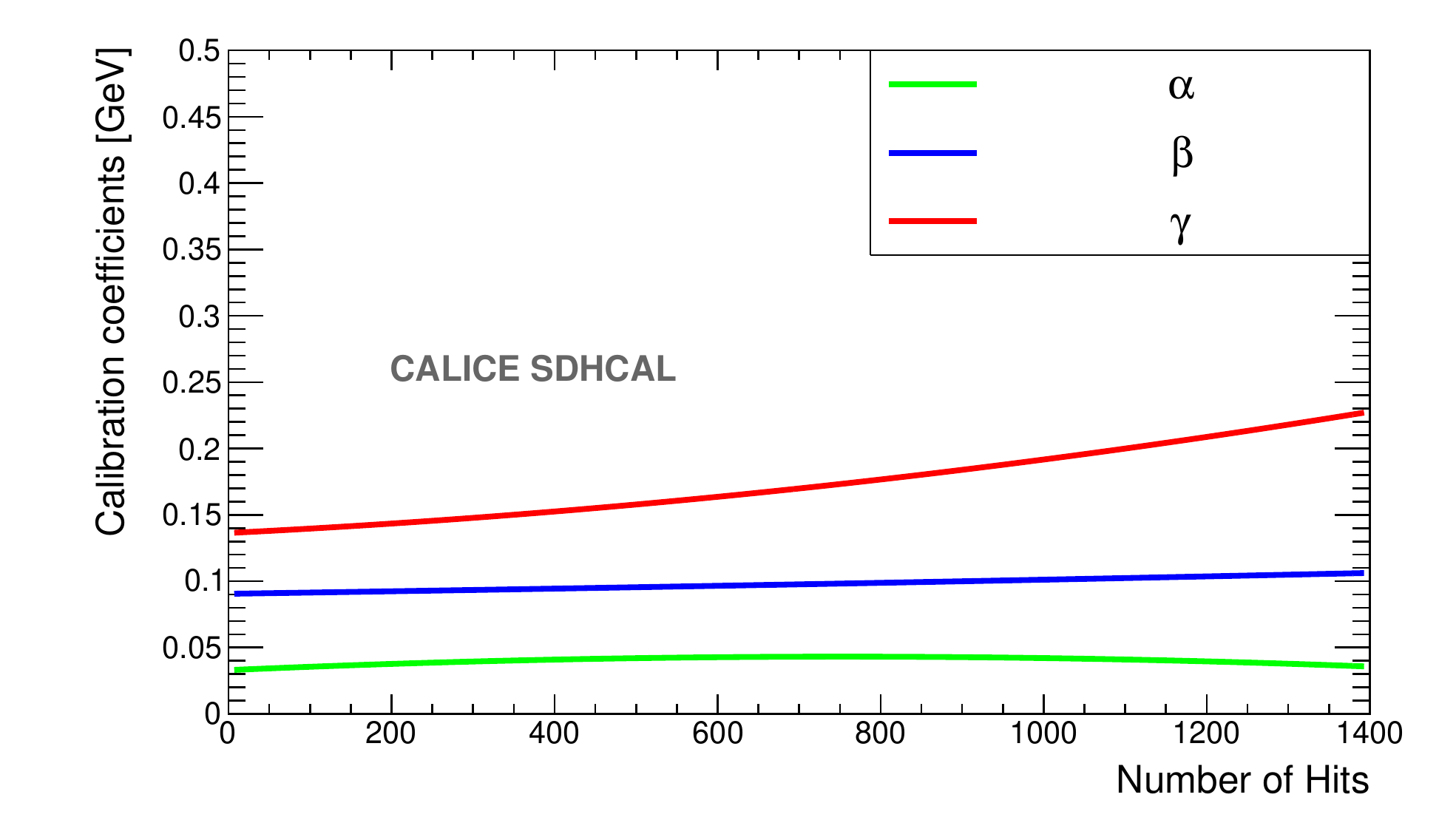}
\caption{Evolution of the coefficient $\alpha$ (green), $\beta$ (blue) and $\gamma$ (red) in terms of the total number of hits.}
\label{fig.coeff}
\end{center}
\end{figure}

\begin{figure}[htp]
\begin{center}
\begin{tabular}{cc}
\includegraphics[width=0.45\textwidth]{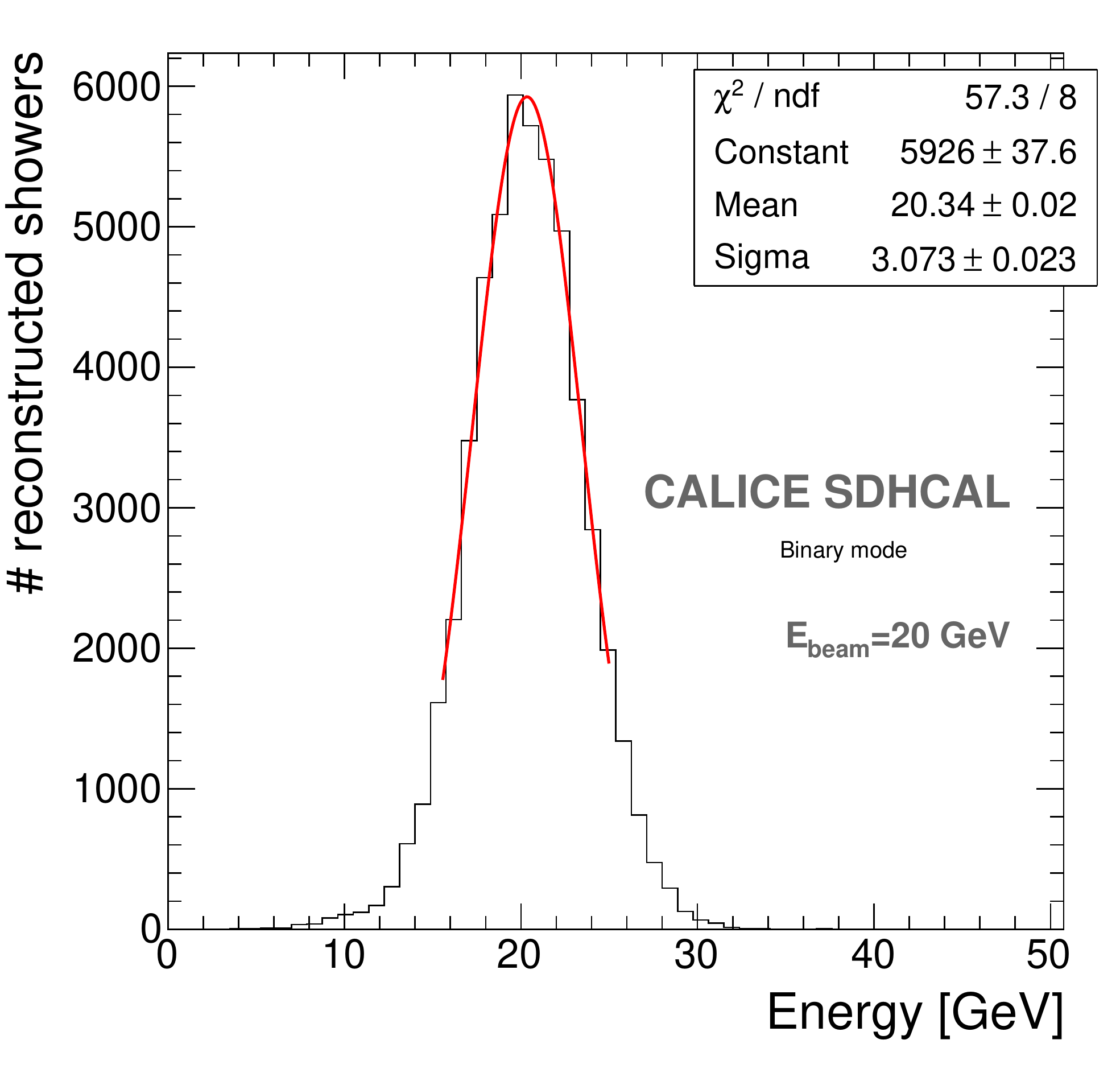} 
\includegraphics[width=0.45\textwidth]{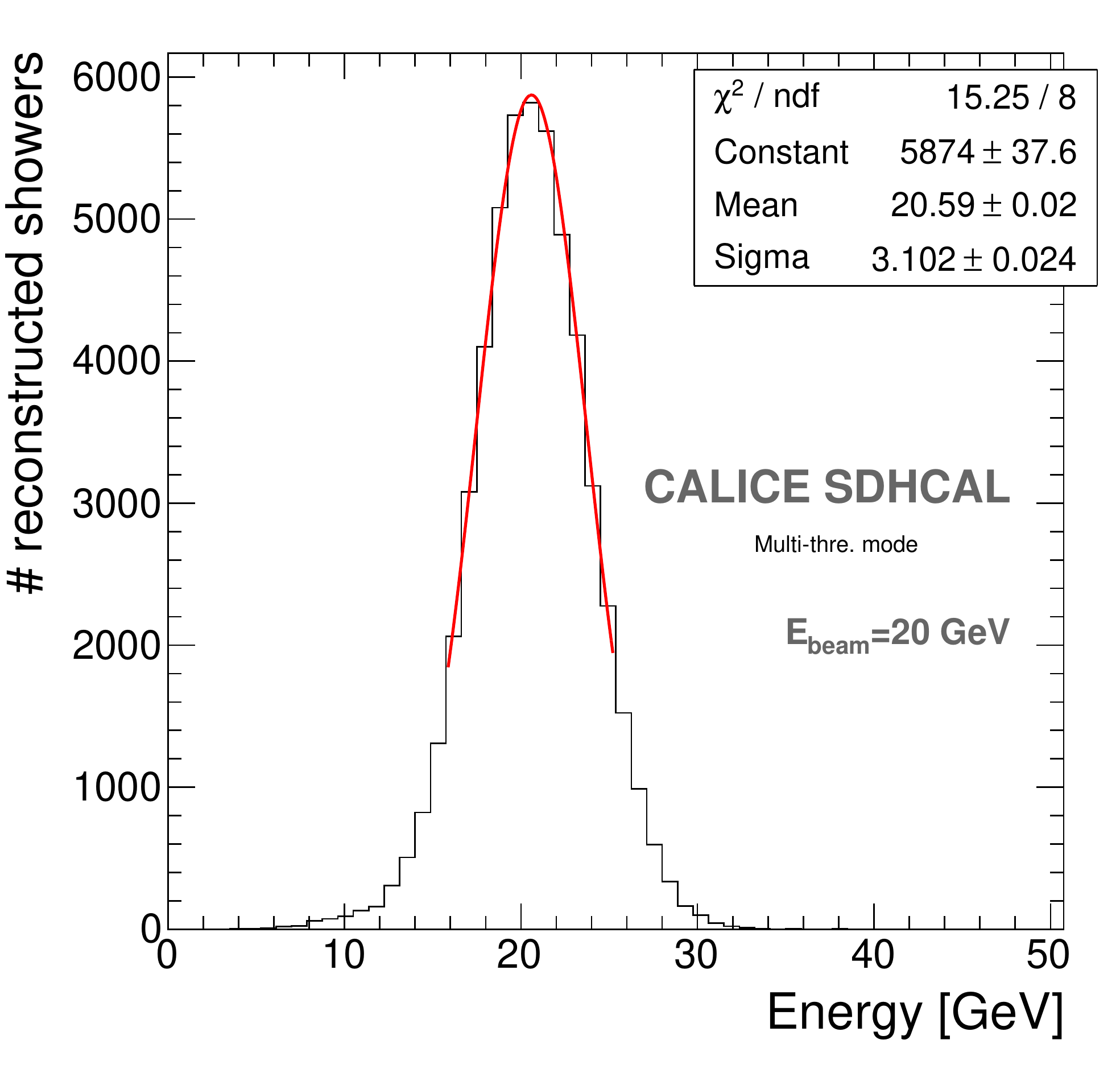} 
\end{tabular}
\caption{Left: histogram of the reconstructed energy for hadron showers of 20~GeV 2012 H6 runs using only the total number of hits (binary mode). Right: histogram of the reconstructed energy for hadron showers using information from three thresholds (multi-threshold mode with energy reconstruction described in section~\protect\ref{sec.MultiThreshMode}).  
The distributions are fitted with a Gaussian function in a $\pm 1.5 \sigma$ range around the mean.
}
\label{fig.DHCAL_Energies_Gauss}
\end{center}
\end{figure}

\begin{figure}[htp]
\begin{center}
\begin{tabular}{cc}
\includegraphics[width=0.45\textwidth]{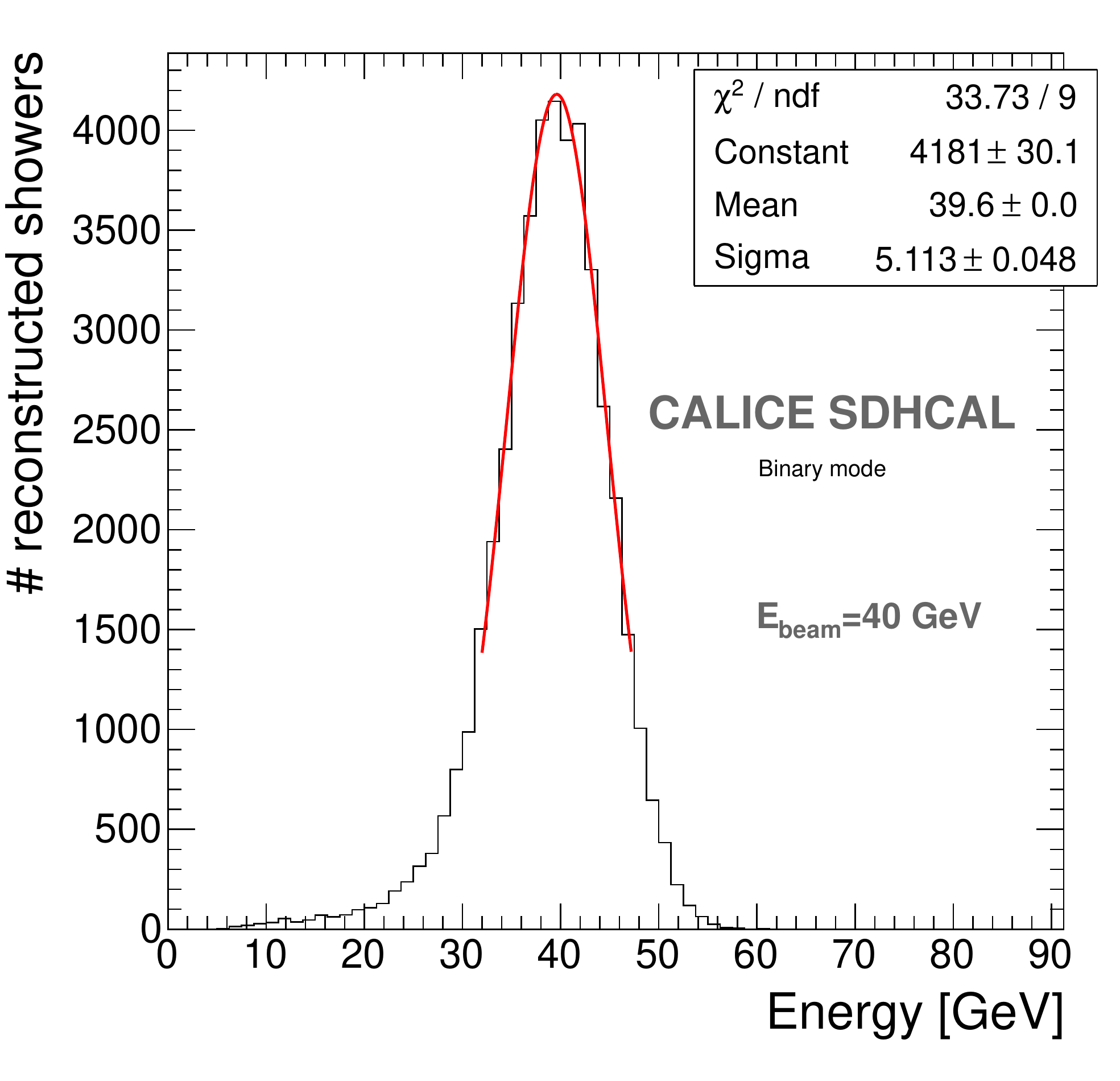} &
\includegraphics[width=0.45\textwidth]{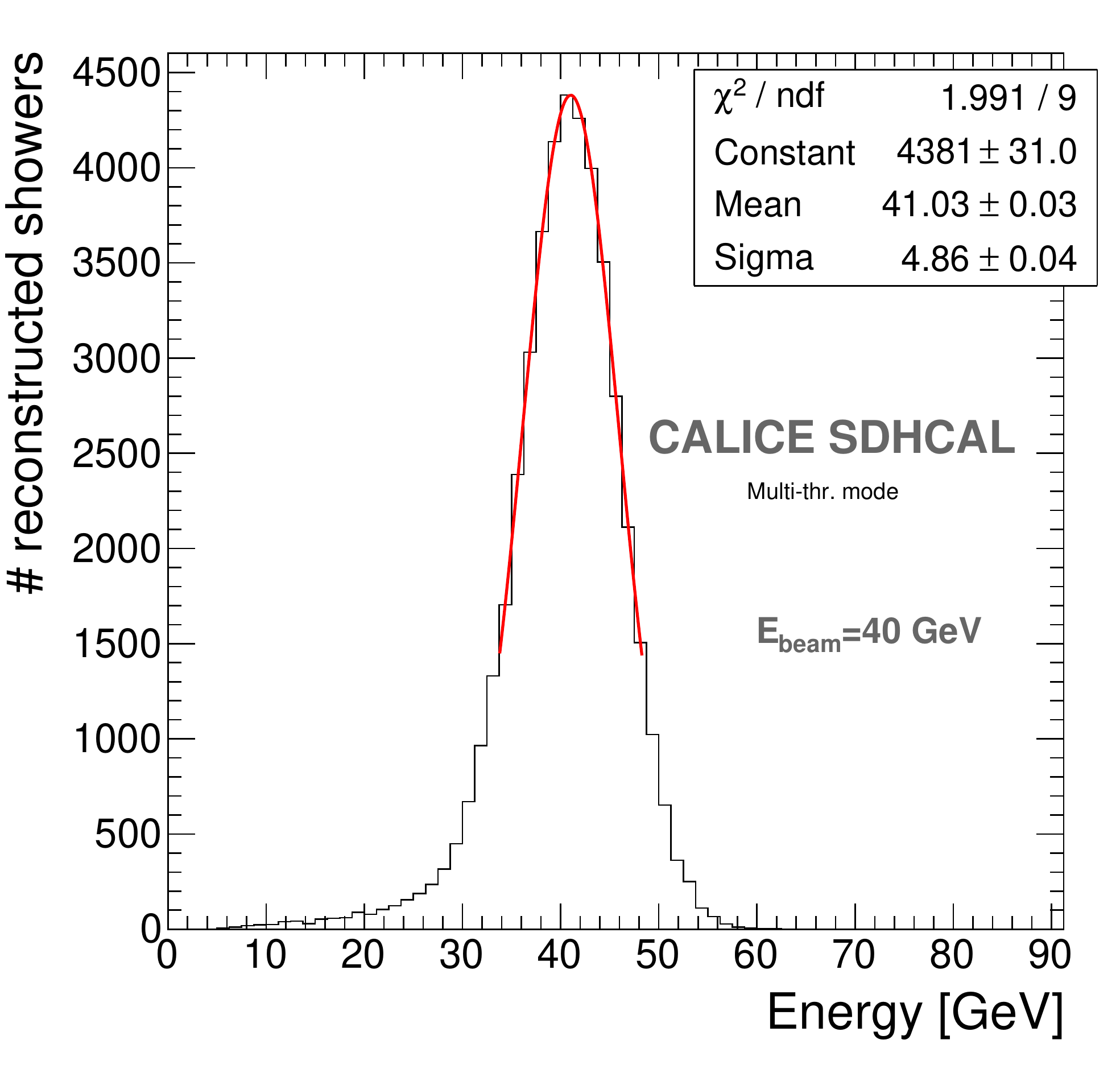} 
\end{tabular}
\caption{Left: histogram of the reconstructed energy for hadron showers of 40~GeV 2012 H6 runs using only the total number of hits (binary mode). Right: histogram of the reconstructed energy for hadron showers using information from three thresholds (multi-threshold mode with energy reconstruction described in section~\protect\ref{sec.MultiThreshMode}).  
The distributions are fitted with a Gaussian function in a $\pm 1.5 \sigma$ range around the mean.
}
\label{fig.DHCAL_Energies_CB}
\end{center}
\end{figure}

\begin{figure}[htp]
\begin{center}
\begin{tabular}{cc}
\includegraphics[width=0.45\textwidth]{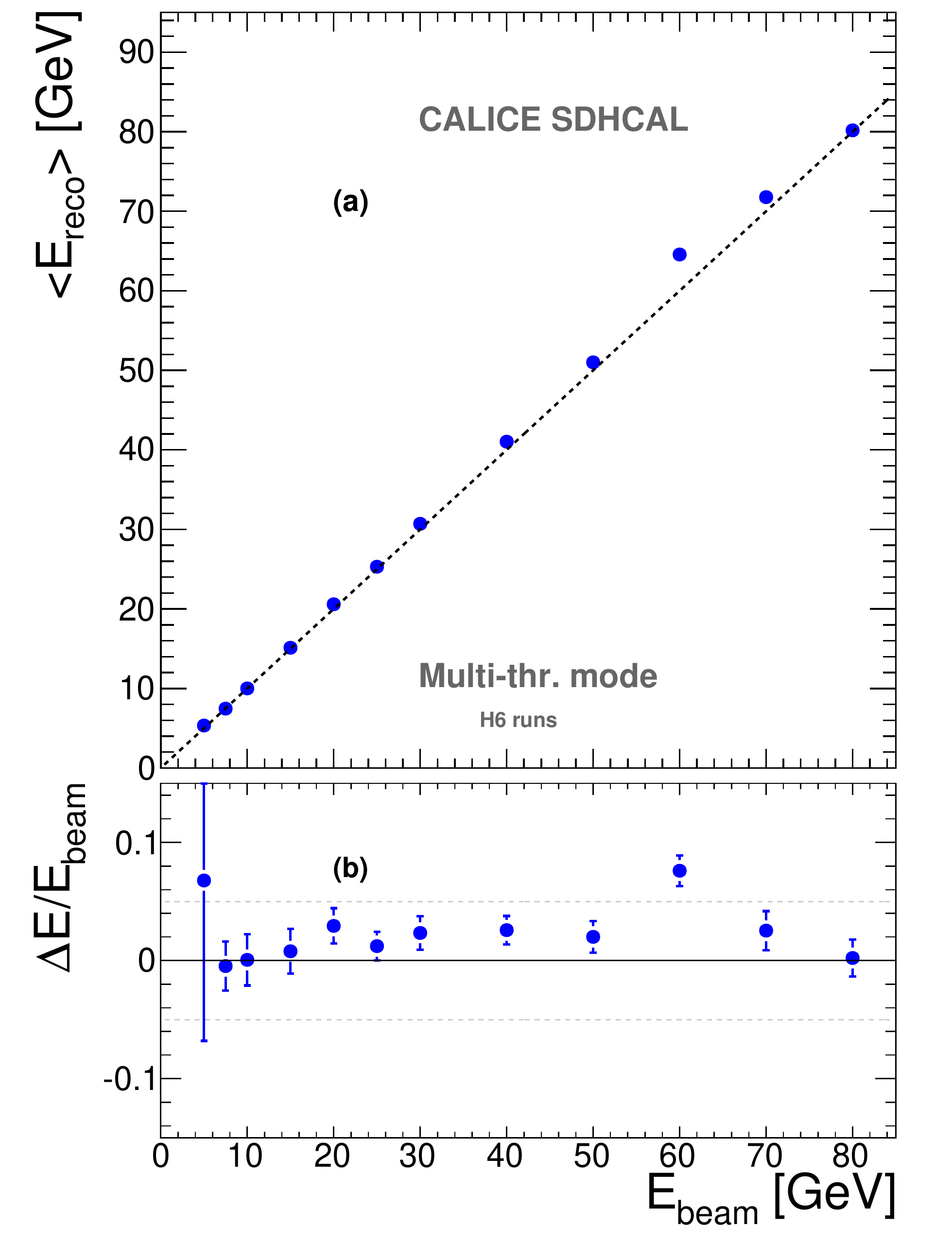} &
\includegraphics[width=0.45\textwidth]{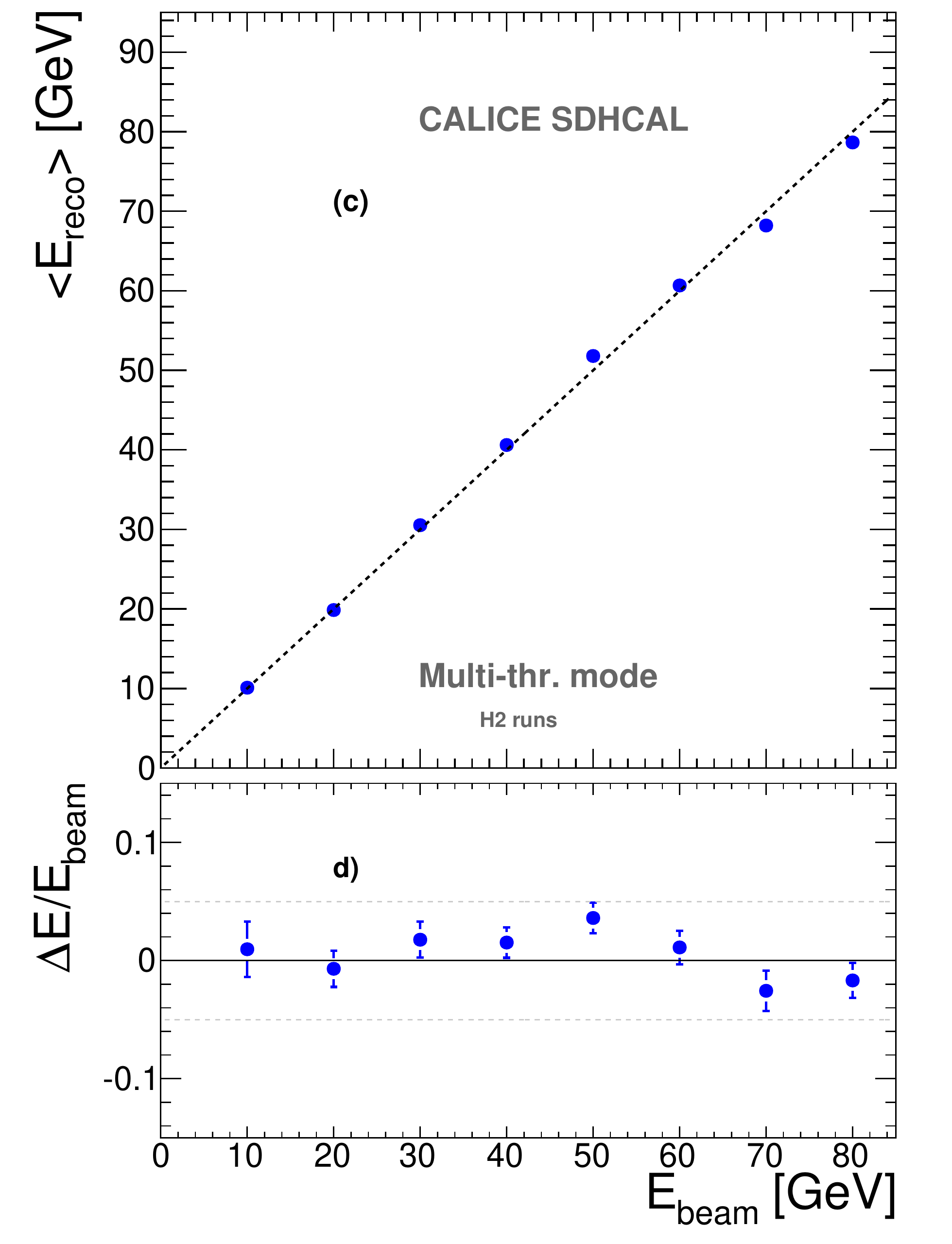}
\end{tabular}
\caption{ Mean reconstructed energy for hadron showers (a) in H6  and (c) in H2 data.  Errors bars are shown but are smaller than the marker points. The dashed line passes through the origin with unit gradient.
Relative deviation of the hadron mean reconstructed energy with respect to the beam energy as a function of the beam energy for hadron showers (b) in H6  and (d) in H2 data. The reconstructed energy is computed using the three thresholds information (multi-threshold mode) and the linearity-restoring algorithm described in section~\protect\ref{sec.MultiThreshMode}. }
\label{fig.SDHCAL_linearity_ecart}
\end{center}
\end{figure}

\begin{figure}[htp]
\begin{center}
\begin{tabular}{cc}
\includegraphics[height=0.32\textheight]{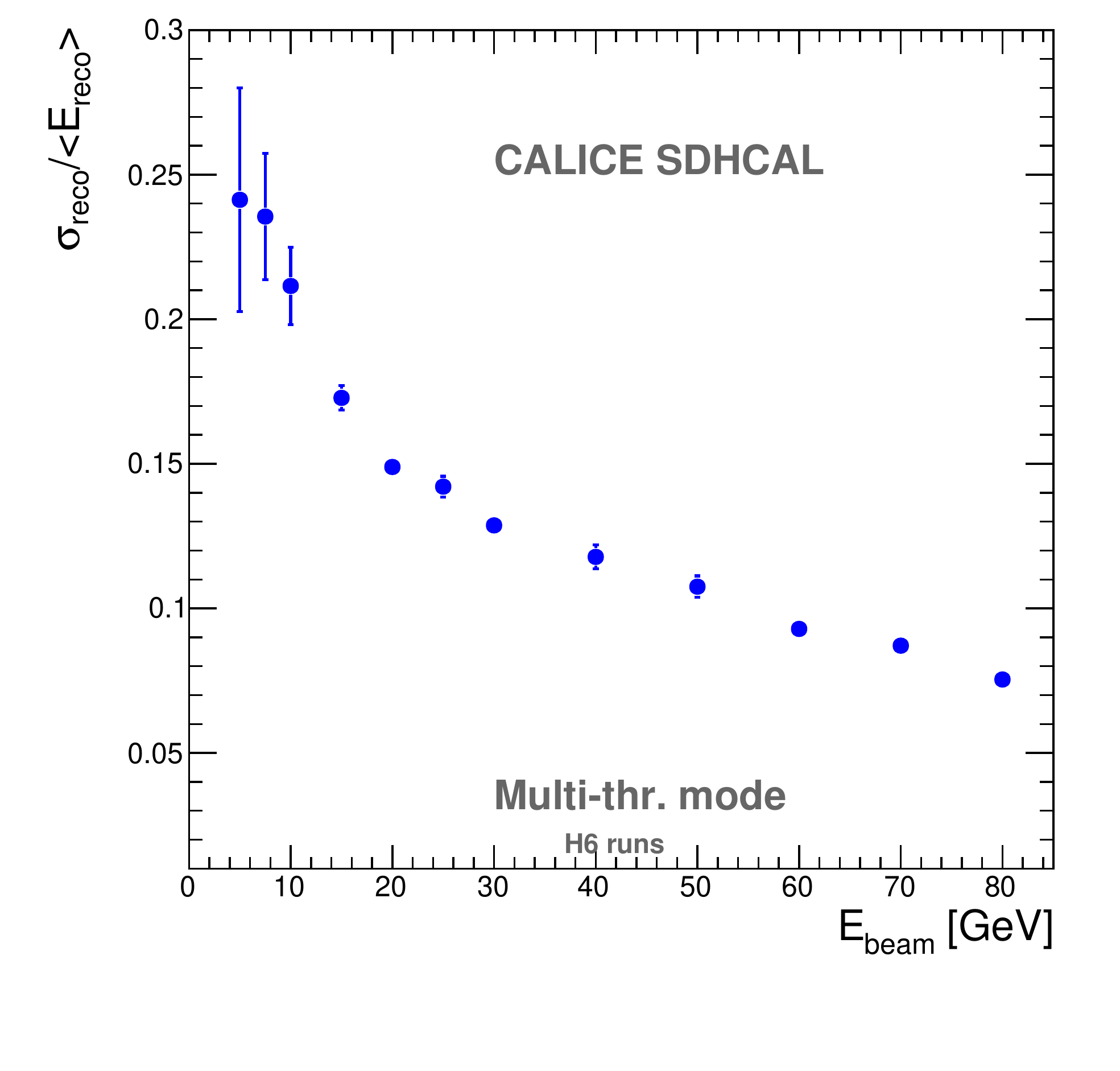} 
\includegraphics[height=0.32\textheight]{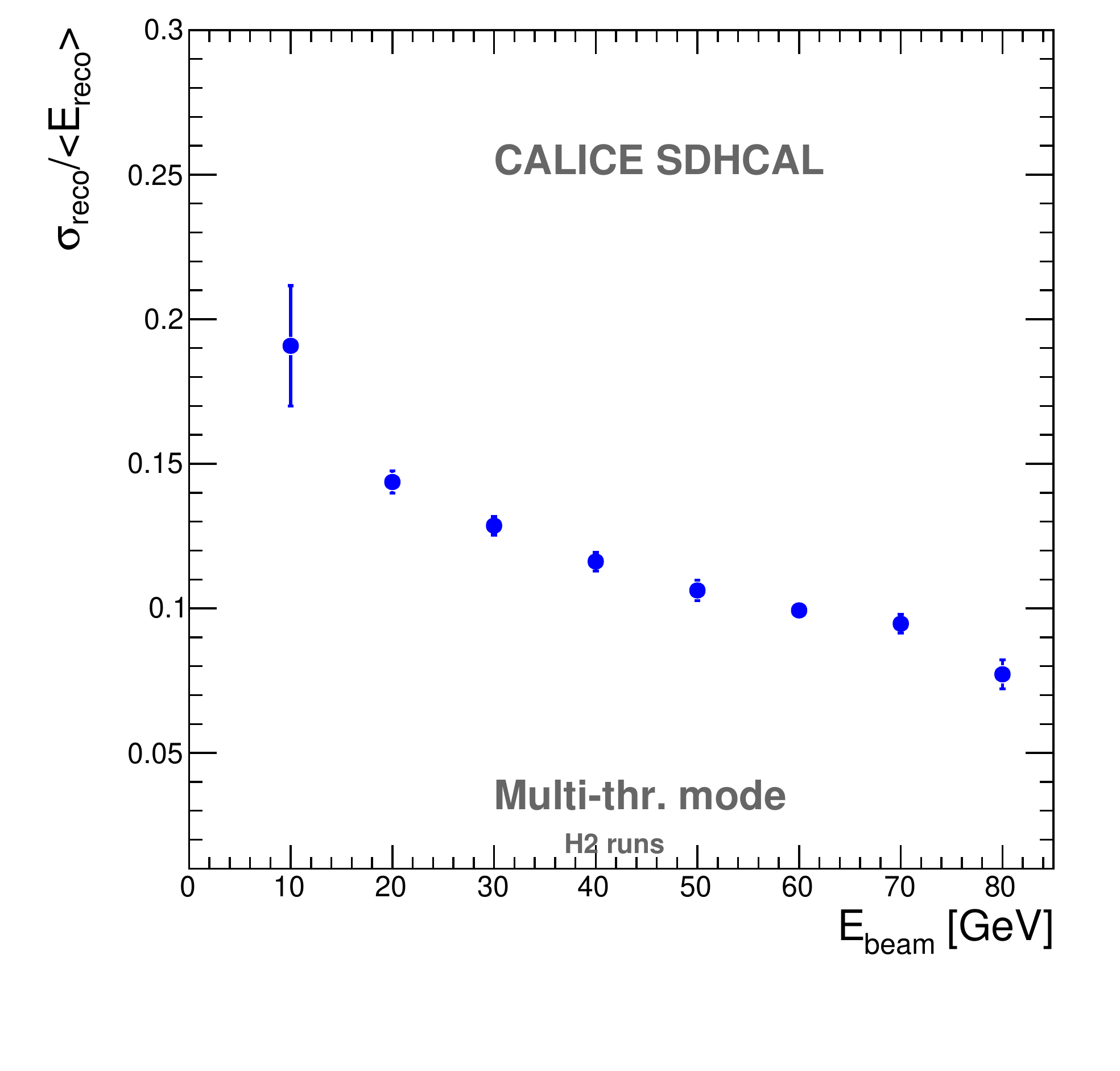} 
\end{tabular}
\caption{$\frac{\sigma_{{\mathrm{reco}}}}{<E_{\mathrm{reco}}>}$ is the relative resolution of the reconstructed hadron energy as a function of the beam energy at H6 (left) and H2 (right) runs.
The reconstructed energy is computed using the three thresholds information as described in section~\protect\ref{sec.MultiThreshMode}. 
}
\label{fig.SDHCAL_Resolution}
\end{center}
\end{figure}

\begin{figure}[htp]
\begin{center}
\includegraphics[width=0.6\textwidth]{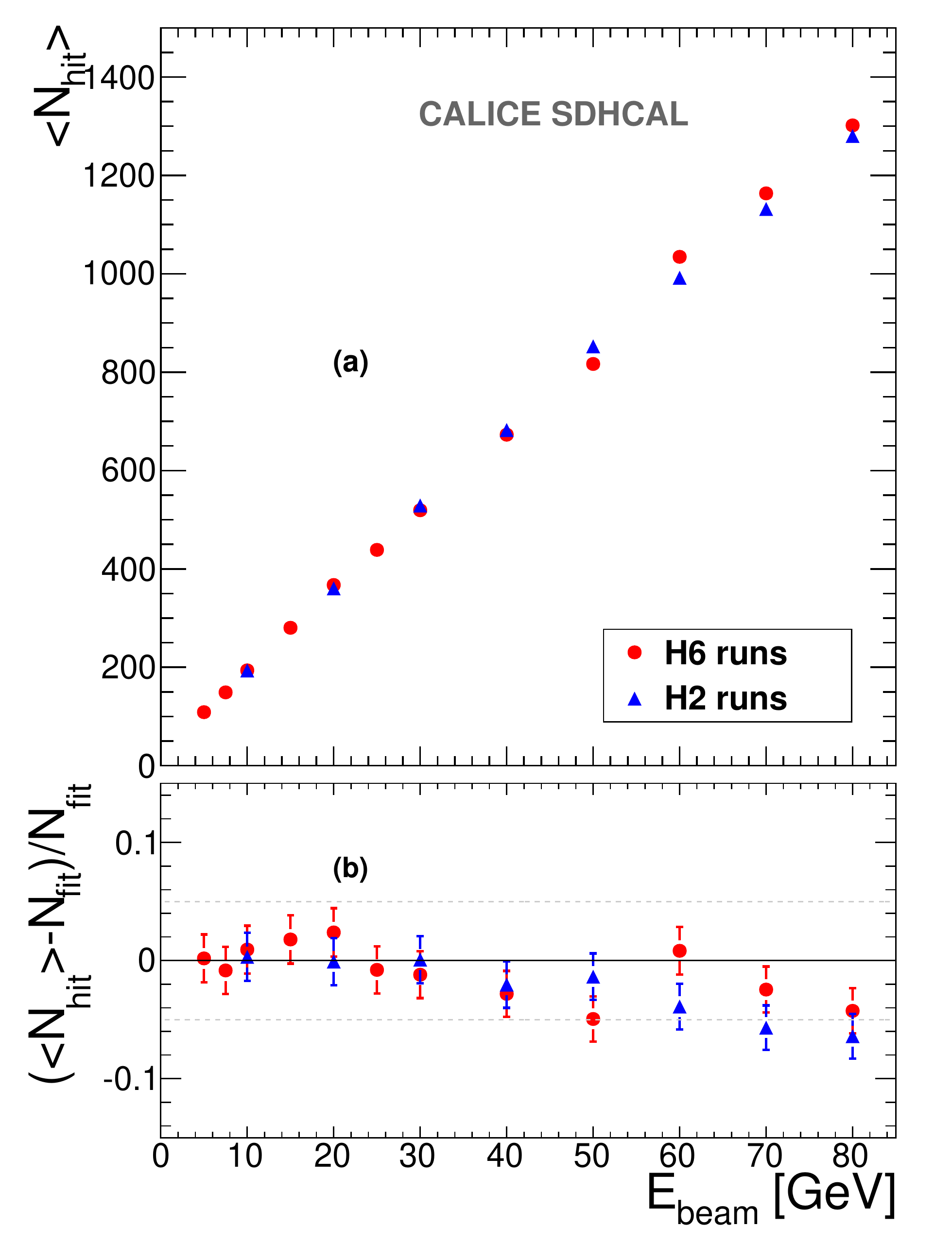}
\caption{ Mean number of hits as a function of the beam energy for reconstructed hadron showers (a) of the 2012 H6 data (red) and for pion showers of the 2012 H2 data (blue).  Relative deviation of the observed mean number of hits (b) of the 2012 H6 data (red) and of the 2012 H2 data (blue) to a linear fit of the H2 data up to 30 GeV as a function of the beam energy for reconstructed hadronic showers.}
\label{fig.Nhit_linearity_ecart_SN}
\end{center}
\end{figure}

\begin{figure}[htp]
\begin{center}
\begin{tabular}{cc}
\includegraphics[width=0.45\textwidth]{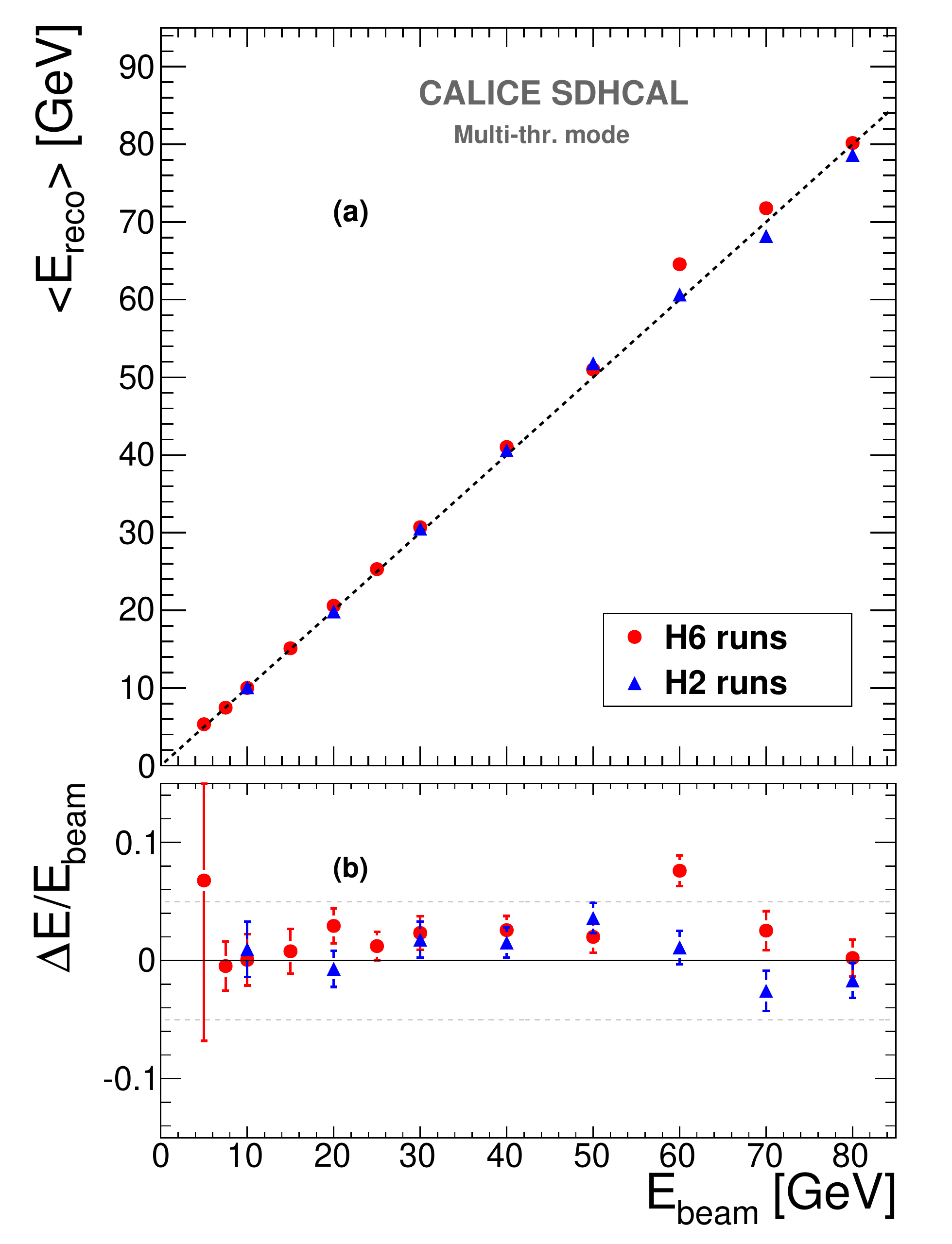} 
\includegraphics[width=0.45\textwidth]{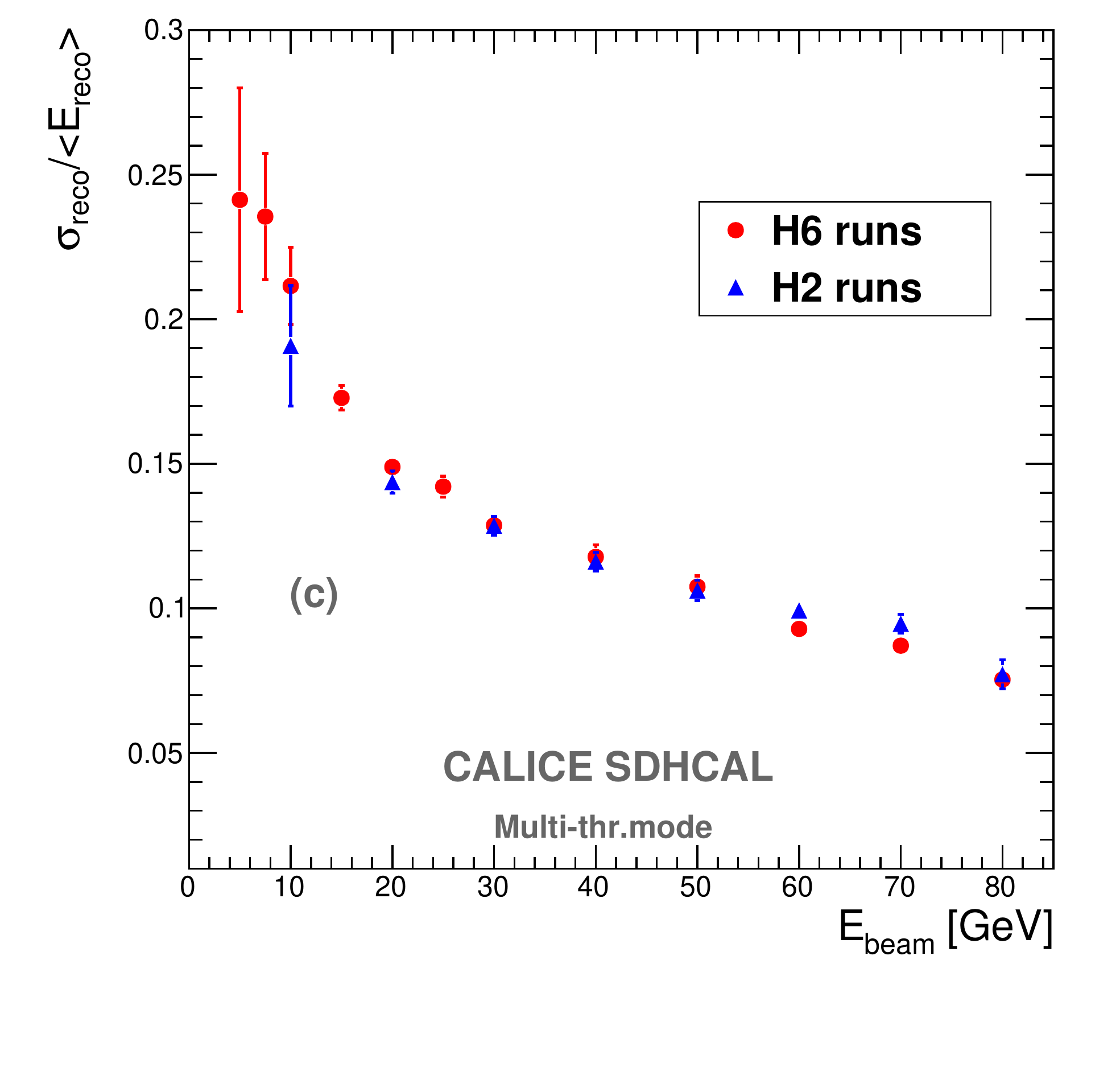}
\end{tabular}
\caption { Mean reconstructed energy for pion showers as a function of the beam energy (a) of the  2012 H2 (blue) and the 2012 H6 (red) data. The dashed line passes through the origin with unit gradient.   Relative deviation of the pion mean reconstructed energy with respect to the beam energy as a function of the beam energy (b) of the  2012 H2 (blue) and the 2012 H6 (red) data. The reconstructed energy is computed using the three thresholds information as described in section~\protect\ref{sec.MultiThreshMode}.  $\frac{\sigma_{{\mathrm{reco}}}}{<E_{\mathrm{reco}}>}$ (c) is the relative resolution of the reconstructed hadron energy as a function of the beam energy of the  2012 H2 (blue) and the 2012 H6 (red) data.}
\label{fig.ALL_linearity_ecart_SN}
\end{center}
\end{figure}

\begin{figure}[!h]
\begin{center}
\includegraphics[width=.48\textwidth]{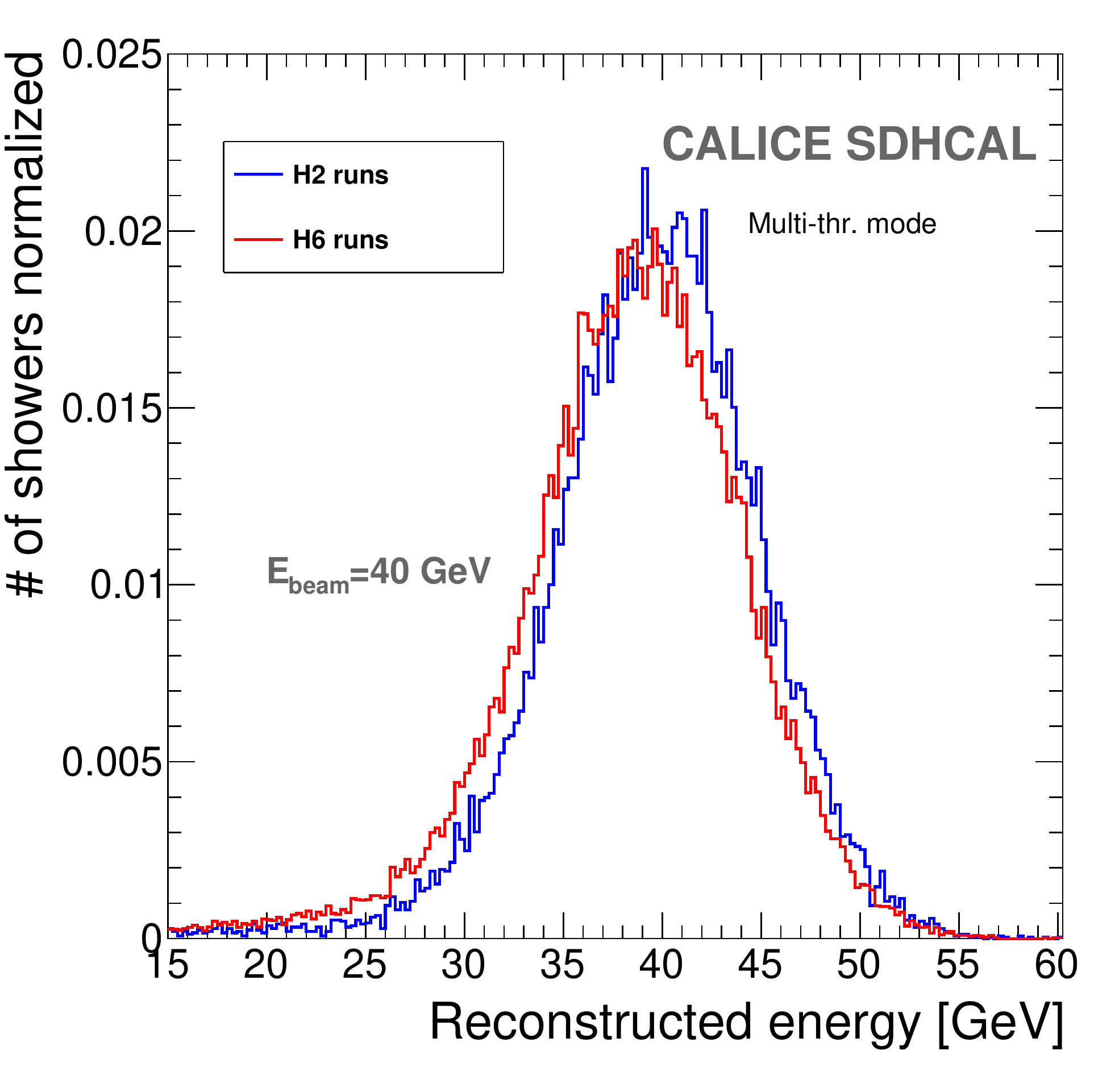}
\includegraphics[width=.48\textwidth]{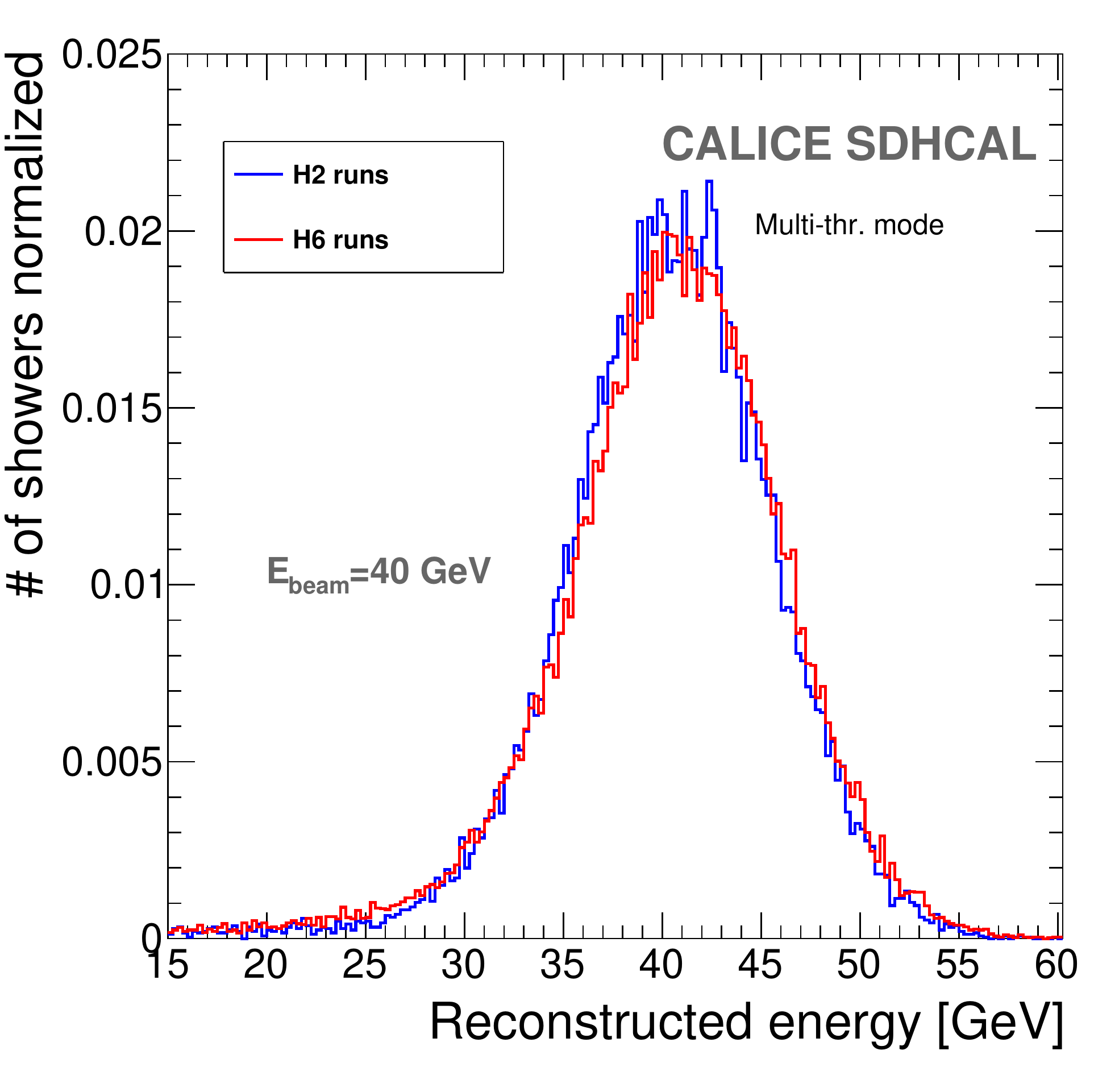}
\caption{Reconstructed energy distributions for H6 (red line) and H2  (blue line) 40~GeV runs before (left) and after (right) beam intensity correction using the multi-threshold mode.}
\label{fig:hits}
\end{center}
\end{figure}

 The threshold information can be useful to understand the shower structure as suggested by these event displays. Nevertheless, here this  will be used  only to improve the energy measurement by expressing the  energy of the hadronic shower as a weighted sum of  $N_1$,  $N_2$ and $N_3$. In Fig.~\ref{fig.N123} the average values of $N_1$,  $N_2$, $N_3$ and of the total number of hits of the selected hadronic showers are shown for the 2012 H6 runs (left) and the 2012 H2 runs (right). 
In Fig.~\ref{fig.N123e}  the same variables are shown for electron samples obtained during the the 2012 H6 beam test. Since electromagnetic showers feature, on average, a more compact and dense structure than hadronic ones, the difference of ratios, between electromagnetic and hadronic showers, of the hits above the second and the third threshold over the total number of hits confirms that more hits above the  second and  third thresholds are present in  high particle density regions within the shower  where more particles are present\footnote{The presence of these hits is however not limited to the high particle density zones only. A few of them are also observed at the end of stopping particles where the $dE/dx$ is high.}. Therefore, using the thresholds information could help to account better for the number of particles created within the shower and improve on its energy estimation.\\ 
Taking an empirical approach, the energy can be reconstructed by a weighted sum:
 \begin{center}
\begin{equation}
 E_{\mathrm{reco}} = \alpha N_1 + \beta N_2 + \gamma N_3.
\end{equation}
\end{center}      
The complexity of the hadronic shower structure and its evolution with energy mean that the optimal values of $ \alpha, \beta $ and $\gamma$ are not constant over a large energy range.
To overcome this difficulty $ \alpha, \beta$ and $\gamma$ are parameterized as functions of the total number of hits 
($\Nhit =    N_1+N_2+N_3$).  
To find the best parameterization, 
a  $\chi^2$-like expression was used for the optimization procedure:
\begin{center}
\begin{equation}
\chi^2= \sum_{i=1}^{N} \frac {(E^{i}_{\mathrm{beam}} -E^{i}_{\mathrm{reco}})^2}  {\sigma^2_{i}}
\end{equation}
\end{center}  
\noindent
where $N$ is the number of events used for the optimisation and $\sigma_i = \sqrt{E^{i}_{\mathrm{beam}}}$\footnote{This choice is suggested by the fact that the calorimeter energy resolution is expected to be approximately proportional to $\sqrt{E_{\mathrm{beam}}}$.}. Different  functions of $\Nhit$ were tested to parameterize the evolution of 
$\alpha, \beta$ and $\gamma$ with $\Nhit$. A polynomial function of second degree was found to give the best results. 
The procedure was applied to only a few energy points using only about a third of the collected data in H2 runs where there is no proton contamination\footnote{
 The same procedure was applied to extract   $A_1$, $A_2$ and $A_3$ in the case of the binary mode.}.   The parameterization of $\alpha, \beta$ and $\gamma$ as a function of $\Nhit$ is presented in Fig.~\ref{fig.coeff}.\\ 
The three coefficients of these polynomial functions  are then used to estimate the energy of all  collected data (H2 and H6 runs) without using the information of the beam energy.  The energy distributions obtained in this way are fitted as before and 
are shown in Figs~\ref{fig.DHCAL_Energies_Gauss} and \ref{fig.DHCAL_Energies_CB} (right) for two energies.   These two figures (left)  also show the energy distributions obtained with the binary mode for comparison. 
 As expected, the multi-threshold method of energy reconstruction of hadronic showers restores linearity over a wide energy range going from 5~GeV up to 80~GeV  as  shown in Figs~\ref{fig.SDHCAL_linearity_ecart} (a,c).  Figs~\ref{fig.SDHCAL_linearity_ecart} (b,d) show the relative deviation of the reconstructed energy with respect to the beam energy.   The use of the three  threshold information has very good impact on the energy resolution (Fig.\ \ref{fig.SDHCAL_Resolution}) at energies higher than 30~GeV as was predicted from our preliminary simulation studies~\cite{calor2010}.   The energy resolution reaches a value of 7.7\% at 80~GeV which is an encouraging result  since the data were collected without using any electronics gain correction to improve the homogeneity of the detector response.  The results obtained with the two data samples with the same  energy points are in a good agreement especially at low energy where the proton contamination of the H6 pion beam is low. For energies higher than 60~GeV, the presence of protons in the H6 data can explain why the reconstructed energies are higher than those of the H2 data. It is important  to mention here that having the same response to hadrons in the energy range for which the proton contamination is low  shows clearly that the behavior of  the SDHCAL prototype is stable between the two periods as can be seen in Fig.~\ref{fig.Nhit_linearity_ecart_SN} and Fig.~\ref{fig.ALL_linearity_ecart_SN} as well as in Table~\ref{tabA} and Table~\ref{tabB}.  The  importance of the  beam intensity correction role in achieving this stability is shown in Fig.~\ref{fig:hits} where the energy  distributions of 40~GeV hadrons from H2 and H6  data  are compared before and after this correction.

\noindent 


\subsection{Multi-threshold compared to binary mode analysis}

\begin{figure}[htp]
\begin{center}
\includegraphics[width=0.48\textwidth]{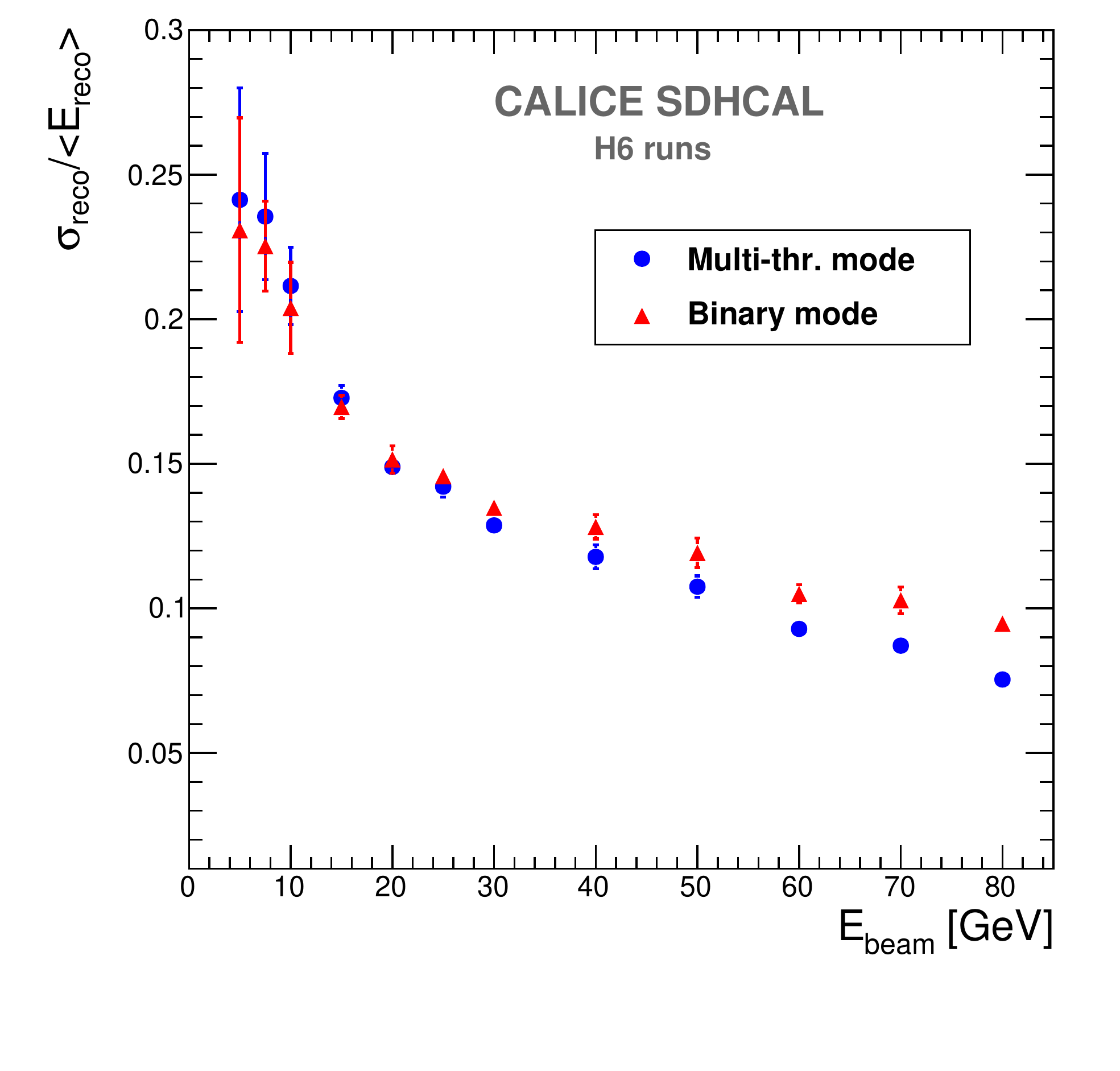}
\includegraphics[width=0.48\textwidth]{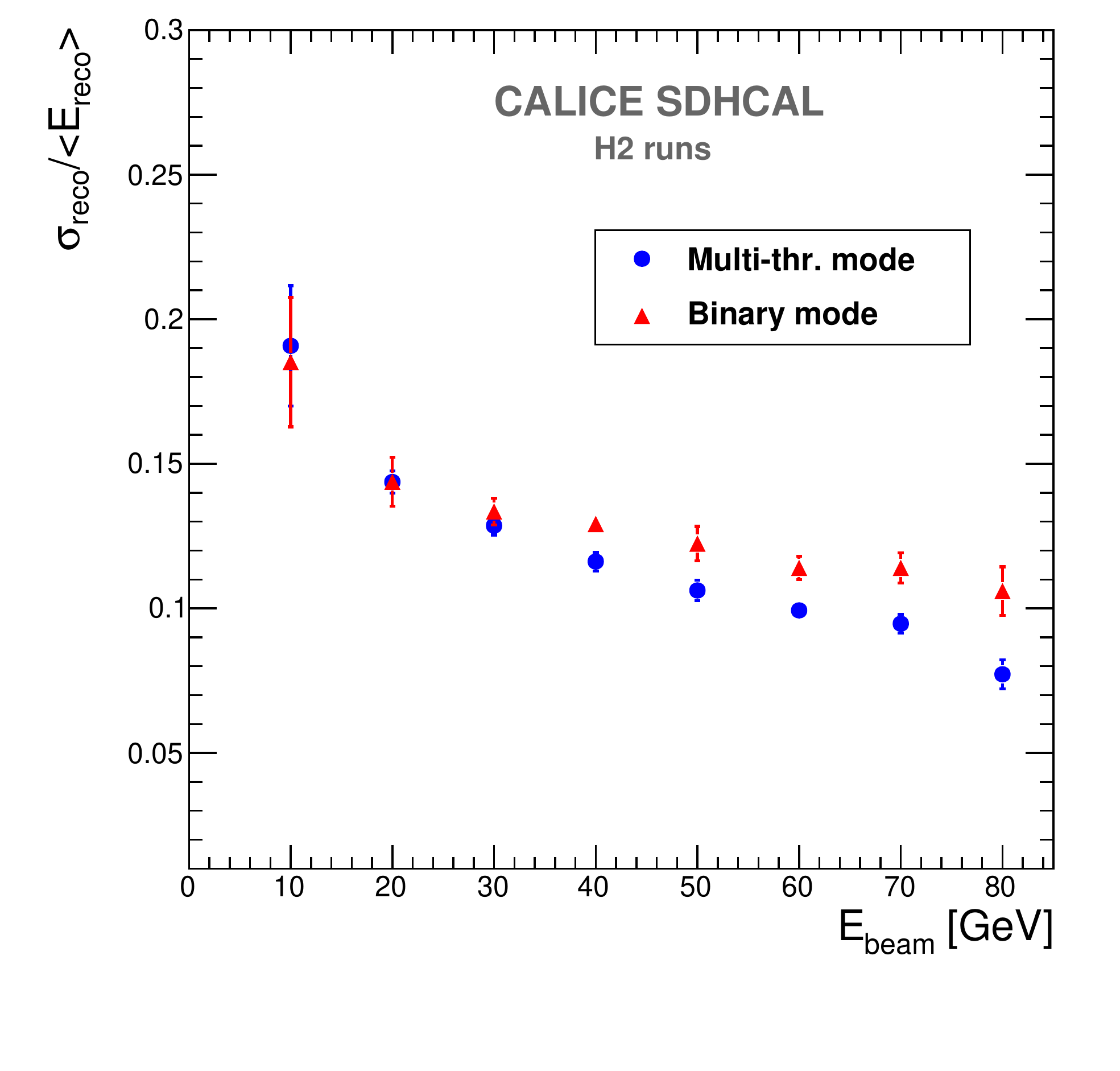}
\caption{$\frac{\sigma_{{\mathrm{reco}}}}{<E_{\mathrm{reco}}>}$ is the relative resolution of the reconstructed  hadron energy as a function of the beam energy of the  2012 H6 (left) and the 2012 H2  (right) data.
For the red triangles graph, the reconstructed energy is computed using only the total number of hits (binary mode).
For the blue circles graph, the reconstructed energy is computed using the three thresholds information (multi-threshold mode).
For both modes, the energy is reconstructed using quadratic functions of the total number of hits.}
\label{fig.Dhcal-Sdhcal}
\end{center}
\end{figure}

\begin{figure}[htp]
\begin{center}
\includegraphics[height=0.3\textheight]{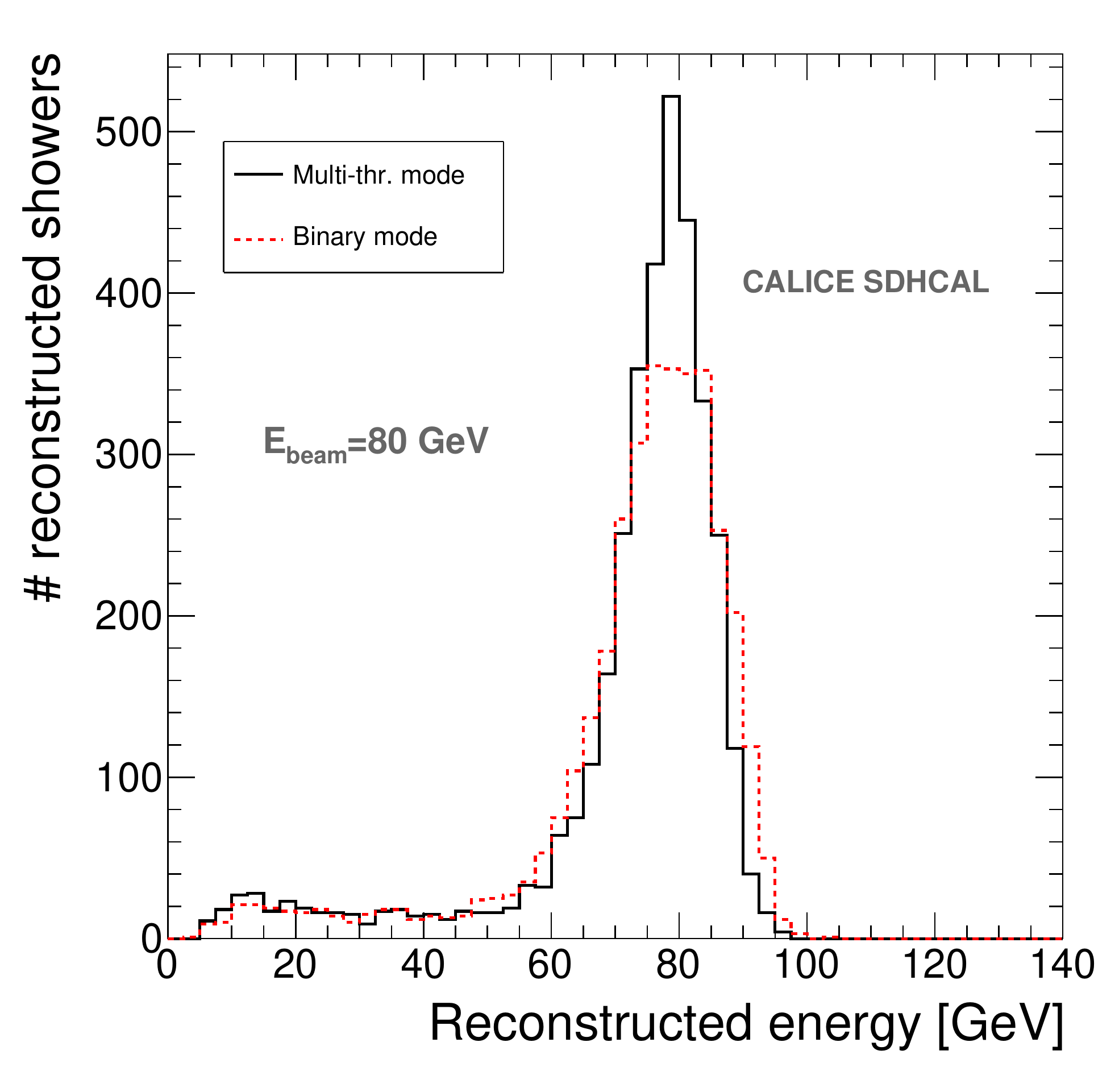}

\includegraphics[height=0.3\textheight]{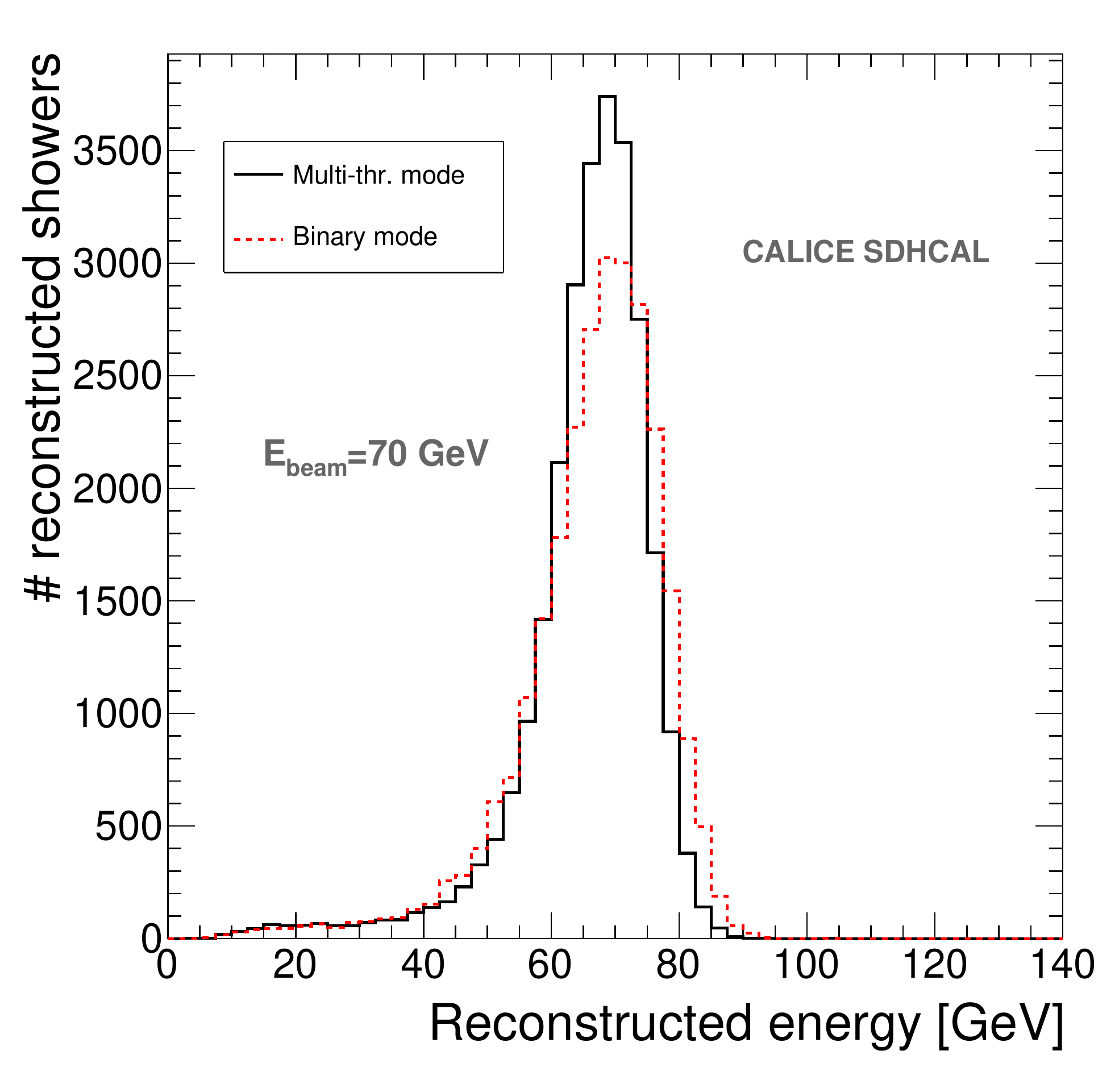}

\includegraphics[height=0.3\textheight]{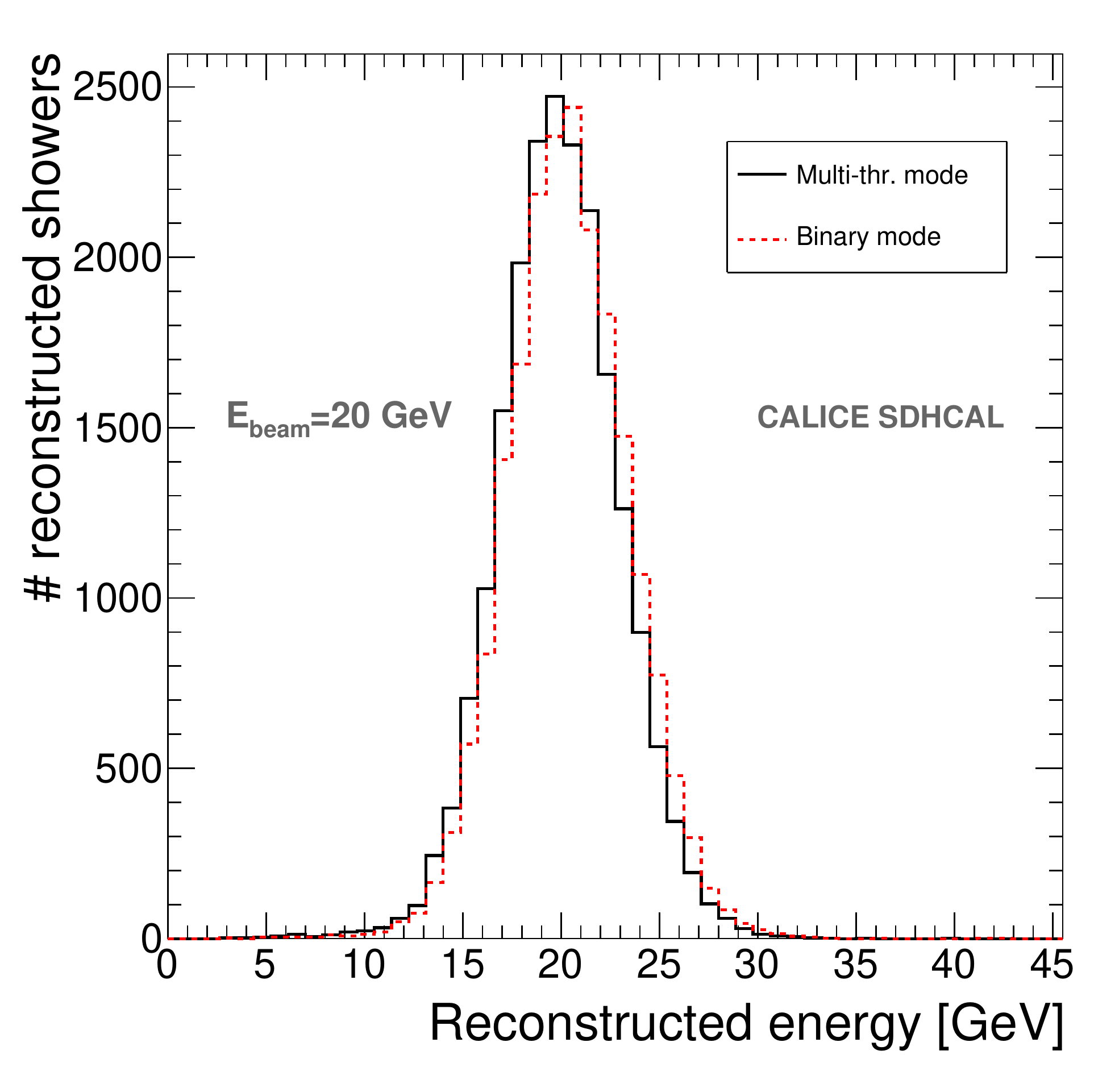}
\caption{Distribution of the reconstructed energy with the SDHCAL binary mode (red dashed line), and with the SDHCAL multi-threshold mode (solid black line) for pions of 80~GeV (top), 70~GeV (middle) and 20~GeV (bottom) of the 2012 H2 data. The tail at low energy observed in the  70 and 80 GeV distributions is due essentially to the remaining radiating muons.}
\label{fig.Dhcal-Sdhcal-distribution}
\end{center}
\end{figure}

Energy reconstruction algorithms using weights that are polynomial functions of the total number of hits have been developed both for the binary and the 
multi-threshold modes of our prototype. Although these algorithms restore linearity over a large energy range, the energy resolution achieved with the  multi-threshold mode was found to be better than that obtained with the binary mode for energies higher than 30~GeV. 
A direct comparison of the two results is shown in Fig.~\ref{fig.Dhcal-Sdhcal}. 
To further compare the two results, the reconstructed energy distribution is shown for both modes in Fig.~\ref{fig.Dhcal-Sdhcal-distribution} 
for 80~GeV, 70~GeV and 20~GeV pions.
At 70~GeV and 80~GeV, the difference of the resolution obtained with the two modes is significant as can be seen from the energy distributions. 


\subsection{Systematic uncertainties}\label{SYS}
Statistical and systematical uncertainties were computed and included in the linearity and energy resolution results presented previously.  The following sources of systematics were included:
\begin{itemize}
\item For all energy points of the different runs, a Gaussian fit  was performed as well as a CB one. The difference of the two results was  used as a source of uncertainty.  This  takes  into account the deviation from an exact Gaussian shape of the studied distributions.
\item Each of the different selection criteria was varied by  an arbitrary  5\% in both directions with respect to the nominal values when this is possible. For each energy point the fitting procedure was then applied to estimate the energy. The maximal deviation with respect to the nominal result  was considered as a systematic uncertainty. The contributions of all the cuts were then added quadratically.
\item The effect of the selection criteria  on the energy reconstruction was estimated using the simulation samples. The relative difference of the estimated energy before and after the selection  was computed and used as an additional source of systematic uncertainties. 
\item The differences of linearity and energy resolution obtained by applying the spill time correction method described in Sect.~\ref{CORRECTION} was also included as an additional uncertainty.
\item Finally a 1\% uncertainty on the  beam energy was added~\cite{BEAM}. This takes into account  the difference of  energies of pions and protons having the same momentum.
\end{itemize}

At low energy, the first three sources are of the same order and  form the main contribution.  At high energy, the first and the third source of systematics, albeit reduced,  continue to be relatively significant  but not the first one while the  fourth source of systematics becomes the largest representing half of the contribution to the total systematic uncertainty.  The contribution of the fifth source was found to be a factor of 10 smaller than the others for all the runs. 

 Although the statistical uncertainties were found to be negligible for almost all the data runs with respect to the systematical ones, contributions of the different uncertainties  were added quadratically. The results are summarized in Tables  \ref{tabA} and \ref{tabB}.


\section{Conclusions}
The data  obtained with the SDHCAL technological prototype during the 2012 test beam runs using the triggerless, power-pulsing mode are analyzed.  Algorithms to linearize the calorimeter response are developed. They result in an energy  response with a  4--5\% deviation from linearity when applied to the raw data over a wide  energy range (5-80 GeV).  The resolution associated to the  linearized energy response  of the same selected data sample is also estimated in both the binary and the multi-threshold modes.  
The multi-threshold capabilities of the SDHCAL clearly improve the resolution at high energy  ($>$30~GeV). 
This relative improvement which reaches 30\% at 80~GeV is probably related to a better treatment of the saturation effect  thanks to the information provided by the second and third thresholds.  The energy resolution reaches  7.7\% at 80~GeV. \\

Finally we think that the exploitation of the topological information provided by such a high-granularity calorimeter to account for  saturation and leakage effects in an appropriate way as well as the application of an  electronic gain correction to improve on the calorimeter response uniformity are likely  to improve the hadronic energy estimation and should be investigated in future works. 

\section{acknowledgements}

We would like to thank the CERN-SPS staff for their availability and precious help during the two beam test periods.   We would like to acknowledge the important support provided by the  F.R.S.-FNRS, FWO (Belgium), CNRS and ANR (France), SEIDI and CPAN (Spain). This work was also supported by the Bundesministerium f\"{u}r Bildung und  Forschung (BMBF), Germany; by the Deutsche Forschungsgemeinschaft  (DFG), Germany; by the Helmholtz-Gemeinschaft (HGF), Germany; by  the Alexander von Humboldt Stiftung (AvH), Germany. 

\vspace{5mm}
\begin{table}
\begin{center}
\begin{tabular}{|c|c|c|c|c|} \hline
  Energy (GeV) & H6 runs     & H2  runs \\\hline
  5	&$0.231 \pm 0.039$ &   \\\hline
  7.5   &$0.225 \pm 0.016$ &   \\\hline
  10    &$0.204 \pm 0.016$ &$0.185 \pm 0.022$  \\\hline
  15    &$0.169 \pm 0.004$ &   \\\hline
  20    &$0.151 \pm 0.005$ &$0.144 \pm 0.008$  \\\hline
  25    &$0.146 \pm 0.002$ &   \\\hline
  30    &$0.135 \pm 0.003$ &$0.133 \pm 0.004$  \\\hline
  40    &$0.128 \pm 0.004$ &$0.129 \pm 0.003$  \\\hline
  50    &$0.119 \pm 0.005$ &$0.122 \pm 0.006$  \\\hline
  60    &$0.105 \pm 0.003$ &$0.114 \pm 0.004$  \\\hline
  70    &$0.103 \pm 0.005$ &$0.114 \pm 0.005$  \\\hline
  80    &$0.095 \pm 0.003$ &$0.106 \pm 0.008$  \\\hline
\end{tabular}
\caption{ List of  the  relative resolutions $\frac{\sigma_{{\mathrm{reco}}}}{<E_{\mathrm{reco}}>}$ observed and associated uncertainties for binary mode.}
\label{tabA}
\end{center}
\end{table}

\begin{table}
\begin{center}
\begin{tabular}{|c|c|c|c|c|} \hline
\end{tabular}
\begin{tabular}{|c|c|c|c|c|} \hline
  Energy (GeV) & H6 runs     & H2 runs \\\hline
  5	&$0.241 \pm 0.039$ & \\\hline
  7.5   &$0.235 \pm 0.022$ & \\\hline
  10    &$0.211 \pm 0.013$ &$0.191 \pm 0.021$  \\\hline
  15    &$0.173 \pm 0.004$ & \\\hline
  20    &$0.149 \pm 0.003$ &$0.144 \pm 0.004$  \\\hline
  25    &$0.142 \pm 0.004$ &  \\\hline
  30    &$0.129 \pm 0.003$ &$0.129 \pm 0.003$  \\\hline
  40    &$0.118 \pm 0.004$ &$0.116 \pm 0.003$  \\\hline
  50    &$0.107 \pm 0.004$ &$0.106 \pm 0.003$  \\\hline
  60    &$0.093 \pm 0.003$ &$0.099 \pm 0.003$  \\\hline
  70    &$0.087 \pm 0.003$ &$0.095 \pm 0.003$  \\\hline
  80    &$0.075 \pm 0.003$ &$0.077 \pm 0.005$  \\\hline
\end{tabular}
\caption{ List of the relative resolutions $\frac{\sigma_{{\mathrm{reco}}}}{<E_{\mathrm{reco}}>}$ observed and associated uncertainties for multi-threshold mode.}
\label{tabB}
\end{center}
\end{table}


\clearpage



\end{document}